\documentclass[prd,aps,twocolumn,a4paper,floatfix]{revtex4}

\usepackage{graphicx,psfrag}
\usepackage{amsmath,amsfonts,amssymb,amsthm}
\usepackage[mathscr]{euscript}
\usepackage{hyperref}

\usepackage{float}

\usepackage{upgreek}

\usepackage{bbm}

\newcommand{\non}{\nonumber}

\newcommand{\p}{\partial}
\newcommand{\Lie}{\mathcal{L}}

\newcommand{\Kn}{K}

\newcommand{\vhat}{\hat{v}}
\newcommand{\Wcs}{W_{c_{s}}}
\newcommand{\cs}{c_{s}}

\newcommand{\pb}{\bot{b}}

\newcommand{\gamN}{{}^{\textrm{\tiny{(N)}}}\!\gamma}
\newcommand{\gamu}{{}^{\textrm{\tiny{(u)}}}\!\gamma}
\newcommand{\gB}{\mathbbmss{g}}
\newcommand{\gBi}{(\mathbbmss{g}^{-1})}
\newcommand{\gBN}{{}^{\textrm{\tiny{(N)}}}\!\mathbbmss{g}}
\newcommand{\gBNi}{{}^{\textrm{\tiny{(N)}}}\!(\mathbbmss{g}^{-1})}

\newcommand{\gBui}{{}^{\textrm{\tiny{(u)}}}\!(\mathbbmss{g}^{-1})}

\newcommand{\s}{{\mathbbmss{s}}}

\newcommand{\qa}{{\mathbbmss{q}_1}}
\newcommand{\qb}{{\mathbbmss{q}_2}}

\newcommand{\shat}{{\hat{\mathbbmss{s}}}}
\newcommand{\qhata}{{\hat{\mathbbmss{q}}_1}}
\newcommand{\qhatb}{{\hat{\mathbbmss{q}}_2}}

\newcommand{\Qa}{{{Q}_1}}
\newcommand{\Qb}{{{Q}_2}}

\newcommand{\qsa}{{{q}_1}}
\newcommand{\qsb}{{{q}_2}}

\newcommand{\orta}{{A}}
\newcommand{\ortb}{{B}}
\newcommand{\ortc}{{C}}

\newcommand{\ortA}{{\mathbbmss{A}}}
\newcommand{\ortB}{{\mathbbmss{B}}}
\newcommand{\ortC}{{\mathbbmss{C}}}

\newcommand{\ortAhat}{{\hat{\mathbbmss{A}}}}
\newcommand{\ortBhat}{{\hat{\mathbbmss{B}}}}
\newcommand{\ortChat}{{\hat{\mathbbmss{C}}}}

\newcommand{\perpq}{{}^{{\tiny{\mathbbmss{q}}}}\!\!\!\perp}
\newcommand{\perpm}{{}^{{\tiny{m}}}\!\!\!\perp}
\newcommand{\perpqs}{{}^{\textrm{{\scriptsize{q}}}}\!\!\!\perp}
\newcommand{\perpQ}{{}^{\textrm{{\tiny{Q}}}}\!\!\!\perp}
\newcommand{\perpQl}{{}^{\textrm{{\tiny{Q}}}_{\lambda}}\!\!\!\perp}

\newcommand{\epsu}{{}^{\textrm{{\tiny{(u)}}}}\!\epsilon}
\newcommand{\epsuS}{{}^{\textrm{{\tiny{(S)}}}}\!\epsilon}

\newcommand{\epsns}{{}^{\textrm{{\tiny{(s)}}}}\!\epsilon}
\begin{document}
%%%%%%%%%%%%%%%%%%%%%%%%%%%%%%%%%%%%%%%%%%%%%%%%%%%%%%%%%%%%%

\title{Revisiting Hyperbolicity of Relativistic Fluids}

\author{Andreas \surname{Schoepe}$^{1}$, David
  \surname{Hilditch}$^{1,2}$ and Marcus \surname{Bugner}$^{1}$}

\affiliation{${}^1$Friedrich-Schiller-Universit\"at Jena, 07743 Jena,
Germany,\\
${}^2$CENTRA, University of Lisbon, 1049 Lisboa, Portugal.}

\date{\today}

\begin{abstract}
Motivated by the desire for highly accurate numerical computations of
compact binary spacetimes in the era of gravitational wave astronomy,
we reexamine hyperbolicity and well-posedness of the initial value
problem for popular models of general relativistic fluids. Our
analysis relies heavily on the dual-frame formalism, which allows us
to work in the Lagrangian frame, where computation is relatively easy,
before transforming to the desired Eulerian form. This general
strategy allows for the construction of compact expressions for the
characteristic variables in a highly economical manner. General
relativistic hydrodynamics, ideal magnetohydrodynamics,
and resistive magnetohydrodynamics are considered in turn. In
the first case, we obtain a simplified form of earlier expressions. In
the second, we show that the flux-balance law formulation used in
typical numerical applications is only weakly hyperbolic and thus does
not have a well-posed initial value problem. Newtonian ideal magnetohydrodynamics is
found to suffer from the same problem when written in 
flux-balance law
form. An alternative formulation, closely related to that of Anile and
Pennisi, is instead shown to be strongly hyperbolic. In the 
final case,
we find that the standard forms of resistive magnetohydrodynamics,
 relying upon a particular
choice of ``generalized Ohm's law'', are only weakly hyperbolic. The
latter problem may be rectified by adjusting the choice of Ohm's law,
but we do not do so here. Along the way, weak hyperbolicity of the
field equations for dust and charged dust is also observed. More
sophisticated systems, such as multifluid and elastic models, are also
expected to be amenable to our treatment.
\end{abstract}

\maketitle

\tableofcontents

%%%%%%%%%%%%%%%%%%%%%%%%%%%%%%%%%%%%%%%%%%%%%%%%%%%%%%%%%%%%%
\section{Introduction}\label{section:Introduction}
%%%%%%%%%%%%%%%%%%%%%%%%%%%%%%%%%%%%%%%%%%%%%%%%%%%%%%%%%%%%%

The first multimessenger observation of gravitational waves from a
binary neutron star merger in Ref.~\cite{AbbAlt17ycollab} marks 
the beginning of
a new era in astronomy. One of the main tasks of numerical relativity
in the coming years will thus be in the {\it accurate}
 construction and
modeling of gravitational waveforms from such spacetimes. 
This work is of course well underway, see, for example, 
Ref.~\cite{LacBerGal16}, but from
the point of view of accuracy suffers from a number of problems in
practice and in principle. As a consequence, numerical relativity
simulations of binary neutron star systems are less accurate than
those of binary black holes. The principal cause of this difference is
presumably the fact of shock formation in the fluid. For this,
sophisticated methods can be employed, see, for example, 
Refs.~\cite{Fon08}
and~\cite{Shi16} for introductions to shock-capturing methods in
numerical relativity, but ultimately there is no avoiding the fact
that a loss of differentiability means forfeiting accuracy. Since
shocks are only expected to occur slightly before merger, we 
may expect
that up until that point the quality of the neutron star data would be
comparable to the vacuum case. This is also not the case, partially
because the additional computational cost of the fluid forces the use
of lower resolution but also because the singular nature of the
fluid equations at the stellar surface, and the numerical hacks to
treat this, serves as a constant source of error.

A mathematically pure approach to the problem would be to first give a
proper analysis of the initial free boundary value problem for the
full system consisting of the Einstein equations coupled to fluid
matter. Unfortunately, such an analysis has not been undertaken 
for the
standard form of the fluid equations in use in numerical relativity,
although see Refs.~\cite{Oli14,Oli17} for interesting work in this
direction. Our view is that this question deserves much more
attention. After all, {\it no} numerical approximation can converge if
the continuum partial differential equation (PDE) 
problem being approximated is ill posed. Such an
analysis is, however, fiendishly difficult, not least because even the
standard expressions for the characteristics of the relativistic Euler
equations are complicated~\cite{FonIbaMar94}. An alternative approach
would be to switch completely to
smoothed-particle hydrodynamics~\cite{Ros14}, although the
mathematically inclined might ask similar questions also in that
context.

Therefore, as a first step in this direction, we reexamine this basic
question of hyperbolicity of general relativistic hydrodynamics (GRHD)
and, relying heavily on insights from the dual-foliation and
slightly more general dual-frame (DF) formalisms, as presented
in Refs.~\cite{Hil15,HilHarBug16,Hil18}, exploit structure 
in the
field equations that simplify the resulting expressions. Consequently
we use the same methodology to give a characteristic analysis of the
standard form of general relativistic magnetohydrodynamics (GRMHD) and
resistive general relativistic magnetohydrodynamics (RGRMHD) as used
in numerical relativity.

The paper is structured as follows. In
Sec.~\ref{section:Motivation}, for motivation, we explain the basic
treatment of the stellar surface in numerical relativity and give
examples of the consequent issues. We then give a brief review of the
relevant PDEs theory and DF formalism as used in the
paper. Section~\ref{section:GRHD} contains our hyperbolicity analysis
of GRHD, and Sec.~\ref{section:GRMHD} contains that of GRMHD. In
Sec.~\ref{section:RGRMHD}, we investigate the hyperbolicity of
RGRMHD. We conclude in Sec.~\ref{section:Conclusion}. We work
in~$3+1$ dimensions; geometric units with~$c=G=1$ and the summation
convention are used throughout. The calculations were performed
primarily with xTensor for {\it Mathematica}~\cite{xAct_web_aastex};
 our notebooks are available in Ref.~\cite{HilWebsite_aastex}.

%%%%%%%%%%%%%%%%%%%%%%%%%%%%%%%%%%%%%%%%%%%%%%%%%%%%%%%%%%%%%
\section{Motivation and Theory overview}
\label{section:Motivation}
%%%%%%%%%%%%%%%%%%%%%%%%%%%%%%%%%%%%%%%%%%%%%%%%%%%%%%%%%%%%%

%%%%%%%%%%%%%%%%%%%%%%%%%%%%%%%%%%%%%%%%%%%%%%%%%%%%%%%%%%%%%
\subsection{Stellar surfaces}\label{section:SF}
%%%%%%%%%%%%%%%%%%%%%%%%%%%%%%%%%%%%%%%%%%%%%%%%%%%%%%%%%%%%%

\begin{figure}[t!]
\centering
\includegraphics[width=0.48\textwidth]{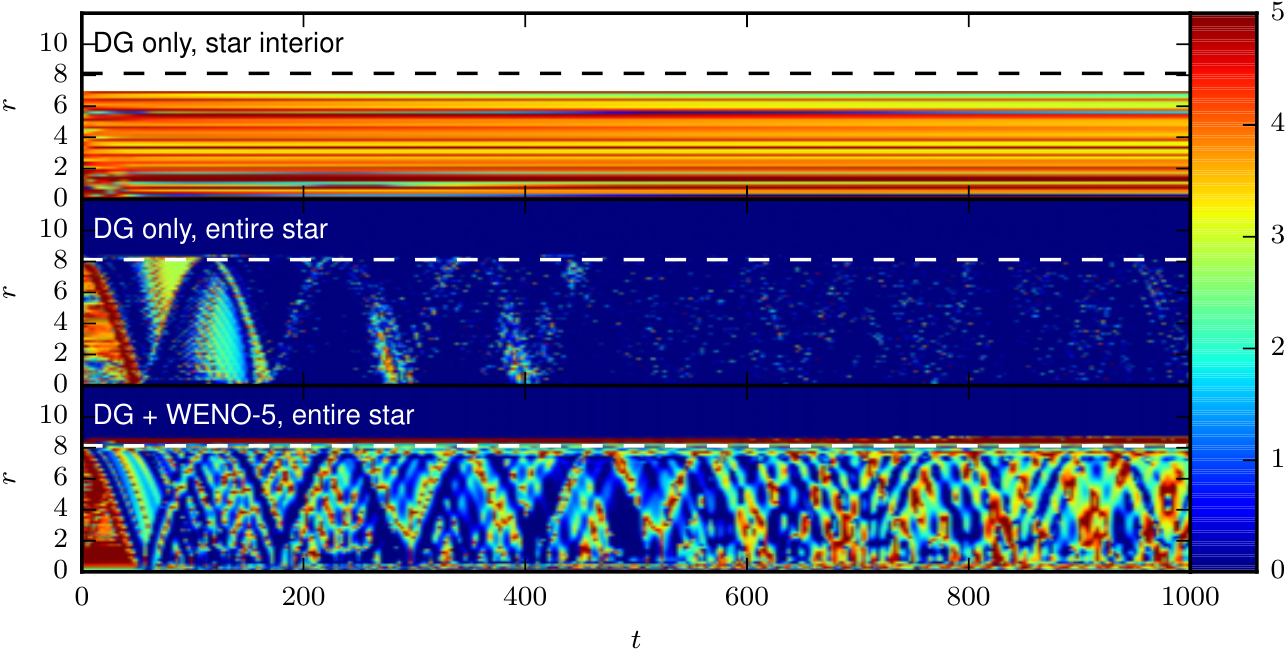}
\caption{Pointwise convergence order (color coding) for TOV star
  simulations with a DG scheme.  Top: Only the interior of the star is
  simulated with a pure DG method and analytic outer boundary
  conditions. The stellar surface (dashed line) is not inside the
  numerical domain.  Middle: Realistic setup of the entire star
  including its surface. It is surrounded by a low density atmosphere
  $\rho_{\rm atm} = 10^{-8} \rho_{\rm max}$.  A pure DG method is used
  for the simulation.  Bottom: Realistic setup of the entire star
  including its surface. The DG method is extended by a WENO-5
  limiting procedure.}
\label{fig:TOV_1D}
\end{figure}

In most numerical approaches for the treatment of relativistic
hydrodynamics, the ``Valencia'' formulation~\cite{BanFonIba97} of the
governing equations is
employed~\cite{FonMilSue98,BaiHawMon04,ThiBerBru11,RadRezGal14}. This
formulation is based on the use of two sets of variables: the
primitive variables, such as the rest mass density~$\rho$, the
pressure~$p$, and the fluid three-velocity~$v^i$, and 
the corresponding
conserved variables~\cite{Fon08}. In practice, the flux-balance law
PDE for the latter set is used for the time evolution. However, the
primitive variables are also required for the flux calculation. The
conserved variables can be expressed as simple functions of the
primitives, whereas the inverse is usually done by a numerical root
finding procedure~\cite{MarMul96,FonMilSue98}. A fundamental problem
of this approach is that this mapping is singular for~$\rho
\rightarrow 0$. Therefore, a low density ``atmosphere'' is introduced
as a threshold to avoid~$\rho = 0$ in numerical schemes. Typically,
this floor value is chosen to be around~$8$--$12$ orders of magnitude
smaller than the maximum density of the star. Although an artificial
atmosphere allows robust simulations of various neutron star setups,
it does not constitute a satisfactory solution to the underlying
problem.  Furthermore, an artificial atmosphere poses a new problem
for high-order schemes. In Fig.~\ref{fig:TOV_1D}, the convergence
results from the simulation of a single stationary, nonrotating
neutron star [Tolman-Oppenheimer-Volkoff (TOV) solution] are shown.

%%%%%%%%%%%%%%%%%%%%%%%%%%%%%%%%%%%%%%%%%%%%%%%%%%%%%%%%%%%%%
%%%%%%%%%%%%%%%%%%%%%%%%%%%%%%%%%%%%%%%%%%%%%%%%%%%%%%%%%%%%%

In this simulation, a discontinuous Galerkin (DG) method of polynomial
order $N=3$ is employed. For the top panel result, only the star
interior with analytical outer boundary conditions was evolved. Almost
perfect pointwise fourth-order convergence can be observed, as
expected. However, if stellar surface and artificial atmosphere are
added as in a realistic simulation of the entire star, the convergence
order rapidly decreases (middle panel), leaving behind no clear
systematic behavior. The application of shock-capturing techniques,
like the weighted essentially non-oscillatory (WENO) limiting
methodology~\cite{LiuOsh94,QiuShu05,BugDieBer15}, partially cures this
problem (bottom panel), and convergence in the~$L_1$ norm does look
somewhat better, although still not satisfactory. In any case, this
strategy can only be seen as a workaround, which is clearly
restricting the potential of high-order methods. It is possible that
with a proper analysis it will turn out that full neutron star
solutions have only a very limited level of regularity and that
high-order schemes will never be of huge use in this context. In any
case, it would be desirable to know so, since then the focus for
developing numerical methods could be placed squarely on obtaining at
least low-order convergence while maintaining perfect scalability.

As mentioned in the Introduction, such a ``proper analysis'' would
require a treatment of the general relativistic initial free boundary
value problem for the fluid models treated in numerical
relativity. Presently we are unable to do so, in part because of the
algebraic complexity of the expressions involved in even the simplest
hyperbolicity analysis of these models. This motivates us in what
follows to revisit that question and look for structure in the
equations that may not have been spotted or used in the past.

%%%%%%%%%%%%%%%%%%%%%%%%%%%%%%%%%%%%%%%%%%%%%%%%%%%%%%%%%%%%%
\subsection{PDE analysis}\label{section:PDEana}
%%%%%%%%%%%%%%%%%%%%%%%%%%%%%%%%%%%%%%%%%%%%%%%%%%%%%%%%%%%%%

In this subsection, we introduce our notation and explain the key
points in showing whether or not a system of PDEs is strongly
hyperbolic. We are concerned purely with first-order PDE systems. The
statements are taken primarily from Ref.~\cite{GusKreOli95}, with only
slight adjustment for our needs.

%%%%%%%%%%%%%%%%%%%%%%%%%%%%%%%%%%%%%%%%%%%%%%%%%%%%%%%%%%%%%
\subsubsection{Well-posedness of hyperbolic equations}
%%%%%%%%%%%%%%%%%%%%%%%%%%%%%%%%%%%%%%%%%%%%%%%%%%%%%%%%%%%%%

We start by considering a quasilinear system of evolution PDEs, with
given time coordinate~$t$, of the form,
\begin{align}
 \p_t\mathbf{U}=\mathbf{A}^p(x^\mu,\mathbf{U})\p_p\mathbf{U}
 +\boldsymbol{\mathcal{S}}(x^\mu,\mathbf{U})\,,\label{equation:PDEsystem}
\end{align}
where in this subsection~$p$ stands for a spatial component index. We
call~$\mathbf{U}$ the state vector,
and~$\mathbf{A}^p(\mathbf{U},x^\mu)$, the coefficient matrices of the
spatial derivatives, are referred to as the principal matrix, although
these are a number of matrices equal to the spatial dimensionality. 
 The
initial value problem (IVP) for~\eqref{equation:PDEsystem} is called
well posed if it admits a unique solution that depends continuously,
in a suitable norm, on the initial data. The particular norm will not
concern us in the present work. The source
vector~$\boldsymbol{\mathcal{S}}(x^\mu,\mathbf{U})$ contains all 
nonprincipal terms. These terms will not contribute to our PDE
 analysis
whatsoever and, when they are included, are present only for
completeness. From now on, the dependence of the principal matrix on
both the solution and the coordinates~$x^\mu$ will be suppressed in
our notation. Let~$s_i$ be a spatial $1$-form normalized so
that~$(m^{-1})^{ij}s_is_j=1$, with~$(m^{-1})^{ij}$ an arbitrary
symmetric uniformly positive definite matrix which is permitted to
depend upon the solution. Contracting the principal matrix with~$s_i$,
we call the resulting matrix
\begin{align}
\mathbf{P}^s\equiv \mathbf{A}^s=\mathbf{A}^ps_p\,,
\end{align}
the principal symbol of the PDE
system~\eqref{equation:PDEsystem} (in the~$s_i$-direction). At each
point in spacetime, the system~\eqref{equation:PDEsystem} is called:
\begin{itemize}
\item \textit{weakly hyperbolic}, if for each such~$s_i$ the
  eigenvalues of $\mathbf{P}^s$ are real;
\item \textit{strongly hyperbolic}, if the system is weakly hyperbolic
  and for each such~$s_i$ the principal symbol has a complete set of
  eigenvectors written as columns in a matrix~$\mathbf{T}_s$ and there
  exists a constant~$K>0$, independent of $s_i$, such that,
  \begin{align}
    |\mathbf T_s|+|\mathbf T_s^{-1}| \leq K;
    \label{equation:regularitycondition}
  \end{align}
\item \textit{strictly hyperbolic}, if the system is weakly hyperbolic
  and if for each~$s_i$ the eigenvalues are distinct;
\item \textit{symmetric hyperbolic}, if there exists a symmetric
  positive definite {\it symmetrizer}~$\mathbf{H}$, independent
  of~$s_i$, such that $\mathbf{H}\,\mathbf{A}^p$ is symmetric for
  each~$p$.
\end{itemize}
Note that if the eigenvectors depend continuously on~$s^i$ then
condition~\eqref{equation:regularitycondition}, which will typically
be the case in physical systems, with the matrix norm~$|\cdot|$ is
automatically fulfilled. In that case, proving strong hyperbolicity at
a point requires then showing that the principal
symbol~$\mathbf{P}^s$ has only real eigenvalues and a complete set of
eigenvectors; i.e.~$\mathbf{P}^s$ is diagonalizable. If a system is
strictly and/or symmetric hyperbolic it is also strongly
hyperbolic~\cite{GusKreOli95,Hil13}. Since the principal symbol is
solution dependent, we note that the precise level of hyperbolicity is
too. For linear constant coefficient problems, strong hyperbolicity is
equivalent to well-posedness of the IVP. In the more general case,
strong hyperbolicity at each point is a necessary condition for
well-posedness; additional smoothness conditions are needed to
guarantee well-posedness. We are interested in the present study in
establishing hyperbolicity of relativistic fluid models in an
efficient manner.

%%%%%%%%%%%%%%%%%%%%%%%%%%%%%%%%%%%%%%%%%%%%%%%%%%%%%%%%%%%%%
\subsubsection{Characteristic variables}
%%%%%%%%%%%%%%%%%%%%%%%%%%%%%%%%%%%%%%%%%%%%%%%%%%%%%%%%%%%%%

Given a strongly hyperbolic systems in the form
of~\eqref{equation:PDEsystem} with principal symbol~$\mathbf{P}^{s}$
and matrix of right eigenvectors~$\mathbf T_s$, the diagonalized form
of~$\mathbf{P}^s$ with its eigenvalues on the diagonal is given by
\begin{align}
\boldsymbol{\Lambda}^s=\mathbf T_s^{-1}\mathbf{P}^{s} \mathbf T_s\,.
\end{align} 
We introduce the orthogonal projector to~$s_i$, that
is,~${\perpm^j{}_i}=\delta^j{}_i-(m^{-1})^{jk}s_ks_i$, and, in this
subsection, use capital letters~$A,B,C$ to denote projected component
indices. We call the components of the transformed state
vector~$d_\mu\mathbf{\hat{U}}=\mathbf{T}^{-1}_s\p_\mu\mathbf{U}$ the
characteristic variables in direction~$s_i$. The~$d$ symbol here
symbolizes the fact that the matrix~$\mathbf{T}^{-1}_s$, which is
generally both position and solution dependent, is {\it not} to be
commuted with the partial derivative. In practice, we may think of the
characteristic variables as being constructed from perturbations to
the solution. When presenting them, we will employ a notation
like~$\delta\varphi$ to denote some derivative of a
component~$\varphi$ of state vector. The characteristic variables have
the property that they satisfy particularly simple equations of motion
if we ignore derivatives transverse to~$\hat{s}^i=(m^{-1})^{ij}s_j$
and the lower-order source terms,
\begin{align}
  d_t\mathbf{\hat{U}}=\boldsymbol{\Lambda}^sd_{\hat{s}}\mathbf{\hat{U}}
  +(\mathbf{T}_s^{-1}\mathbf{A}^A\mathbf
  T_s)d_A\mathbf{\hat{U}}+\mathbf
  {T}^{-1}_s\boldsymbol{\mathcal{S}}\,.
\end{align}
In the linear constant coefficient approximation, dropping the
aforementioned terms leaves just decoupled advection equations
propagating with speeds determined by the eigenvalues
of~$\mathbf{P}^s$.

%%%%%%%%%%%%%%%%%%%%%%%%%%%%%%%%%%%%%%%%%%%%%%%%%%%%%%%%%%%%%
\subsubsection{General relativity with matter}
%%%%%%%%%%%%%%%%%%%%%%%%%%%%%%%%%%%%%%%%%%%%%%%%%%%%%%%%%%%%%

In this paper, we will study different types of matter in full general relativity (GR). We
are interested in solutions to the IVP for the Einstein equations,
\begin{align}
G_{\mu\nu}=8\pi T_{\mu\nu}\,,\label{equation:Einsteinequation}
\end{align}
which contain derivatives up to second order in space and time for the
metric components~$g_{\mu\nu}$ on the left-hand side with the
energy-momentum tensor~$T_{\mu\nu}$ as a source term on the right-hand
side. These equations are supplemented with additional evolution
equations for the matter variables. The latter may be fluid and/or
electromagnetic variables, depending on the physical system under
consideration. To treat the metric variables, we may proceed to use a
first-order reduction and construct a suitably hyperbolic
reformulation of the Einstein equations. In this way, one can 
write the
principal symbol schematically as
\begin{align}
\mathbf{P}^s=\begin{pmatrix} \mathbf{P}_{\text{g}}^s&
\mathbf{P}^s_{\text{g}\times\text{m}}\\ \mathbf{P}^s_{\text{m}\times\text{g}}
&\mathbf{P}_{\text{m}}^s \\
\end{pmatrix},\label{equation:principalsymbolEinstein}
\end{align}
with the principal symbols for the metric~$\mathbf{P}_{\text{g}}^s$
and matter variables~$\mathbf{P}_{\text{m}}^s$. If the evolution
equations for the matter variables contain no second-order derivatives
of the metric, the matrix~$\mathbf{P}^s_{\text{m}\times\text{g}}$ can
be set to zero by replacing first derivatives with reduction
variables. If furthermore the energy-momentum tensor contains no
derivatives of the fluid variables, the equations of motion of
 which we assume to be first-order, we
have~$\mathbf{P}^s_{\text{g}\times\text{m}}=0$, and the statement
above that~$T_{\mu\nu}$ serves as a source term is justified from the
PDEs point of view. In such a case, we may perform the
characteristic analysis separately for~$\mathbf{P}_{\text{g}}^s$
and~$\mathbf{P}_{\text{m}}^s$. Thus, taking a strongly hyperbolic
first-order formulation for the metric variables, one needs only to
study the properties of~$\mathbf{P}^s_{\text{m}}$. In the following,
the index~m will be dropped. We assume such a minimal coupling
throughout the work.

In the following sections, we write the equations of motion in various
forms similar to~\eqref{equation:PDEsystem}, but for convenience
instead of the partial derivative operator~$\p_\mu$, 
we use the spacetime
covariant derivative~$\nabla$, the Lie derivative~$\Lie_n$, and various
other operators to be introduced momentarily. The assumption of
minimal coupling allows us to ignore first derivatives of the metric
that appear in these expressions by implicitly assuming that they are
replaced by the metric reduction variables. This approach is
appropriate for any minimally coupled metric-based theory of gravity.
Note that care is sometimes needed in avoiding violating the
condition, which may render the analysis appropriate only in the
Cowling approximation, in which the metric is simply given and only
the matter variables must be evolved.

%%%%%%%%%%%%%%%%%%%%%%%%%%%%%%%%%%%%%%%%%%%%%%%%%%%%%%%%%%%%%
\subsection{Dual-frame formalism}\label{section:DF}
%%%%%%%%%%%%%%%%%%%%%%%%%%%%%%%%%%%%%%%%%%%%%%%%%%%%%%%%%%%%%

In this subsection, we give a brief review of the DF approach
of Refs.~\cite{Hil15,HilHarBug16}. Since only some quantities 
and relations
of the formalism will be given, Ref.~\cite{Hil15} is required 
reading for
deeper insights and a full understanding of the construction. Note
that, despite the naming of the formalism, we will in fact here 
use two
frames, only one of which defines a coordinate tensor basis.

%%%%%%%%%%%%%%%%%%%%%%%%%%%%%%%%%%%%%%%%%%%%%%%%%%%%%%%%%%%%%
\subsubsection{Index notation} 
%%%%%%%%%%%%%%%%%%%%%%%%%%%%%%%%%%%%%%%%%%%%%%%%%%%%%%%%%%%%%

Throughout the paper, we use the Latin letters~$a$--$e$ as abstract
indices. We also use~$p$ as an abstract index, placing it always on
the spatial derivative appearing on the right-hand side of our 
first-order PDE system. The inverse four-metric~$g^{ab}$ is the
 only object
permitted to raise and lower indices. Greek indices run
from~$0$ to $3$ and denote the components of tensors in the
coordinate basis associated with our
coordinates~$x^{\mu}=(t,x^i)$. Latin indices~$i$--$k$ run
from~$1$ to $3$ and stand for the spatial components in the same
basis. The symbol~$\p_a$ stands for the flat covariant derivative
naturally defined by~$x^\mu$. Indices~$n$, $N$, $u$, $V$, $S$, $\Qa$,
$\Qb$, $s$, $q_1$, $q_2$, $\s$, $\qa$, $\qb$, and~$z$ label
contraction in that slot with~$n^a$ or~$n_a$ and so on,
respectively. We take capital Latin letters~$A$--$C$ as abstract
indices denoting appliance of the projection operators~$\perpQ$
or~$\perpqs$, to be defined later.  Similarly, we use
indices~$\ortA$--$\ortC$ and~$\ortAhat$--$\ortChat$ to denote
the application of the projection operator~$\perpq$ over a vector or
dual-vector, respectively. This will become clear later. For
products of different projectors, we write for
instance~$\perpq^a{}_\ortBhat\perpQ^\ortb{}_c\equiv\perpq^a{}_b\perpQ^b{}_c\,$.
Please note that in our notebooks the index notation convention
differs somewhat from that used here (see~{README.txt}
accompanying the notebooks).

%%%%%%%%%%%%%%%%%%%%%%%%%%%%%%%%%%%%%%%%%%%%%%%%%%%%%%%%%%%%%
\subsubsection{Basic idea and objects}
%%%%%%%%%%%%%%%%%%%%%%%%%%%%%%%%%%%%%%%%%%%%%%%%%%%%%%%%%%%%%

The basic idea of the DF approach is to describe a region of spacetime
in two different frames, called the lower- and the uppercase
frames. In this paper, the lowercase frame is Eulerian, that is, a
coordinate frame associated with coordinates~$x^\mu$, as is standard
in numerical relativity. It consists of the four
vectors~$\p_\mu^a$. The associated coframe
is~$\nabla_ax^\mu$. Associated with the lowercase frame is also the
usual future pointing timelike unit normal to spatial slices of
constant~$t$, which is, as usual, denoted by~$n^a$. Tensors orthogonal
to~$n^a$ are called lowercase spatial, or just lowercase. The
 uppercase frame consists of a future pointing timelike unit 
 vector~$N^a$,
which in our application will be identified with the fluid
four-velocity~$u^a$, plus any three linearly independent vector fields
orthogonal to~$N^a$. The latter vectors will be chosen for
convenience. Tensors orthogonal
to~$N^a$ are called uppercase spatial, or just uppercase. We also
 employ a further frame, consisting of~$n^a$ plus
three linearly independent lowercase vectors which are to be
fixed as and when required. The future pointing unit vectors of the
lower- and uppercase frames can be mutually~$3+1$ decomposed as
\begin{align}
n^a=W(N^a+V^a)\,,\qquad N^a=W(n^a+v^a)\,,
\end{align}
with the Lorentz factor~$W=(1-V^aV_a)^{-1/2}=(1-v^av_a)^{-1/2}
=(1+\hat{v}^a\hat{v}_a)^{1/2}$. The vectors~$v^a=\hat{v}^a/W$
and~$V^a$ are the boost vectors orthogonal to~$n^a$ and~$N^a$,
respectively. We also define projection operators by
\begin{align} 
\gamma^b{}_a=\delta^b{}_a+n^bn_a\,,\qquad
\gamN^b{}_a=\delta^b{}_a+N^bN_a\,,
\end{align}
which are obviously orthogonal to their associated normal
vectors,~$\gamma^b{}_an_b=0$, $\gamN^b{}_aN_b=0$. The projection
operator~$\gamma^b{}_a$ becomes the natural induced
metric~$\gamma_{ab}$ on slices of constant~$t$ when both indices are
lowered. We call~$\gamN_{ab}$ and~$\gamma_{ab}$ the upper- and
lowercase spatial metrics, respectively. Projecting the uppercase
spatial metric with~$\gamma^b{}_a$ yields
\begin{align}
\gB_{ab} :=\gamma^c{}_a\gamma^d{}_b \gamN_{cd}
=\gamma_{ab}+\vhat_{a}\vhat_{b}\,,
\end{align}
with inverse
\begin{align}
\gBi^{ab} =\gamma^{ab}-v^av^b\,,
\end{align}
which we call the boost metric and inverse boost metric, respectively.
In the same way but projecting the lowercase projector~$\gamma^b{}_a$
with~$\gamN^b{}_a$, we define the uppercase boost metric and its
inverse,
\begin{align}
\gBN_{a b} :&= \gamN^{ c}_{\ a} \gamN^{ d}_{\ b} \gamma_{c d}=\gamN_{a
  b}+W^2V_{a}V_{b}\,,\non\\ \gBNi^{a b} &=\gamN^{a b}-V^{a}V^{b}.
\end{align}
These various relations are collected in
Table~\ref{tab:threeplusonedecomp}.

%%%%%%%%%%%%%%%%%%%%%%%%%%%%%%%%%%%%%%%%%%%%%%%%%%%%%%%%%%%%%
\begin{table}[t]
\caption{Overview of the relationship between the upper- and
 lowercase quantities.~\label{tab:threeplusonedecomp}}
\centering
\begin{tabular}{l|c|c}
\hline\hline
  & Uppercase & Lowercase\\
\hline
Unit normal   & $N^{a}=W(n^a+v^a)$ & $n^a=W(N^a+V^a)$ \\ 
\ \ \ vector  &  &  \\ 
Boost vector & $V^a$ & $\qquad v^a=\hat{v}^a/W$ \\ 
Lorentz factor& $W=(1-V^aV_a)^{-1/2}$ & $W=(1-v^av_a)^{-1/2}$ \\ 
Projector & $\gamN^a{}_b=g^a{}_b+N^aN_b$ & $\gamma^a{}_b=g^a{}_b+n^an_b$\\ 
Boost metric & $\gBN_{ab} :=\gamN_{ab}+W^2 V_aV_b$ & $\gB_{ab} :=\gamma_{ab}+\vhat_a\vhat_b$\\ 
Inverse boost & $\gBNi^{ab} =\gamN^{ab}-V^aV^b$ & $\gBi^{ab} =\gamma^{ab}-v^av^b$\\
\ \ \  metric &  & \\
\hline
\end{tabular}
\end{table}
%%%%%%%%%%%%%%%%%%%%%%%%%%%%%%%%%%%%%%%%%%%%%%%%%%%%%%%%%%%%%

The vector~$n^a$ is by construction hypersurface orthogonal. The 
lapse function~$\alpha$, shift vector~$\beta^a$, and time
vector~$t^a\equiv \p_t^a$ are defined and related via
\begin{align}
&\alpha=(-\nabla_a t\nabla^a t)^{-\frac{1}{2}}\,,&\qquad
&n^a=-\alpha\nabla^a t\,,\non\\
&\beta^a=\gamma^a{}_bt^b=t^a-\alpha n^a\,.
\end{align}

The spacetime metric can be expanded in the lowercase frame as 
\begin{align}
g_{\mu \nu}=\begin{pmatrix}
-\alpha^2+\beta_k\beta^k & \beta_i\\
\beta_j & \gamma_{ij}
\end{pmatrix}\,,
\end{align}
with inverse
\begin{align}
g^{\mu \nu}=\begin{pmatrix}
-\alpha^{-2}& \alpha^{-2}\beta^i\\
\alpha^{-2}\beta^j & \gamma^{ij}-\alpha^{-2}\beta^i\beta^j
\end{pmatrix}.
\end{align}
The intrinsic covariant derivative operator, defined by projection of
the spacetime covariant derivative acting on spatial tensors, is
denoted by~$D$ and has connection~$\Gamma$. Finally, the extrinsic
curvature~$K_{ab}$ is defined using the standard numerical relativity
sign convention, by
\begin{align}
\Kn_{ab}=-\gamma^c{}_a\nabla_c n_b\,.
\end{align}
In the present work, we need not define any such connection variables
associated with the uppercase frame, since it will be used
exclusively in an algebraic manner to simplify the various matrices
that appear in our analysis. The key idea is that by using the DF
formalism we may express the equations of motion in a Lagrangian frame
that is, for fluid matter, in some sense preferred. This allows us to
exploit structure in the field equations that is otherwise not
obvious and consequently makes the computation necessary to 
analyze hyperbolicity relatively straightforward.

%%%%%%%%%%%%%%%%%%%%%%%%%%%%%%%%%%%%%%%%%%%%%%%%%%%%%%%%%%%%%
\subsubsection{2+1 decomposition} 
%%%%%%%%%%%%%%%%%%%%%%%%%%%%%%%%%%%%%%%%%%%%%%%%%%%%%%%%%%%%%

%%%%%%%%%%%%%%%%%%%%%%%%%%%%%%%%%%%%%%%%%%%%%%%%%%%%%%%%%%%%%
\begin{table}[t]
\caption{Summary of the various unit spatial vectors appearing in our
$2+1$ decomposed equations, plus their associated projection
operators.~\label{tab:spatialvectors}} \centering
\begin{tabular}{l|cc|c}
\hline\hline
& Uppercase & Lowercase & Lowercase \\
\hline
Unit normal  & $N^a$ & $n^a$ & $n^a$  \\
\ \ \ vector & & \\
Spatial 1-form & $S_a$ & $\s_a$ & $s_a$\\ 
Spatial vector &$S^a=\gamN^{ab}S_b$ &$\shat^a=\gBi^{ab}\s_{b}$ & $s^a=\gamma^{ab}s_{b}$\\ 
Norm &$S_aS^a=1$ &$\s_a\gBi^{ab}\s_{b}=1$ & $s_as^a=1$ \\
Orthogonal & $\perpQ_{\ a}^{ b}=$ & $\perpq_{\ a}^{ b}=$ & $\perpqs_{\ a}^{ b}=$ \\
\ \ \ projector  & $\gamN^{ b}{}_{a}-S^bS_a$ & $\gamma_{\ a}^{ b}-\shat^b\s_a$ & $\gamma_{\ a}^{ b}-s^bs_a$\\
Index notation & $\perpQ^\ortb \!\!_{ \orta}$  & $\perpq^\ortB \!\!_{ \ortAhat}$ & $\perpqs^\ortb \!\!_{ \orta}$ \\ 
\hline
\end{tabular}
\end{table}
%%%%%%%%%%%%%%%%%%%%%%%%%%%%%%%%%%%%%%%%%%%%%%%%%%%%%%%%%%%%%

In our analysis, we not only split the equations in a~$3+1$ 
manner against
the future pointing unit timelike vectors~$n^a$ and~$N^a$, but
we furthermore decompose the two spatial projectors~$\gamma^a{}_b$
and~$\gamN^a{}_b$ against various arbitrary unit spatial vectors. The
spatial vectors and associated orthogonal projectors are collected
together in Table~\ref{tab:spatialvectors}. Please note that
$g_{cb}\perpq_{\ a}^{ b}$ is not symmetric. Therefore, we
distinguish between the abstract indices~$\ortA$ and~$\ortAhat$
of~$\perpq$ when applied on a tensor.

%%%%%%%%%%%%%%%%%%%%%%%%%%%%%%%%%%%%%%%%%%%%%%%%%%%%%%%%%%%%%
\subsubsection{PDE notation and characteristic analysis}
\label{subsubsection:PDEnotation}
%%%%%%%%%%%%%%%%%%%%%%%%%%%%%%%%%%%%%%%%%%%%%%%%%%%%%%%%%%%%%

Starting from a four-dimensional formulation of a quasilinear 
first-order system,
\begin{align}
\boldsymbol{\mathcal{A}}^a\p_a\mathbf{U}+\boldsymbol{\mathcal{S}}=0\,,
\label{equation:firstorderquasifourdimform}
\end{align}
we may~$3+1$ split the equations against~$N^a$ or~$n^a$ by
inserting~$\delta^{ b}{}_{ a}=\gamN^{ b}{}_{ a} -N^bN_a=\gamma^{
b}{}_{ a}-n^bn_a$ between~$\boldsymbol{\mathcal{A}}^a$ and the derivative
operator~$\p_a$. We then obtain two potentially equivalent evolution
systems for~$\mathbf{U}$ in terms of~$n^a$ and~$N^a$. These are
\begin{align}
\boldsymbol{\mathcal{A}}^n\p_n\mathbf{U}=&
\boldsymbol{\mathcal{A}}^a\gamma_{\ a}^{ b}\p_b\mathbf{U}+\boldsymbol{\mathcal{S}}\,,\non\\
\boldsymbol{\mathcal{A}}^N\p_N\mathbf{U}=&
\boldsymbol{\mathcal{A}}^a\gamN_{\ a}^{b}\p_b\mathbf{U}+\boldsymbol{\mathcal{S}}\,.
\end{align}
To denote clearly the properties of the matrices, we make the
following definitions,
\begin{align}
\boldsymbol{\mathcal{A}}^n&\equiv\mathbf{A}^{\textrm n},
&\ \ \boldsymbol{\mathcal{A}}^a\gamma^{b}{}_a&\equiv\mathbf{A}^b,
&\ \ \mathbf{A}^bn_b=0\,,\non\\
\boldsymbol{\mathcal{A}}^N&\equiv\mathbf{B}^{\textrm N},
&\ \ \boldsymbol{\mathcal{A}}^a\gamN^{b}{}_a&\equiv\mathbf{B}^b,
&\ \ \mathbf{B}^bN_b=0\,.
\end{align}
Let~$S^{a}$ be an arbitrary unit uppercase spatial vector
against~$N^{a}$, so~$S^{a}S_{a}=1, S^{a}N_{a}=0$, and let~$\s_{a}$ be
an arbitrary lowercase spatial $1$-form against~$n^{a}$,
$\s_{a}n^{a}=0$, normalized against the inverse boost
metric~$\s_{a}\gBi^{ab}\s_{b}=1$. The eigenvalue problems of these
systems in directions~$\s_a$ and~$S_a$ read
\begin{align}
\mathbf{l}^{\text n}_{\lambda}((\mathbf{A}^{\textrm n})^{-1}\mathbf{A}^{\s}
&-\mathbbm{1}\lambda)=0\,,
\non\\
\mathbf{l}^{\text N}_{\lambda_{\text N}}((\mathbf{B}^{\textrm N})^{-1}\mathbf{B}^{S}
&-\mathbbm{1}\lambda_{\text N})=0\,,
\end{align}
with principal symbols~$(\mathbf{A}^{\textrm n})^{-1}\mathbf{A}^{\s}$
and~$(\mathbf{B}^{\textrm N})^{-1}\mathbf{B}^S$, left
eigenvectors~$\mathbf{l}^{\text n}_{\lambda}$ and~$\mathbf{l}^{\text
  N}_{\lambda_{\text N}}$, and eigenvalues~$\lambda$
and~$\lambda_{\text N}$ for lowercase and uppercase, respectively.
 Please
note that we place on the lowercase eigenvalues no index~$\text{n}$.
 The eigenvalues will in general depend on
 the spatial vector chosen to obtain the principal symbol.
 Dependencies on spatial
vectors will sometimes be explicitly indicated by square brackets.

Introducing the 
four-vectors~$\boldsymbol{\phi}^a$,~$\tilde{\boldsymbol{\phi}}^a$,
we could also write
the eigenvalue problems as
\begin{align}
\mathbf{l}^{\text n}_{\lambda}\boldsymbol{\mathcal{A}}^a
\boldsymbol{\phi}_a&=0,\
& \boldsymbol{\phi}_a&=-\lambda n_a+\s_a\,, \non \\
\mathbf{l}^{\text N}_{\lambda_{\text N}}\boldsymbol{\mathcal{A}}^a
\tilde{\boldsymbol{\phi}}_a&=0,\
& \tilde{\boldsymbol{\phi}}_a&=-\lambda_{\text N} N_a+S_a.
\label{equation:EigenvalueProblemCovPhi}
\end{align}

%%%%%%%%%%%%%%%%%%%%%%%%%%%%%%%%%%%%%%%%%%%%%%%%%%%%%%%%%%%%%
\subsection{Frame independence of strong hyperbolicity}
\label{section:ProofFrameIndepHyp}
%%%%%%%%%%%%%%%%%%%%%%%%%%%%%%%%%%%%%%%%%%%%%%%%%%%%%%%%%%%%%

In Ref.~\cite{Hil15}, it is shown that strong hyperbolicity is
 unaffected by a switch of coordinates, provided that the boost 
 vector is sufficiently
small. Following that result, we will prove that strong hyperbolicity
is independent of the choice of frame provided that a specific
estimate on the boost vector is satisfied. This estimate will depend
 on the maximum eigenvalue of the system. We start with the system
of equations for the state vector~$\mathbf{U}$ in the uppercase
frame,
\begin{align}
\p_N \mathbf{U}=\mathbf{B}^p \p_p \mathbf{U}+
\boldsymbol{\mathcal{S}}\,,\label{equation:SystemUpperFrame}
\end{align}
and suppose that it is strongly hyperbolic there, so that there is a
complete set of (left) eigenvectors in all uppercase spatial
directions.  Expressing~$N_a$ and~$\gamN^b_{\ a}$ in terms of the
lowercase quantities, the same system can be written as
\begin{align}
  W\left(\mathbbm{1}+\mathbf{B}^V\right)& \p_n \mathbf{U}=
  \label{equation:SystemLowerFrame}\\
  &\,\left[\mathbf B^a \left( \gamma_{\ a}^{ p}+\vhat^p V_a\right) -\left(
  \mathbbm{1}+\mathbf{B}^V \right) \vhat^p \right] \p_p
\mathbf{U}+\boldsymbol{\mathcal{S}}\,,\non
\end{align}
where we have to first investigate the invertibility
of~$\mathbf{A}^{\text n}=W\left(\mathbbm{1}+\mathbf{B}^V\right)$. Let the
uppercase boost vector be written as~$V^a=|V|S_V^a$ with
norm~$|V|=(V^aV_a)^{1/2}$ and unit vector~$S_V^a$ in the direction of
$V^a$. Since~$\mathbf{B}^{S_V}$ is diagonalizable with diagonal
form~$\boldsymbol{\Lambda}^{S_V}$, it has a complete set of right
eigenvectors written as columns in the matrix~$\mathbf{T}_{S_{V}}$
and~$\mathbf{T}_{S_{V}}$ is invertible. Performing a similarity
transformation, we obtain
\begin{align}
(\mathbf{T}_{S_{V}})^{-1} \left(\mathbbm{1}+\mathbf{B}^V
\right)\mathbf{T}_{S_{V}} =
\mathbbm{1}+|V|\boldsymbol{\Lambda}^{S_V}\,,
\end{align}
and invertibility of~$\mathbbm{1}+\mathbf{B}^V$ is guaranteed if for
each eigenvalue~$\lambda_{\text N}[S_V^a]$ the inequality
\begin{align}
1+|V|\lambda_{\text N}[S_V^a]>0\,
\label{equation:InequalityLambdaVequation}
\end{align}
for arbitrary unit~$S_V^a$ holds. This condition will be guaranteed by
assumption in the proof that follows.

Let~$S^a$ be an arbitrary unit uppercase spatial vector. The
eigenvalue problem in direction $S^a$ corresponding to the PDE system
in~\eqref{equation:SystemUpperFrame} in the uppercase frame can be written
as
\begin{align}
\mathbf{l}_{\lambda_{\text N}}^{\text N} \left( \mathbf{B}^S-
\mathbbm{1} \lambda_{\text N}[S^a] \right)=0\,,
\label{equation:EVproblemupperframe1}
\end{align}
where~$\mathbf{l}_{\lambda_{\text N}}^{\text N}$ is the uppercase
left eigenvector for the principal symbol~$\mathbf{B}^S$ with
eigenvalue~$\lambda_{\text N}[S^a]$.

The eigenvalue problem for direction~$\s_a$ in the lowercase frame for the
PDE system~\eqref{equation:SystemLowerFrame} may be written as
\begin{align}
\mathbf{l}_{\lambda}^{\text n}
(\mathbbm{1}+\mathbf{B}^V)^{-1}\left[ \mathbf{B}^S -
(\mathbbm{1}+\mathbf{B}^V)(\vhat^{\s}+W \lambda
)\right]=0 \label{equation:EVproblemlowerframe1}
\end{align}
for lowercase left eigenvector~$\mathbf{l}_{\lambda}^{\text n}$ with
eigenvalue~$\lambda$. The associated principal symbol is
\begin{align}
\mathbf{P}^{\s}=\frac{1}{W}\left[ (\mathbbm{1}+\mathbf{B}^V)^{-1}\mathbf{B}^S -
\mathbbm{1}\vhat^{\s}\right],
\end{align}
and the lowercase spatial $1$-form~$\s_a$ is related to the 
uppercase one by~$\s_a=S_a+W^2V^S(N_a+V_a)$; see also
Table~\ref{tab:RelationSpatialVectors}. The projectors given in
Table~\ref{tab:spatialvectors}
satisfy~${\perpq^a\!_b=\gBi^{a c}\,\perpQ_{cd}\gamma^d{}_b}$.

%%%%%%%%%%%%%%%%%%%%%%%%%%%%%%%%%%%%%%%%%%%%%%%%%%%%%%%%%%%%%
\begin{table}
\caption{The relationship between upper- and lowercase unit spatial
vectors.~\label{tab:RelationSpatialVectors}}
\begin{tabular}{l|cc}
  \hline\hline
  & Uppercase & Lowercase\\
  \hline
Unit normal vector  & $N^a$ & $n^a$   \\
Boost vector & $V^a$ & $v^a$\\
Spatial vector &$S^a=\gamN^{ab}S_b$ &$\shat^a=\gBi^{ab}\s_{b}$ \\
Spatial 1-form &$S_a=$ & $\s_a=$ \\ 
  & $\s_a +v^{\s}n_a$ & $S_a+W^2V^S(N_a+V_a)$ \\
& $\gamN_{a b} \gBi^{bc}\s_c$ & $\gamma^{ b}{}_aS_b$\\
\hline
\end{tabular}
\end{table}
%%%%%%%%%%%%%%%%%%%%%%%%%%%%%%%%%%%%%%%%%%%%%%%%%%%%%%%%%%%%%

Introducing the modified lowercase left
eigenvector~$\mathbf{L}_{\lambda}^{\text
  n}=\mathbf{l}_{\lambda}^{\text n} (\mathbbm{1}+\mathbf{B}^V)^{-1}$
and collecting terms of~$\mathbf{B}$, we rewrite
Eq.~\eqref{equation:EVproblemlowerframe1} as
\begin{align}
\mathbf{L}_{\lambda}^{\text n}\left[\mathbf{B}^{S- V(\vhat^{\s}+W \lambda )}
- \mathbbm{1}(\vhat^{\s}+W \lambda )\right]=0\,.
\end{align}
By defining the new uppercase unit spatial vector 
\begin{align}
&S_{\lambda}^{ a}[S^b,\lambda]:=\frac{1}{N}\left( S^a- V^a(\vhat^{\s}+W
  \lambda)\right)\,,
\end{align}
with normalization,
\begin{align}
N&=\left[\left(
S^a- V^a(\vhat^{\s}+W \lambda)\right)\left( S_a- V_a(\vhat^{\s}+W
\lambda)\right) \right]^{1/2}\non \\
&=\sqrt{W^2(\lambda-W V^S)^2+1+(V^S)^2W^2-\lambda^2},\non\\
&=\sqrt{W^2(\lambda+ v^{\s})^2+1+(v^{\s})^2-\lambda^2}\,,
\label{equation:frameconnectionSvectors}
\end{align}
we finally arrive at the eigenvalue problem
\begin{align}
\mathbf{L}_{\lambda}^{\text n}\left[\mathbf{B}^{S_{\lambda}}-
\mathbbm{1}\frac{1}{N} (\vhat^{\s}+W \lambda)\right]=0\,,
\label{equation:frameconnectionEigenvvalueProblem}
\end{align} 
for the redefined lowercase left
eigenvector~$\mathbf{L}_{\lambda}^{\text n}$, principal
symbol~$\mathbf{B}^{S_{\lambda}}$, and eigenvalue~$(\vhat^{\s}+W
\lambda)/N$ in the direction of~ $S_{\lambda}^a$. The
relation~$WV^S=-v^{\s}$ follows by using relations given in
Tables~\ref{tab:threeplusonedecomp}
and~\ref{tab:RelationSpatialVectors}.  The lowercase eigenvalue
problem~\eqref{equation:frameconnectionEigenvvalueProblem} for 
fixed~$\lambda$ is the same
eigenvalue problem as for the uppercase system for eigenvalue
~$(\vhat^{\s}+W \lambda)/N$ in~\eqref{equation:EVproblemupperframe1} 
where the spatial direction~$S^a$ is replaced by~$S_{\lambda}^a$. 
Therefore,
\begin{align}
\frac{1}{N} (\vhat^{\s}+W \lambda) = \lambda_{\text N}[S^a_{\lambda}]
\label{equation:frameconnectionEigenvalues}
\end{align}
must hold. 

Equation~\eqref{equation:frameconnectionEigenvalues} is a strong
result, since we are now able to calculate the lowercase frame
eigenvalues from knowledge of the uppercase results. Nevertheless,
solving for~$\lambda$ may be hard since both~$N$ and~$\lambda_{\text
  N}$ contain polynomials in~$\lambda$. The lowercase left
eigenvector to eigenvalue~$\lambda$ is then simply given by
\begin{align}
\mathbf{l}_{\lambda}^{\text n}[\s_b]
=&\mathbf{l}_{\lambda_{\text N}}^{\text
  N}[S^a_{\lambda}]
\left(\mathbbm{1}+\mathbf{B}^V
\right)\,,\label{equation:frameconnectionLeftEigenvectors}
\end{align}
and the right eigenvectors are given by
\begin{align}
\mathbf{r}_{\lambda}^{\text n}[\s_b]=
\mathbf{r}_{\lambda_{\text N}}^{\text
 N}[S^a_{\lambda}]\,.\label{equation:frameconnectionRightEigenvectors}
\end{align}

The proof is as follows. We know that for arbitrary unit
spatial~$S^a$ the principal symbol $\mathbf P^S$ has:
\begin{itemize}
\item[$(1)$] real eigenvalues~$\lambda_{\text N}$\,,
\item[$(2)$] a complete set of left and right eigenvectors
obeying~$|\mathbf T_S|+|\mathbf T_S^{-1}|\leq K$, where~$\mathbf
T_S$ is the matrix of right (or left) eigenvectors written as
columns (or rows) and~$K$ is independent of~$S^a$.
\end{itemize}
We assume furthermore that:
\begin{itemize}
\item[$(3)$] all uppercase eigenvalues fulfill the inequality
~$1-|\lambda_{\text N}||V|>0$, for all uppercase unit spatial~$S^a$.
\end{itemize}
This assumption automatically guarantees the
condition~\eqref{equation:InequalityLambdaVequation} for the
invertibility of~$\mathbbmss{1}+\mathbf{B}^V$.

%%%%%%%%%%%%%%%%%%%%%%%%%%%%%%%%%%%%%%%%%%%%%%%%%%%%%%%%%%%%%
\paragraph*{The lowercase eigenvalues are real.}
We start by showing that the lowercase system is at least weakly
hyperbolic. By use of~\eqref{equation:frameconnectionEigenvalues}, we
obtain
\begin{align}
\lambda=\frac{W^3V^S(1-\lambda_{\text N}^2)+\lambda_{\text N}
W\sqrt{1+\lambda_{\text N}^2(1/W^2-1+(V^S)^2)}}{W^2(1-\lambda_{\text N}^2
(1-1/W^2))}
\end{align}
for given~$\lambda_{\text N}$. The only danger is that the terms
within the square root are negative, but considering these, we have
\begin{align}
1+\lambda_{\text N}^2(1/W^2-1+(V^S)^2)&=1-\lambda_{\text N}^2(|V|^2-
(V^S)^2)\non\\
&\geq 1-\lambda_{\text N}^2|V|^2>0\,,
\end{align}
where we have used assumptions~$(1)$ and~$(3)$. Therefore, all
 lowercase eigenvalues are real.

%%%%%%%%%%%%%%%%%%%%%%%%%%%%%%%%%%%%%%%%%%%%%%%%%%%%%%%%%%%%%
\paragraph*{The lowercase eigenvectors are linearly independent.}
Take a lowercase eigenvalue~$\lambda$ with algebraic
multiplicity~$k$. Then, by
Eq.~\eqref{equation:frameconnectionEigenvalues}, the corresponding
uppercase eigenvalue~$\lambda_{\text N}[S^a_{\lambda}]$ has also
algebraic multiplicity~$k$. Thus, by assumptions~$(2)$, which ensures
that we can find~$k$ linearly independent eigenvectors to the
associated eigenvalue
problem~\eqref{equation:frameconnectionEigenvvalueProblem}, and~$(3)$,
which guarantees the invertibility of~$\mathbbm{1}+\mathbf{B}^V$, and
the use of Eq.~\eqref{equation:frameconnectionLeftEigenvectors},
we know that we can find~$k$ linearly independent lowercase left
eigenvectors in the eigenspace of~$\lambda$. This statement holds also
for the right eigenvectors. Therefore, the lowercase principal symbol
is diagonalizable.

%%%%%%%%%%%%%%%%%%%%%%%%%%%%%%%%%%%%%%%%%%%%%%%%%%%%%%%%%%%%%
\paragraph*{Show necessary regularity conditions.}
Let us label the left and right eigenvectors and eigenvalues, making duplicates to
account for their multiplicity if necessary, with an index,
writing~$\mathbf{l}_{\lambda_{(i)}}$,~$\mathbf{r}_{\lambda_{(i)}}$,
and~$\lambda_{(i)}$, respectively. Please note that only in this proof indices~$i,j$ label characteristic quantities and do not stand for spatial tensor basis components. We denote~$\mathbf{T}_\s$ as the
matrix of lowercase right vectors, where the~$i$th column
of~$\mathbf{T}_\s$ is~$\mathbf{r}_{\lambda_{(i)}}$. We order so that
the~$i$th row of~$\mathbf{T}_\s^{-1}$
is~$\mathbf{l}_{\lambda_{(i)}}$.
Thus,~$\mathbf{l}_{\lambda_{(i)}}\mathbf{r}_{\lambda_{(j)}}=\delta_{ij}$.
By Eqs. ~\eqref{equation:frameconnectionLeftEigenvectors}
and~\eqref{equation:frameconnectionRightEigenvectors}, we can express
for each~$i$ the lowercase
eigenvectors~$\mathbf{l}_{\lambda_{(i)}},\mathbf{r}_{\lambda_{(i)}}$
as~$\mathbf{l}^{\text N}_{\lambda^{\text
    N}_{(i)}}[S^a_{\lambda_{(i)}}]\left(\mathbbm{1}+\mathbf{B}^V
\right)$ and~$\mathbf{r}^{\text N}_{\lambda^{\text
    N}_{(i)}}[S^a_{\lambda_{(i)}}]$, respectively. The uppercase
principal symbol is diagonalizable by assumption (2), so for each~$i$,
we may extend each such left or right eigenvector with the remaining
linearly independent eigenvectors of the uppercase principal symbol
for spatial vector~$S^a_{\lambda_{(i)}}$. We denote
by~$\mathbf{T}_{S_{\lambda_{(i)}}}$ the matrices of those completed
sets of eigenvectors expanding the chosen~$\mathbf{r}^{\text
  N}_{\lambda^{\text N}_{(i)}}[S^a_{\lambda_{(i)}}]$ (and
$\mathbf{l}^{\text N}_{\lambda^{\text N}_{(i)}}[S^a_{\lambda_{(i)}}]$)
written as columns (rows). The chosen~$i$th right (left) eigenvector
is placed in the~$i$th column (row). By assumption (2), we then have
\begin{align}
|\mathbf{T}_{S_{\lambda_{(i)}}}^{-1}|+|\mathbf{T}_{S_{\lambda_{(i)}}}|
\leq K_{(i)}\label{equation:estmateofTslambdai}
\end{align}
for each~$i$.

Define now the square diagonal quadratic matrix~$\mathbf{D}_{(i)}$,
which has in the~$i$th entry of its diagonal~$1$ and otherwise
zeros,
\begin{align*}
{\mathbf{D}_{(i)}}:=\text{diag}(\underbrace{0,\ldots,0}_{i-1\ \text{times}},1,0,
\ldots,0);\,\sum_{i}\mathbf{D}_{(i)}=\mathbbmss{1}.
\end{align*}
Their norm is~$|\mathbf{D}_{(i)}|=\underset{|\mathbf{y}|=1}{\max}|\mathbf{y}_{(i)}|=1$, where~$\mathbf{y}_{(i)}$ is the $i$th component of~$\mathbf{y}$.
Then, with the above definitions,
\begin{align}
\mathbf T_\s&=\sum_{i}\mathbf{T}_{S_{\lambda_{(i)}}} \mathbf{D}_{(i)},\non\\
\mathbf T_\s^{-1}&=\sum_{i}\mathbf{D}_{(i)} \mathbf{T}^{-1}_{S_{\lambda_{(i)}}}
\left(\mathbbmss{1}+\mathbf{B}^V\right),\label{equation:ExpressTswithTslambda}
\end{align}
and we can give the estimate
\begin{align}
|\mathbf T_\s^{-1}|+|\mathbf T_\s| &\leq \sum_{i}
\left(|\mathbf{T}^{-1}_{S_{\lambda_{(i)}}}||\mathbbmss{1}+\mathbf{B}^V|
+|\mathbf{T}_{S_{\lambda_{(i)}}}|\right)\non \\
&\leq \sum_{i}\left(|\mathbf{T}^{-1}_{S_{\lambda_{(i)}}}|
+|\mathbf{T}_{S_{\lambda_{(i)}}}|\right)\max\{1,|\mathbbmss{1}+\mathbf{B}^V|\}\non\\
&\leq \sum_{i}K_{(i)}\max\{1,|\mathbbmss{1}+\mathbf{B}^V|\}\equiv K.
\end{align}
In the first step, we inserted~\eqref{equation:ExpressTswithTslambda}
for the matrices and used the submultiplicity of the norm. In the
second, we estimated the prefactors and finally, in the last step, 
we used
assumption (2) given by~$\eqref{equation:estmateofTslambdai}$ for
each~$i$. We thus arrive at the
inequality~\eqref{equation:regularitycondition}, which together with
the above properties gives strong hyperbolicity in the lowercase
frame.

%%%%%%%%%%%%%%%%%%%%%%%%%%%%%%%%%%%%%%%%%%%%%%%%%%%%%%%%%%%%%
\paragraph*{Multiplicity and degeneracies.} The definition of
strong hyperbolicity does not require that the multiplicity of the
eigenvalues be constant as the spatial direction is varied. In the
literature on relativistic fluids, special cases in which the
algebraic multiplicity of a particular eigenvalue increases when
looking in particular special directions are called degeneracies of
the system. All such possible degeneracies must be taken into account
in the demonstration of strong hyperbolicity since diagonalizability
of the principal symbol is required in all directions. Note that the
relationship between the occurrence of degeneracies in the uppercase
and lowercase systems is, however, not trivial. The key point is that
when transforming from the lowercase system to the associated
 uppercase eigenvalue
problem~\eqref{equation:frameconnectionEigenvvalueProblem} we consider
the latter only for a fixed eigenvalue. For different eigenvalues, we
naturally assign {\it different} uppercase eigenvalue
problems. Therefore, it may be that, for example, uppercase
degeneracies always occur in pairs, while the same is not true in the
lowercase frame. Indeed, we will see that this is the case for a
particular formulation of GRMHD. The relationship between the
degeneracies plays no role in the foregoing proof of the equivalence
of strong hyperbolicity across the two frames.

%%%%%%%%%%%%%%%%%%%%%%%%%%%%%%%%%%%%%%%%%%%%%%%%%%%%%%%%%%%%%
\paragraph*{Discussion.} All systems we study in relativistic physics
will satisfy, by construction, that the boost velocities~$v_a$ are
always smaller than the speed of light. We will furthermore
immediately reject any equation of state that results in wave speeds,
that is, eigenvalues of the principal symbol, that are greater than
the speed of light. This is reasonable in the current study since we
are concerned exclusively with relativistic fluid models. On the other
hand however, one should not get the false impression that this must
always be the case in relativistic physics. Theories with gauge
freedom, such as the electromagnetism and GR, do admit hyperbolic
formulations with superluminal speeds. In GR, the obvious example of
such a gauge is the popular moving-puncture family. In that case, when
the boost vector becomes too large, uppercase strong hyperbolicity
will not be sufficient to guarantee the same in the lowercase frame,
since the crucial inequality~$|\lambda_{\text N}||V|<1$ can 
be violated. In
fact, since GRMHD also inherits some gauge freedom from the Maxwell
equations, the same could be said for that model. Such subtleties will
not affect us in practice.

%%%%%%%%%%%%%%%%%%%%%%%%%%%%%%%%%%%%%%%%%%%%%%%%%%%%%%%%%%%%%
\subsection{Variable independence of strong hyperbolicity}
\label{section:ProofStrHypVariables}
%%%%%%%%%%%%%%%%%%%%%%%%%%%%%%%%%%%%%%%%%%%%%%%%%%%%%%%%%%%%%

Let~$\mathbf{U}$ be a state vector for which the principal
symbol~$\mathbf{P}_{\mathbf{U}}^s$ is diagonalizable for each unit
spatial $1$-form~$s_a$. Let~$\mathbf{V}$ be another state vector of
the same dimension, the components of which depend smoothly on the components
of~$\mathbf{U}$. Derivatives of the two state vectors are then related
by the Jacobian~$\mathbf J$,
\begin{align}
\p_a\mathbf{V}=\mathbf J \p_a \mathbf{U}\,.
\end{align}
The principal symbol for~$\mathbf V$ is then 
\begin{align}
\mathbf{P}_{\mathbf V}^s=\mathbf{J}\mathbf{P}_{\mathbf U}^s\mathbf{J}^{-1}.
\end{align}
Since this transformation is nothing more than a similarity
transformation, the eigenvalues remain the same, and the (left) right
eigenvectors for~$\mathbf{V}$ are just modified by a matrix
multiplication with the (inverse) Jacobian. Thus, as is well-known,
strong hyperbolicity is independent of the choice of evolved
variables. Note that during the hyperbolicity analysis, one choice of
variables may make the practical computations very much easier than
another.

%%%%%%%%%%%%%%%%%%%%%%%%%%%%%%%%%%%%%%%%%%%%%%%%%%%%%%%%%%%%%
\subsection{Recovering the eigenvalues and eigenvectors of the
lowercase frame}\label{section:RecoverEVandEVecsGeneral}
%%%%%%%%%%%%%%%%%%%%%%%%%%%%%%%%%%%%%%%%%%%%%%%%%%%%%%%%%%%%%

In this subsection, we explain how we use the results of
Sec.~\ref{section:ProofFrameIndepHyp}. As mentioned before, the
uppercase frame will be chosen as the frame of a comoving observer
 with the fluid, so we take
\begin{align}
N^a\equiv u^a,\qquad \gamu^a{}_{ b}=g^a{}_{ b}+u^au_b\,,
\end{align} 
with the four-velocity of the fluid~$u^a$. Despite that we are in the
fluid frame or so-called Lagrangian frame, we never set one of the
boost vectors to zero.

Since we obtain all our results using computer algebra, it is
convenient to introduce a basis to obtain scalar quantities as entries
in the matrices. The various basis vectors are given in Table
\ref{tab:BasisVectors}. Given a spatial vector~$s^a$ of unit magnitude
with respect to some metric, we consider a set~$\{s^a,\qsa^a,\qsb^a\}$
forming a right-handed orthonormal basis with respect to the same
metric.

%%%%%%%%%%%%%%%%%%%%%%%%%%%%%%%%%%%%%%%%%%%%%%%%%%%%%%%%%%%%%
\begin{table}
\caption{Overview of the upper- and lowercase basis
vectors.~\label{tab:BasisVectors}}
\begin{tabular}{l||cc|cc}
  \hline\hline
 & Uppercase & & Lowercase & \\
\hline
Unit normal vector  & $N^a$ & $N^a$ & $n^a$  & $n^a$ \\
Spatial 1-form & $S^{\lambda}_a$  & $S_a$ & $\s_a$ & $s_a$ \\ 
Spatial vector& $S^a_{\lambda}$ & $S^a$ & $\shat^a$ & $s^a$\\
Orthogonal basis& $\Qa_a^{\lambda},\Qb_a^{\lambda}$ & $\Qa_a,\Qb_a$  
& $\qa_a,\qb_a$ & $\qsa_a, \qsb_a$ \\
\ \ \  1-forms & & & & \\
Orthogonal basis & $\Qa^a_{\lambda},\Qb^a_{\lambda}$  & $\Qa^a,\Qb^a$  
& $\qhata^a,\qhatb^a$ & $\qsa^a, \qsb^a$\\
\ \ \ vectors & & & & \\
Normalized$/$&  $\gamu^{ab}$ & $\gamu^{ab}$ & $\gBi^{ab}$ & $\gamma^{ab}$\\
\ \ \ orthogonal via  & & & & \\
\hline
\end{tabular}
\end{table}
%%%%%%%%%%%%%%%%%%%%%%%%%%%%%%%%%%%%%%%%%%%%%%%%%%%%%%%%%%%%%

Let $S^a$ be an arbitrary unit uppercase spatial vector.
Given a strongly hyperbolic system of PDEs in the form 
of Eq.~\eqref{equation:SystemUpperFrame}
with~$N^a\equiv u^a$, we write the principal symbol 
as~${\mathbf P}^S$. We denote the known eigenvalues 
of~${\mathbf P}^S$ by~$\lambda_{\text u}[S^a]$ and the 
known complete set of left
eigenvectors, obtained by Eq.~\eqref{equation:EVproblemupperframe1},
 by $\mathbf{l}_{\lambda_{\text u}}^{\text u}[S^a]$.
Then, the lowercase
eigenvalues are given by
Eq.~\eqref{equation:frameconnectionEigenvalues}, and the 
lowercase left eigenvectors~$\mathbf{l}_{\lambda}^{\text n}$ for
eigenvalue~$\lambda$ are given by
Eq.~\eqref{equation:frameconnectionLeftEigenvectors}, that is, for
a specific choice of a basis,
\begin{align}
\left.\mathbf{l}_{\lambda}^{\text n}\right|_{\s}
=&\left.\mathbf{l}_{\lambda_{\text u}}^{\text
  u}[S^a_{\lambda}]\right|_{\mathbf{S}}
\left(\mathbbm{1}+\left.\mathbf{B}^V
\right|_{\mathbf{S}}\right)\non\\ =& \left.\mathbf{l}_{\lambda_{\text
    u}}^{\text
  u}[S^a_{\lambda}]\right|_{\mathbf{S}_{\lambda}}\mathbf{T}_{\lambda}
\left(\mathbbm{1}+\left.\mathbf{B}^V \right|_{\mathbf{S}}\right)\non
\\ =&\left.\mathbf{l}_{\lambda_{\text u}}^{\text
  u}[S^a_{\lambda}]\right|_{\mathbf{S}_{\lambda}}
\left(\mathbbm{1}+\left.\mathbf{B}^V
\right|_{\mathbf{S}_{\lambda}}\right)\mathbf{T}_{\lambda}\,,
\label{equation:ObteinLeftEigenvectors}
\end{align}
and the lowercase right eigenvectors~$\mathbf{r}_{\lambda}^{\text n}$
are obtainable by
\begin{align}
\left.\mathbf{r}_{\lambda}^{\text n}\right|_{\s}
=&\left.\mathbf{r}_{\lambda_{\text u}}^{\text
  u}[S^a_{\lambda}]\right|_{\mathbf{S}}
\non\\
=&  (\mathbf{T}_{\lambda})^{-1} \left.\mathbf{r}_{\lambda_{\text u}}^{\text
  u}[S^a_{\lambda}]\right|_{\mathbf{S}_{\lambda}}\label{equation:ObteinRightEigenvectors}
\end{align}
for given uppercase right eigenvector~$\mathbf{r}_{\lambda_{\text
 u}}^{\text u}[S^a_{\lambda}]$. We denote by~$\mathbf{T}_{\lambda}$
the transformation matrix between bases associated to $S^a$ and
$S_{\lambda}^a$ on the level of eigenvectors and matrices.
  
Two opportunities to obtain the lowercase eigenvectors are possible:
Either we take the uppercase principal symbol
$\left.\mathbf{B}^{{S}_{\lambda}} \right|_{\mathbf{S}}$ in a basis
associated to~$S^a$ and calculate for given~$\lambda_{\text
  u}[S_{\lambda}^a]$ the new uppercase eigenvectors or we take the
uppercase eigenvectors to
$\left.\mathbf{B}^{{S}}\right|_{\mathbf{S}}$ in a basis associated
to~$S^a$ and make the
replacement~$\boldsymbol{S}\rightarrow\boldsymbol{S}_{\lambda}
=(S^a_{\lambda},\Qa^a_{\lambda},\Qb^a_{\lambda})^T$ which naturally
defines a SO(3)-transformation~$\mathbf R$. Using the first way, the
left and right eigenvectors are given by the formulas in the first
line of Eqs.~\eqref{equation:ObteinLeftEigenvectors}
and~\eqref{equation:ObteinRightEigenvectors}. However, the principal
symbol might lose its easy form which could be especially crucial for
a high number of evolved variables. Therefore, we chose the second
procedure in our notebooks, where the second (and/or third) lines of
Eqs.~\eqref{equation:ObteinLeftEigenvectors}
and~\eqref{equation:ObteinRightEigenvectors} are used to obtain the
lowercase eigenvectors.

The recovery will be explicitly shown for the system of GRMHD in
Sec.~\ref{section:GRMHD}. For the analysis of GRHD, the procedure is
given in the corresponding notebook~\cite{HilWebsite_aastex} but not in the paper.

For the sake of clarity, we finally want to relate all our
explanations with the covariant form of characteristic analysis using
the vector~$\boldsymbol{\phi}_a$ and the eigenvalue problem as
in~\eqref{equation:EigenvalueProblemCovPhi}. Taking the four-vector of
the form~$\boldsymbol{\phi}_a=-\lambda n_a+\s_a$ 
with~$\lambda=\lambda[\s_b]$ 
and writing the lowercase vectors in terms 
of~$u^a$,~$V^a$, and~$S^a$, we obtain
\begin{align}
\boldsymbol{\phi}_a&=-\lambda n_a+\s_a \non \\
&=-\lambda(Wu_a+WV_a)+S_a+W^2V^S(u_a+V_a)\non\\
&=(W^2V^S-W\lambda)u_a+S^a+(W^2V^S-W\lambda)V^a\non\\
&=N\left(-\lambda_{\text
 u}[S_{\lambda}^b]u_a+ S^{\lambda}_a\right)\non\\
&\propto -\lambda_{\text
 u}[S_{\lambda}^b]u_a+ S^{\lambda}_a\,. \label{equation:ConnectionPhiToLiterature}
\end{align}
The last step is done since~$\boldsymbol{\phi}_a$ is defined up to an
arbitrary scalar factor and we always consider unit spatial vectors
for the characteristic analysis.

%%%%%%%%%%%%%%%%%%%%%%%%%%%%%%%%%%%%%%%%%%%%%%%%%%%%%%%%%%%%%
\section{Hyperbolicity of GRHD}\label{section:GRHD}
%%%%%%%%%%%%%%%%%%%%%%%%%%%%%%%%%%%%%%%%%%%%%%%%%%%%%%%%%%%%%

We now start applying the formalism of the last section to concrete
examples of fluid matter models. We begin with the simple case of an
ideal fluid. Because a full characteristic analysis has been nicely
given in Ref.~\cite{FonIbaMar94}, the calculations here serve first 
as a sanity check in a nontrivial example, but second as a proof of
principle that the DF approach to the analysis results in an economic
treatment. Thus, we consider the energy-momentum tensor of an ideal
fluid,
\begin{align}
T^{ab}=\rho_0 h u^a u^b+ p g^{ab}\,,
\end{align}
with the four-velocity of the fluid elements~$u^a$, rest mass
density~$\rho_0$, specific enthalpy~$h$, and pressure~$p$. The
specific enthalpy~$h$ can be expressed in terms of~$\rho_0, p$ and the
specific internal energy~$\varepsilon$ as
\begin{align}
h = 1+\varepsilon +\frac{p}{\rho_0}\,.
\end{align}
The evolution equations of the system are the conservation of
energy momentum
\begin{align}
\nabla_{a}(T^{a b})=0\,, \label{equation:ConservationOfEnergyMomentum}
\end{align}
and the conservation of particle number
\begin{align}
\nabla_{a}(\rho_0 u^{a})=0\,. \label{equation:ConservationOfParticles}
\end{align}
Projecting Eq.~\eqref{equation:ConservationOfEnergyMomentum}
along and perpendicular to the fluid four-velocity~$u^a$, we get the
equations
\begin{align}
\rho_0h \nabla_au^a+u^a
\nabla_a(\rho_0+\varepsilon\rho_0)=0\,
\label{equation:ProjectuConservationOfEnergyMomentum}
\end{align}
and
\begin{align}
\rho_0  h \gamu^{c}{}_{ b}u^{a}\nabla_{a}u^{b}+\gamu^{c a}\nabla_{a}
p=0\,,\label{equation:ProjectUpsilonConservationOfEnergyMomentum}
\end{align}
respectively. We choose an arbitrary equation of state (EOS) of the
form
\begin{align}
p=p(\rho_0,\varepsilon)\,. \label{equation:EOSHD}
\end{align}
Equations~(\ref{equation:ConservationOfParticles})--(\ref{equation:EOSHD})
provide us with six equations for the six unknown quantities~$(\rho_0,
\varepsilon, p, \vhat_a)$. By~\eqref{equation:EOSHD}, we only need to
evolve the state vector~$\textbf{U}=(p,\vhat_a,\varepsilon)^T$.  The
components of~$\mathbf{U}$, expanded in our lowercase (Eulerian)
tensor basis, may be viewed as a slightly modified version of the {\it
  primitive variables} $\rho_0,\varepsilon,v_i$ commonly used in the
literature. The characteristic analysis will be performed on the
system of
equations~(\ref{equation:ConservationOfParticles})--(\ref{equation:ProjectUpsilonConservationOfEnergyMomentum}),
the state vector~$\textbf{U}$, in particular in a 
non-flux-balance law
form. Since there is no gauge freedom in the system, the analysis
applies unambiguously even after a change of variables, 
for example, to
the conservative variables~$D,\tau,S_i$ defined in, for
example, Ref.~\cite{FonMilSue98}. This is assured by the proof in
Sec.~\ref{section:ProofStrHypVariables}.

%%%%%%%%%%%%%%%%%%%%%%%%%%%%%%%%%%%%%%%%%%%%%%%%%%%%%%%%%%%%%
\subsection{Lowercase formulation}\label{section:LowerCaseHD}
%%%%%%%%%%%%%%%%%%%%%%%%%%%%%%%%%%%%%%%%%%%%%%%%%%%%%%%%%%%%%

We split the 
Eqs.~(\ref{equation:ConservationOfParticles})--(\ref{equation:ProjectUpsilonConservationOfEnergyMomentum}) 
now
against~$n^{a}$ and~$\gamma^a{}_b$ to get a system of first-order
partial differential equations for the
variables~$(p,\vhat_a,\varepsilon)$. Doing so, it is easy to show that
the system of equations can be rewritten as
\begin{align}
\Lie_np =& (\cs^2-1)\Wcs^2v^a D_a p-\cs^2\rho_0 h
\frac{\Wcs^2}{W} \gBi^{ab}D_a\vhat_b\non \\ &+\cs^2\rho_0 h
\Wcs^2\gBi^{ab} K_{ab}\,, \label{equation:systemLowerCaseHDa}\\
\gamma^b{}_a\Lie_n\vhat_b =& -\frac{1}{W\rho_0
  h}\left(\gamma^{c}{}_a+\cs^2 \Wcs^2  v^cv_a
\right) D_c p -v^cD_c \vhat_a \non\\
&+\cs^2\Wcs^2 v_a \gBi^{bc} D_b \vhat_c - \cs^2 \Wcs^2\gBi^{bc}
K_{bc} \vhat_a\non \\
&-WD_a\ln\alpha \,,\\
\Lie_n\varepsilon =& \frac{p}{\rho_0^2 h}\frac{\Wcs^2}{W^2}v^aD_a
p-\frac{p}{\rho_0}\frac{\Wcs^2}{W} \gBi^{ab}D_a\vhat_b
\non\\
&-v^a D_a \varepsilon
+\frac{p}{\rho_0}\Wcs^2\gBi^{ab}K_{ab}\,,
\label{equation:systemLowerCaseHDc}
\end{align}
with~$\Wcs=1/ \sqrt{1-\cs^2v^2}$, where~$\cs$ is the local speed of
sound and
\begin{align}
\cs^2=\frac{1}{h}\left( \chi+\frac{p}{\rho_0^2}\kappa
\right),\ \ \chi= \left( \frac{\p p}{\p \rho_0} \right)_{\varepsilon},\
\kappa=\left(\frac{\p p}{\p
\varepsilon}\right)_{\rho_0}.\label{equation:localspeedofsound}
\end{align}
Unless otherwise stated, we consider only  matter or EOS with speed 
of sound~${0<\cs\leq1}$.
As one can see, we have used the Lie derivative~$\Lie_n$ along the
timelike unit normal vector~$n^a$ instead of~$\partial_t$ and have
written the covariant derivative~$D_a$ associated to the intrinsic
metric~$\gamma_{ab}$ instead of~$\partial_i$, but as discussed in
Sec.~\ref{section:Motivation} this makes no difference to our
analysis. Writing the
system~(\ref{equation:systemLowerCaseHDa})--(\ref{equation:systemLowerCaseHDc})
as a vectorial equation of the form
\begin{align}
&\mathbf{A}^{\textrm n} \Lie_n\mathbf{U}=\mathbf{A}^p D_p \mathbf U + \boldsymbol{\mathcal{S}}\,,
\label{equation:systemLowerCaseHDMatrix}
\end{align}
we can identify
\begin{align}
&\qquad \qquad \mathbf{A}^{\textrm n}=\begin{pmatrix}
1 & 0 & 0 \\
0 & \gamma^b{}_a & 0 \\
0 & 0 & 1 \\
\end{pmatrix}, \non\\
&\mathbf{A}^p=\begin{pmatrix}
(\cs^2-1)\Wcs^2v^p & -\cs^2\rho_0 h \frac{\Wcs^2}{W} \gBi^{pc} & 0 \\
  -\frac{1}{W\rho_0 h}f^p{}_a &
  \cs^2 \Wcs^2 \gBi^{pc}v_a -v^p\gamma^{c}{}_{a}  & 0 \\
  \frac{p}{\rho_0^2 h}\frac{\Wcs^2}{W^2}v^p &
  -\frac{p}{\rho_0}\frac{\Wcs^2}{W} \gBi^{pc} & -v^p\\
\end{pmatrix},
\label{equation:PrincipalPartLowerCaseHD}
\end{align}
with shorthand~$f^p{}_a=\gamma^p{}_a+\cs^2 \Wcs^2
v^p v_a$ and can write the source vector here as
\begin{align}
\boldsymbol{\mathcal{S}}=\begin{pmatrix}
\cs^2\rho_0 h \Wcs^2\gBi^{ab} K_{ab}\\
- \cs^2 \Wcs^2\gBi^{bc} K_{bc} \vhat_a-W D_a\ln\alpha\\
\frac{p}{\rho_0}\Wcs^2\gBi^{ab} K_{ab}
\end{pmatrix}.
\end{align}
Note that written in this form the principal parts of special and
general relativistic hydrodynamics take an almost identical form.
Let~$\s_a$ be an arbitrary lowercase spatial 1-form, normalized
against the inverse boost metric so that~$\gBi^{ab}\s_a\s_b=1$, and
let~${\perpq^{b}{}_{a}}:={\gamma^b{}_{a}}-\gBi^{bc}\s_c\s_a$ 
be the orthogonal
projector. Recalling the definition of~$\shat^a=\gBi^{ab}\s_b$ given
in Table~\ref{tab:spatialvectors}, we
write~${\gamma^a{}_b}=\shat^a\s_b+{\perpq^a\!{}_b}$. Inserting this relation
into~\eqref{equation:systemLowerCaseHDMatrix} and expanding leads to
\begin{align}
\left(\Lie_n\mathbf{U}\right)_{\shat,\,\ortAhat}\simeq\mathbf{P}^{\s}
\left(D_{\hat{\s}}\mathbf{U}\right)_{\shat,\,\ortBhat},
\end{align}
with the principal symbol~$\mathbf{P}^{\s}=\mathbf{A}^{\s}=$ 
\begin{align}
\begin{pmatrix}
\Wcs^2(\cs^2-1)v^{\s} &-\frac{\Wcs^2}{W}\cs^2 \rho_0 h & 0^{\ortB}& 0\\
-\frac{W^2+\cs^2 (v^{\s})^2 \Wcs^2}{W^3 \rho_0 h}
&-\frac{W^2-\cs^2\Wcs^2}{W^2}v^{\s} &0^{\ortB} & 0\\
-\frac{\cs^2\Wcs^2}{W \rho_0 h} v_{\ortAhat}v^{\s}
& \cs^2\Wcs^2v_{\ortAhat}&
-v^{\s}\,\perpq^{\ortB}\!_{\ortAhat}& 0_{\ortAhat}\\
\frac{p \Wcs^2}{W^2 \rho_0^2 h} v^{\s} &-\frac{p \Wcs^2}{W \rho_0}
&0^{\ortB} & -v^{\s}\\ 
\end{pmatrix}.\label{equation:PrincipalSymbolLowerCaseHD}
\end{align}
The symbol~$``\simeq"$ denotes equality up to transverse principal and
source terms. For any derivative operator~$\delta$ and 
vector~$z^a$, we
write~$(\delta\vhat)_z \equiv z^a \delta\vhat_a$, and for the state
vector~$\left(\delta\mathbf{U}\right)_{\shat,\,\ortAhat}=\left( \delta
p , (\delta\vhat)_{\shat} , (\delta\vhat)_{\ortAhat},
\delta\varepsilon \right)^T$. As explained earlier in
Sec.~\ref{section:DF}, we introduce here furthermore the
indices~$\ortA$ and~$\ortAhat$, which are abstract but which indicate
application of the orthogonal projector~${\perpq^b\!_a}$,
meaning~$z_{\ortAhat}={\perpq^a\!_{\ortAhat}}z_a$
and~$z^{\ortA}={\perpq^\ortA\!_b}z^b$ for any object~$z$.
Then, for example, we get~${\gamma^a{}_b(\delta\vhat)_a=\s_b(\delta
\vhat)_{\shat}+{\perpq^{\ortA}\!_b}(\delta \vhat)_{\ortAhat}}\,$.

Before we proceed with the characteristic analysis,
 a comment should be
made. By the use of~$\vhat_a$ in the state vector, the inverse boost
metric arose in the principal
part~\eqref{equation:PrincipalPartLowerCaseHD}. By taking~$\s_a$
normalized by~$\gBi^{ab}$, we were able to get rid of this
complication in the principal symbol, which became ``easy'', in the
sense that it is highly structured. The principal symbol as well as
the eigenvalues and eigenvectors for a state vector~$(p, v_a,
\varepsilon)$ can be found in the Appendix~\ref{section:HDforv}. 
Since we
normalize the spatial 1-form~$\s_a$ against the 
inverse boost metric, the
eigenvalues and vectors take a form that is slightly modified in
comparison with the literature, but these differences are purely
superficial.

Solving the characteristic polynomial, one gets the five real
eigenvalues
\begin{align}
&\lambda_{(0,1,2)}=-v^{\s}, \non\\
&\lambda_{(\pm)}= -\frac{1}{1-\cs^2v^2}\left((1-\cs^2)v^{\s}
\pm\frac{\cs}{W}\sqrt{1- \cs^2v_{\perp}^2}\right),
\end{align}
with the
shorthand~$v_{\perp}^2:=v^{\ortA}v_{\ortAhat}$.

Please note that all eigenvalues in this paper have the opposite sign
in comparison to the literature by our definition of the principal
symbol. In the one-dimensional limit, i.e.,~$v_{\perp}=0$, the
eigenvalues~$\lambda_{(\pm)}$ reduce to
\begin{align*}
\lambda_{(\pm)}=-\frac{v^{\s}\pm W\cs}{1\pm \frac{\cs v^{\s}}{W}}\,,
\end{align*}
which, as noted elsewhere~\cite{Alc08}, is just the special
relativistic addition of two velocities multiplied with~$W$. Due to
our choice of a three-basis normalized by the inverse boost metric,
the eigenvalues are slightly different as compared to the results 
in the Appendix~\ref{section:HDforv}. 

The left eigenvectors of the principal symbol with our variable choice
for the respective eigenvalues~$\{ \lambda_{(0,1,2)}\text{,}
\lambda_{(\pm)} \}$ are 
\begin{align}
&\begin{pmatrix}
-\frac{p}{\cs^2 \rho_0^2 h}&0 &0^{\ortA} &1
\end{pmatrix},\qquad
\begin{pmatrix}
\frac{1}{\rho_0 h}\vhat_{\ortChat}&0 &\perpq^\ortA\!_{\ortChat} &0 
\end{pmatrix},\non \\
&\begin{pmatrix}
\pm \frac{\sqrt{1- \cs^2v_{\perp}^2}}{\cs \rho_0 h}&1 &0^{\ortA} &0 
\end{pmatrix},
\end{align}
respectively. The associated right eigenvectors are 
\begin{align}
\begin{pmatrix}
0 \\
0 \\
0_{\ortBhat} \\
1
\end{pmatrix},
\begin{pmatrix}
0 \\
0 \\
\perpq^\ortC\!_\ortBhat \\
0
\end{pmatrix},
\begin{pmatrix}
\frac{\cs^2 \rho_0^2 h}{p} \\
\pm \frac{\cs \rho_0}{p } \sqrt{1- \cs^2v_{\perp}^2}\\
-\frac{\cs^2 \rho_0}{p } \vhat_{\ortBhat}\\
1
\end{pmatrix}\,,
\end{align}
respectively. Since there is a complete set of
eigenvectors for each~$\s_a$ which depend furthermore continuously
on~$\s_a$, the system is strongly hyperbolic. The characteristic
variables corresponding to the speeds $\{
\lambda_{(0,1,2)},\lambda_{(\pm)}\}$ are given by
\begin{align}
&\hat{\text{U}}_0= \delta \varepsilon-\frac{p}{\cs^2\rho_0^2h}\delta
 p\,,\ \hat{\text{U}}_{\ortAhat} =(\delta\vhat)_{{\ortAhat}}+\frac{1}{\rho_0
 h}\vhat_{\ortAhat}\delta p\,, \non \\
&\hat{\text{U}}_{\pm}=(\delta\vhat)_{\hat{\s}}
\pm\frac{\sqrt{1-\cs^2v_{\perp}^2}}{\cs \rho_0 h}\delta p\,.
\end{align}

%%%%%%%%%%%%%%%%%%%%%%%%%%%%%%%%%%%%%%%%%%%%%%%%%%%%%%%%%%%%%
\subsection{Dust}\label{section:Dust}
%%%%%%%%%%%%%%%%%%%%%%%%%%%%%%%%%%%%%%%%%%%%%%%%%%%%%%%%%%%%%

A special case for the EOS~\eqref{equation:EOSHD} is
that of dust, in which the pressure is identically zero
everywhere,~$p\equiv0$, and the energy density coincides with the rest
mass density,~$\varepsilon=0$. It follows that the fluid elements then
follow timelike geodesics and that the conservation of the number of
particles~\eqref{equation:ConservationOfParticles} is automatically
fulfilled by the conservation of energy momentum in 
equation~\eqref{equation:ConservationOfEnergyMomentum}. 
For the analysis of
hyperbolicity, we use in this
subsection~$\mathbf{U}=(\rho_0$,$\vhat_a)$ as the state vector.

Using Eqs.~\eqref{equation:ProjectuConservationOfEnergyMomentum}
and~\eqref{equation:ProjectUpsilonConservationOfEnergyMomentum}
with~$\varepsilon=p=0$ and splitting the equations against~$n^a$
and~$\gamma^a_{\ b}$, the PDE system can be written as
\begin{align}
\Lie_n\rho_0=&-v^aD_a\rho_0-\frac{\rho_0}{W}\gBi^{ab}D_a\vhat_b\non \\
&+\rho_0\gBi^{ab}\Kn_{ab}\,,\non\\
\gamma^b{}_a\Lie_n\vhat_b=&-v^bD_b\vhat_a-W D_a\ln\alpha\,.
\label{equation:DustsystemLowerCase}
\end{align}

Using again an arbitrary spatial 1-form~$\s_a$ as in
Sec.~\ref{section:LowerCaseHD}, one ends up with the principal
symbol~$\mathbf{P}^{\s}$
for~$\left(\delta\mathbf{U}\right)_{\shat,\,\ortAhat}$ as
\begin{align}
\mathbf{P}^{\s}=\begin{pmatrix}
-v^{\s} &-\frac{\rho_0}{W} & 0^{\ortB}\\
 0& -v^{\s}& 0^{\ortB} \\
 0_{\ortAhat}&0_{\ortAhat} &-v^{\s}\, \perpq^\ortB\!_\ortAhat
\end{pmatrix},
\end{align} 
which evidently contains a Jordan block. The principal symbol is thus
missing an eigenvector. The system is only weakly hyperbolic and hence
the IVP is ill posed.

%%%%%%%%%%%%%%%%%%%%%%%%%%%%%%%%%%%%%%%%%%%%%%%%%%%%%%%%%%%%%
\subsection{Uppercase formulation}\label{section:UpperCaseHD}
%%%%%%%%%%%%%%%%%%%%%%%%%%%%%%%%%%%%%%%%%%%%%%%%%%%%%%%%%%%%%

We start again with Eqs.~\eqref{equation:ConservationOfParticles},~\eqref{equation:ProjectuConservationOfEnergyMomentum},
and~\eqref{equation:ProjectUpsilonConservationOfEnergyMomentum} but
split them against~$u^a$ and~$\gamu^b{}_a$. Using the definition of
the local speed of sound~\eqref{equation:localspeedofsound}, we derive
after some algebra the following PDEs for the components of the state
 vector:
\begin{align}
\nabla_u p&= -\cs^2 \rho_0 h \gamu^{b}{}_{d}\gBi^{d c} \nabla_b
\vhat_c \non\\ &-\cs^2 W \rho_0 h \gamu^{b}{}_{d}\gBi^{d c} \nabla_b
n_c\,, \label{equation:systemUpperCaseHDa}\\
\gamu_{a b}\gBi^{b c}
\nabla_u \vhat_c&=-\frac{1}{\rho_0 h}\gamu^{b}{}_a\nabla_b p \non\\
&-W\gamu_{a b}\gBi^{b c} \nabla_u n_c\,,\\
\nabla_u \varepsilon&=
-\frac{ p}{\rho_0} \gamu^{b}{}_{d}\gBi^{d c} \nabla_b \vhat_c \non
\\
&-\frac{W p}{\rho_0} \gamu^{b}{}_{d}\gBi^{d c} \nabla_b
n_c\,.\label{equation:systemUpperCaseHDc}
\end{align}
Here we have used the
relationship~$\gamu_{ab}\gBi^{bc}=\gBui_{ab}\gamma^{bc}$. Proceeding
as when splitting against the lowercase frame, we write the
system~(\ref{equation:systemUpperCaseHDa})--(\ref{equation:systemUpperCaseHDc})
as an equation for the state vector~$\textbf{U}$,
\begin{align}
\mathbf{B}^{\textrm u} \nabla_u\mathbf{U}=\mathbf{B}^p \nabla_p \mathbf U
+ \boldsymbol{\mathcal{S}},
\label{equation:systemUpperCaseHDMatrix}
\end{align} 
and identify
\begin{align}
&\qquad \mathbf{B}^{\textrm u}=\begin{pmatrix}
1 & 0 & 0 \\
0 & \gamu_{a b}\gBi^{b c}  & 0 \\
0 & 0 & 1 \\
  \end{pmatrix}\,
\end{align}
and
\begin{align}
 \mathbf{B}^p &= 
\begin{pmatrix}
0 & -\cs^2 \rho_0 h \gamu^{p}{}_{d}\gBi^{d c} & 0\\
-\frac{1}{\rho_0 h}\gamu^{p}{}_a &0 & 0 \\
0 & -\frac{ p}{\rho_0} \gamu^{p}{}_{d}\gBi^{d c} & 0 \\
\end{pmatrix}\,.
\end{align}
The source vector is written as
\begin{align}
\boldsymbol{\mathcal{S}}=\begin{pmatrix}
-\cs^2 W \rho_0 h \gamu^{b}{}_{d}\gBi^{d c} \nabla_b n_c\\
-W\gamu_{a b}\gBi^{b c}  \nabla_u n_c\\
-\frac{W p}{\rho_0} \gamu^{b}{}_{d}\gBi^{d c} \nabla_b n_c
\end{pmatrix}.
\end{align}
It is straightforward to verify that~$\mathbbm{1}+\mathbf{B}^V$ 
is invertible for
all~$v_av^a<1$. Therefore, as long as the various speeds in the system
are not superluminal, that is,~$|\lambda|\leq1$, expected since we are
considering here a fluid model with no gauge freedom, by the argument
of Sec.~\ref{section:ProofFrameIndepHyp}, we may analyze strong
hyperbolicity equivalently in the upper- or lowercase frames.

Let~$S_a$ be an arbitrary uppercase spatial vector, normalized
against~$\gamu_{ab}$ so that~$S_aS^a=1$, and
let~${\perpQ^b{}_a}=\gamu^b{}_a-S^bS_a$ be the orthogonal
projector. Decomposing~$\gamu^a{}_{ b}$
against~$S^a$ and using relations in
Table~\ref{tab:RelationSpatialVectors} to~$\s_a$, we write
Eq.~\eqref{equation:systemUpperCaseHDMatrix} as
\begin{align}
   \left(\nabla_u\mathbf{U}\right)_{\shat,\,\ortAhat}\simeq
   \mathbf{P}^{S}\left(\nabla_S\mathbf{U}\right)_{\shat,\,\ortBhat},
\end{align}
with principal symbol
\begin{align}
\mathbf{P}^{S}=\mathbf{B}^S=\begin{pmatrix}
0&-\cs^2 \rho_0 h & 0^{\ortb} & 0\\
-\frac{1}{ \rho_0 h}&0 & 0^{\ortb} & 0\\
0_{\orta}& 0_{\orta} &0^{\ortb}\!_{\orta} & 0_{\orta}\\
0& -\frac{p}{\rho_0} &0^{\ortb} & 0\\ 
\end{pmatrix}.\label{equation:PrincipalSymbolUpperCaseHD}
\end{align}
Since the uppercase projector is pushed through the lowercase
inverse boost metric, we have ~$S_aS_b\gBi^{b c}\left(\delta
\vhat\right)_c=S_a\shat^c\left(\delta \vhat\right)_c=S_a\left(\delta
\vhat\right)_{\shat}$, and for the orthogonal
projector~${\perpQ^\ortA\!_b}\left(\delta \vhat\right)_{\ortAhat}=
{\perpQ^\orta\!_b}\gamu_{\orta c}\gBi^{cd}\left(\delta\vhat\right)_{d}$.

By employing the uppercase frame, the principal symbol has become much
simpler than before, see~\eqref{equation:PrincipalSymbolLowerCaseHD},
exhibiting now essentially the same shape as that of a simple wave
equation. In the present example, the extra structure is not required
to complete the analysis, because in practice, computer algebra tools
can already manage the more complicated form. In more sophisticated
models, however, additional structure may become crucial if we wish to
perform such an analysis. An obvious question to ask is:  why is
the uppercase form of the principal symbol so much cleaner?  The
reason, which in hindsight is obvious, is that the 
four-dimensional form
of the fluid equations of motion contains the fluid four-velocity, and
so any frame adapted to that fact naturally annihilates many terms in
the principal symbol, uncovering the beautiful structure
of~\eqref{equation:PrincipalSymbolUpperCaseHD}. The five eigenvalues
of~$\mathbf{P}^S$ are
\begin{align}
\lambda_{(0,1,2)}=0\,,\qquad \lambda_{(\pm)}= \pm \cs\,,
\end{align}
with the corresponding left eigenvectors
\begin{align}
&\begin{pmatrix}
-\frac{p}{\cs^2 \rho_0^2 h} &0&0^{\orta}&1 
\end{pmatrix},
\begin{pmatrix}
0&0&\perpQ^{\orta}\!\!_{\ortc}&0
\end{pmatrix},\non\\
&\begin{pmatrix}
\mp\frac{1}{\cs \rho_0 h} &1 &0^{\orta} &0 
\end{pmatrix};
\end{align}
right eigenvectors
\begin{align}
\begin{pmatrix}
0\\
0 \\
0_{\ortb} \\
1
\end{pmatrix},
\begin{pmatrix}
0\\
0 \\
\perpQ^{\ortc}\!\!_{\ortb}\\
0
\end{pmatrix},
\begin{pmatrix}
\frac{\cs^2 \rho_0^2 h}{p}\\
\mp\frac{\cs \rho_0}{p} \\
0_{\ortb}\\
1
\end{pmatrix};
\end{align}
and
characteristic variables
\begin{align}
\hat{\text{U}}_0=\delta \varepsilon-\frac{p}{\cs^2 \rho_0^2 h}\delta
p\,,\ \hat{\text{U}}_{\orta}=(\delta\vhat)_{{\ortAhat}}
\,,\ \hat{\text{U}}_{\pm}=(\delta\vhat)_{\hat{\s}}\mp \frac{\delta p}{\cs \rho_0
  h}\,.
\end{align}
Using the recovery procedure described in
Sec.~\ref{section:RecoverEVandEVecsGeneral} gives the same
results for eigenvalues and eigenvectors and characteristic variables
as in our lowercase analysis. For details, see the notebooks~\cite{HilWebsite_aastex} that 
accompany the paper.

%%%%%%%%%%%%%%%%%%%%%%%%%%%%%%%%%%%%%%%%%%%%%%%%%%%%%%%%%%%%%
\section{Hyperbolicity of GRMHD}\label{section:GRMHD}
%%%%%%%%%%%%%%%%%%%%%%%%%%%%%%%%%%%%%%%%%%%%%%%%%%%%%%%%%%%%%

In this section, we investigate whether or not two different
formulations of GRMHD are strongly hyperbolic. The field equations
will be expressed for a set of eight variables corresponding to those
evolved numerically. The first characteristic analysis for RMHD was
done by Ref.~\cite{AniPen87}. They worked covariantly and 
considered an
augmented system of {\it ten} evolved variables, assuming implicitly a
``free-evolution'' style~\cite{Hil13} to treat the two
additional algebraic constraints,~$u^au_a=-1, u^ab_a=0$, thus
introduced, as well as the Gauss constraint besides. The analysis was
then reviewed and expanded in Ref.~\cite{Ani90a}. 
The conclusion was that
the augmented formulation of RMHD is strongly hyperbolic. Another
augmented system for RMHD using ten variables was later derived
in~Ref.\cite{Put91}. On the basis of Ref.~\cite{AniPen87},
several authors,
e.g., Refs.~\cite{Kom99,AntMirMar10}, reexamined the 
characteristic analysis
and treated degeneracies. In particular, a very detailed discussion is
given in Ref.~\cite{AntMirMar10}.

For numerical implementation, a flux-balance law form of the equations
was needed, as shocks can arise, and used in slightly different forms
by, for example, 
Refs.~\cite{Kom99,Bal01a,GamMcKTot03,AntZanMir05,GiaRez07,
  AntMirMar10,ZanFamDum15}. A detailed overview is given in t
  he review
of Ref.~\cite{Fon08}. In the flux-balance law form considered here, 
a total of eight variables including the magnetic field
are evolved. It is important to stress that changing the number of
variables can cause a breakdown of hyperbolicity, so in general, it is
not enough to know that there is {\it some} good form of the system
being treated. Rather, it is required that the {\it particular}
formulation being employed should itself be at least strongly
hyperbolic. The analysis of Ref.~\cite{Ani90a} therefore does not
necessarily apply to the system in use in applications.

Our analysis begins with two observations that motivate a careful
reconsideration of GRMHD. First, when numerical schemes to treat GRMHD
are constructed, one sometimes sees that the longitudinal component of
the magnetic field is ignored in evaluating the fluxes. This is
ultimately because the approximation works by repeated application of
a one-dimensional scheme, which is of course a 
perfectly legitimate approach. It is,
however, easy to overlook the fact that when performing hyperbolicity
analysis we are not free to discard any variable and must find a
complete set of eigenvectors of the principal symbol, including that
associated with the Gauss constraint. We must therefore be careful not
to be misled by tricks that apply only to the method, rather than the
system of equations itself.

Second, even if we can show strong hyperbolicity for a formulation of
GRMHD that requires the evolution of only eight variables, we still
may not claim that the flux-balance law formulation used in
applications satisfies the same property. Like the field equations of
GR and electrodynamics, those of GRMHD have a gauge freedom, which,
from the free-evolution point of view is just the freedom to add
combinations of the constraint to the evolution equations. Different
choices of this addition affect the level of hyperbolicity of the
formulation.

Neither of these subtleties has been completely taken care of by the
earlier analyses, and indeed a first indication that the system of
GRMHD used, for example, in Refs.~\cite{AntZanMir05,IbaCorCar15}
 differs from that used in the analysis
of Ref.~\cite{AniPen87} is the fact that the eigenvalues 
associated with
the Gauss constraint differ between the two
systems. In Ref.~\cite{AniPen87} the ``entropy eigenvalue'' is 
found with
multiplicity 2. Of these, one corresponds to the Gauss
constraint. In Ref.~\cite{IbaCorCar15}, 
for the system of eight variables,
the entropy eigenvalue has only multiplicity 1, and the constraint
eigenvalue is zero. We suppose that the reason these points have not
been carefully unraveled before is chiefly that the lowercase
principal symbol of GRMHD is a complicated matrix of which the
 structure is
very difficult to spot. Remarkably, there is enough structure in the
symbol so that the calculation of the eigenvalues and eigenvectors is
possible in closed form, but the expressions are {\it very} long. For
example, before developing the DF approach to the problem, which we
will see simplifies matters greatly, we attempted a brute force
treatment; the magnetosonic eigenvalues arrived with more than~$10^4$
terms!

This section is structured as follows. In
Sec.~\ref{section:BasicsGRMHD}, we recapitulate the basic
definitions and equations for GRMHD
following Refs.~\cite{Ani90a,AntMirMar10}. 
Afterward, we 3+1 decompose the
PDEs and derive the evolution equations, where in each multiples of
the Gauss constraint are manually added 
(see Sec.~\ref{section:3plus1PDEsysofGRMHD}).
 We then analyze the characteristic
structure of the principal symbol, taking all constraint addition
coefficients to zero, which forms a set of PDEs that is in some sense
analogous to the set of equations in Ref.~\cite{AniPen87}, 
but with their
algebraic constraints explicitly imposed; see
Sec.~\ref{section:PrototypeConstraintfreesystemGRMHD}. In
Secs.~\ref{section:UpperCaseGRMHD}
and~\ref{section:LowerCaseGRMHD}, we do the analysis in the upper- and
lowercase frames and give some comments about how the eigenvectors
have to be rescaled to take account of degeneracies. Finally, in
Sec.~\ref{section:FluxConsGRMHD}, we take a different choice of
constraint addition coefficients to obtain a set of equations equal to
the flux-balance law system, comparing explicitly
with Ref.~\cite{IbaCorCar15}, and show that this formulation 
of GRMHD which 
is used
in numerical relativity is only weakly hyperbolic.

%%%%%%%%%%%%%%%%%%%%%%%%%%%%%%%%%%%%%%%%%%%%%%%%%%%%%%%%%%%%%
\subsection{Basics of GRMHD}\label{section:BasicsGRMHD}
%%%%%%%%%%%%%%%%%%%%%%%%%%%%%%%%%%%%%%%%%%%%%%%%%%%%%%%%%%%%%

In this subsection, we give a brief review about the basic definitions
and equations of GRMHD following Refs.~\cite{Ani90a,AntMirMar10}.
 However,
this will be done in a primarily mathematical fashion, suppressing
some important physical insights and statements. We use
Lorentz-Heaviside units for electromagnetic quantities
with~$\varepsilon_0=\mu_0=1$ throughout, where~$\varepsilon_0$ is the
vacuum permittivity (or electric constant) and~$\mu_0$ is the vacuum
permeability (or magnetic constant).

%%%%%%%%%%%%%%%%%%%%%%%%%%%%%%%%%%%%%%%%%%%%%%%%%%%%%%%%%%%%%
\subsubsection{Faraday tensor and Ohm's law}
%%%%%%%%%%%%%%%%%%%%%%%%%%%%%%%%%%%%%%%%%%%%%%%%%%%%%%%%%%%%%

We start by introducing the Faraday electromagnetic tensor field (or
for short field strength tensor)~$F^{ab}$. For a generic observer with
four-velocity~$\mathtt N^a$, the field strength tensor and its 
dual can
be expressed via the electric and magnetic four-vectors,~$\mathtt
E^a$, $\mathtt B^a$, as
\begin{align}
F^{ab}=\mathtt N^a \mathtt E^b-\mathtt N^b \mathtt
E^a+\epsilon^{abcd}\mathtt N_c \mathtt
B_d\,,\nonumber\\
{^*F^{ab}}=\mathtt
N^a \mathtt B^b-\mathtt N^b \mathtt B^a-\epsilon^{abcd}\mathtt N_c
\mathtt E_d\,,\label{equation:DefinitionFieldStrengthTensor}
\end{align}
with the Levi-Civit\`a tensor,
\begin{align}
\epsilon^{abcd}=-\frac{1}{\sqrt{-g}}\left[ a b c d \right]\,,
\end{align}
where~$g$ is the determinant of the spacetime metric~$g_{ab}$, $\left[
  a b c d \right]$ is the completely antisymmetric 
  Levi-Civit\`a symbol
and~$2 {^*F^{ab}}=-\epsilon^{abcd}F_{cd}$ holds. We use here the sign
convention of Ref.~\cite{AlcDegSal09}. Both the electric and magnetic
fields satisfy the orthogonality relations~$\mathtt E^a \mathtt
N_a=\mathtt B^a \mathtt N_a=0$.

Using the field strength tensor and its dual, Maxwell's equations are
written as
\begin{align}
\nabla_b^*F^{ab}=0\,,\qquad \nabla_bF^{ab}=\mathcal{J}^a\,.
\end{align}  
According to Ohm's law (see Sec.~\ref{section:RGRMHD}), the
electric four-current~$\mathcal{J}^a$ can be expressed as
\begin{align}
\mathcal{J}^a=\rho_{\text{el}}u^a+\sigma F^{ab} u_b\,,
\end{align}
with the proper charge density~$\rho_{\text{el}}$ measured by the
comoving observer with~$u^{a}$ and~$\sigma$ the electric
conductivity.

%%%%%%%%%%%%%%%%%%%%%%%%%%%%%%%%%%%%%%%%%%%%%%%%%%%%%%%%%%%%%
\subsubsection{Ideal MHD condition}
%%%%%%%%%%%%%%%%%%%%%%%%%%%%%%%%%%%%%%%%%%%%%%%%%%%%%%%%%%%%%

In the limit of infinite conductivity~$\sigma$ but finite current,
the electric field~$e^a$ measured by the comoving observer~$u^a$, has
to vanish,
\begin{align}
e^a=F^{ab} u_b\equiv 0\,.\label{equation:IdealMHDCondition}
\end{align}
This equality holds by use of
expression~\eqref{equation:DefinitionFieldStrengthTensor}
taking~$\mathtt N^a=u^a$, $\mathtt B^a=b^a$ and $\mathtt E^a=e^a$.

%%%%%%%%%%%%%%%%%%%%%%%%%%%%%%%%%%%%%%%%%%%%%%%%%%%%%%%%%%%%%
\subsubsection{Energy-momentum tensor}
%%%%%%%%%%%%%%%%%%%%%%%%%%%%%%%%%%%%%%%%%%%%%%%%%%%%%%%%%%%%%

The total energy-momentum tensor of magnetohydrodynamics (MHD)
 is expressed as the sum of the
ideal fluid part,
\begin{align}
T^{ab}_{\text{fluid}}=\rho_0 h u^a u^b+g^{ab} p\,,
\end{align}
plus the standard electromagnetic energy-momentum tensor,
\begin{align}
T^{ab}_{\text{em}}=F^{ac}F^{b}{}_{c}-
\frac{1}{4}g^{ab}F_{cd}F^{cd}\,.
\label{equation:EnergyMomentumTensorEM_MHD}
\end{align}
Using the ideal MHD condition~\eqref{equation:IdealMHDCondition} and
expressing the field strength tensor
via~\eqref{equation:DefinitionFieldStrengthTensor}, the
 electromagnetic
energy-momentum tensor in terms of the magnetic field is
\begin{align}
T^{ab}_{\text{em}}=\left(u^au^b+\frac{1}{2}g^{ab}
\right)b^2-b^ab^b\,,\label{equation:EnergyMomentumTensorEM_MHD2}
\end{align}
and the total energy-momentum tensor is given by
\begin{align}
T^{ab}=\rho_0 h^* u^a u^b + p^* g^{ab}-b^a b^b\,,
\label{equation:EnergyMomentumTensorMHD}
\end{align}
with~$h^*=h+b^2/ \rho_0$ and~$p^*=p+b^2/2$. In
Eq.~\eqref{equation:EnergyMomentumTensorEM_MHD2}, we used as
a shorthand~$b^2=b^ab_a$.

%%%%%%%%%%%%%%%%%%%%%%%%%%%%%%%%%%%%%%%%%%%%%%%%%%%%%%%%%%%%%
\subsubsection{Covariant PDE system of GRMHD}
\label{section:PDEsysofGRMHD}
%%%%%%%%%%%%%%%%%%%%%%%%%%%%%%%%%%%%%%%%%%%%%%%%%%%%%%%%%%%%%

The equations of GRMHD are the conservation of the number of particles
\begin{align}
\nabla_a(\rho_0 u^a)=0\,,
\end{align}
the conservation of energy-momentum
\begin{align}
\nabla_bT^{ab}=0\,,
\end{align}
and the Maxwell equations
\begin{align}
\nabla_b {^*F}^{ab}=0\,.
\end{align}

%%%%%%%%%%%%%%%%%%%%%%%%%%%%%%%%%%%%%%%%%%%%%%%%%%%%%%%%%%%%%
\subsection{3+1 decomposition of the PDE system}
\label{section:3plus1PDEsysofGRMHD}
%%%%%%%%%%%%%%%%%%%%%%%%%%%%%%%%%%%%%%%%%%%%%%%%%%%%%%%%%%%%%

The 3+1 decomposition needs a bit more care since we have a
constrained system. For convenience, we will use~$\gamu^b{}_a, u^a$ to
decompose the equations given in
Sec.~\ref{section:PDEsysofGRMHD}. Afterward, we will add to
each equation some parametrized combination of the Gauss constraint. A
concrete choice of the constraint addition parameters results in a set
of evolution equations which we call a {\it formulation of GRMHD}. We
will focus here on two specific formulations. The first of these is
essentially that of Ref.~\cite{AniPen87}, but without the artificial
expansion of variables through the definition of the algebraic
constraints~$u^au_a=-1$ and~$u^ab_a=0$, which are 
satisfied {\it a priori}
in our approach. The second formulation corresponds to the
flux-balance law system used in numerics
by Refs.~\cite{AntMirMar10,IbaCorCar15}. 
We arrive at the second by matching
the values of the formulation parameters with the literature to obtain
the desired form of the field equations. We also want to stress that
we neither consider in this work formulations using the magnetic
four-potential instead of the magnetic field as
in Refs.~\cite{GiaRezBai10,EtiPasHaa15} nor systems with 
divergence cleaning
as in Ref.~\cite{ZanFamDum15}.

The eight equations determining the time evolution of the GRMHD system
are
\begin{align}
\nabla_a(\rho_0 u^a)=0\,,\quad
&\gamu_{ab}\nabla_cT^{bc}=0\,,\nonumber\\
u_b\nabla_cT^{bc}=0\,,\quad
&\gamu_{ab}\nabla_c {^*F}^{bc}=0\,,\label{equation:PDEMHD}
\end{align}
together with an equation of state~$p=p(\rho_0,\varepsilon)$ and the
Gauss constraint
\begin{align}
0=u_c\nabla_b {^*F}^{bc}=\gamu^{bc}\nabla_bb_c\,.\label{equation:MWGLConstraint}
\end{align}

The magnetic four-vector~$b^a$ can be split in the lowercase as
\begin{align}
n_ab^a=-(v_a\hat{b}^a)\,, \qquad \gamma^a{}_b\ b^b=\hat{b}^{a}\,,
\end{align}
and we have~$b^a=(\hat{b}^c v_c)n^a+\hat{b}^{a}$ with~$n_a\hat{b}^a=0$.
Furthermore, we introduce the Eulerian magnetic field vector~$B^a$ as
\begin{align}
\hat{b}_a&= \frac{1}{W}\gB_{ab}B^b=\frac{1}{W} B_a+(B^b\vhat_b)v_a\,,\non\\
 B^a&=W\gBi^{ab}\hat{b}_b=W\hat{b}^a-(\hat{b}^c\vhat_c)v^a\,,
\end{align}
where the lowercase Gauss constraint reads
\begin{align}
\gamma^{ab}\nabla_aB_b=0\,.
\end{align}

Taking Eqs.~\eqref{equation:PDEMHD}, a straightforward
calculation similar to that for GRHD in Sec.~\ref{section:UpperCaseHD}
provides evolution equations for the pressure,
\begin{align}
\nabla_u p=& -\cs^2 \rho_0 h \gamu^{d}{}_c\gBi^{c e} \nabla_d \vhat_e +S^{(p)}\non \\
&+\omega^{(p)}\left( \gamu^{d}{}_c\gBi^{c e}\nabla_{d}\pb_{e}+S^{(\text{c})}\right)\,;
\label{equation:systemUpperCaseMHDp}
\end{align}
the boost vector,
\begin{align}
\gamu_{ab}&\gBi^{b c} \nabla_u \vhat_c =-\left(
\frac{b^db_a}{\rho_0^2 h h^*}+\frac{\gamu^{d}{}_a}{\rho_0 h^*}\right)
\nabla_d p\non \\ &+ \frac{2}{\rho_0 h^*}\gamu^{[
    b}{}_{a}b^{d]}\gamu_{bc}\gBi^{c e}\nabla_{d}\pb_{e}+S^{(\mathbf{\vhat})}_a
\non \\ &+\omega^{(\vhat)}_a\left(
\gamu^{d}{}_c\gBi^{c e}\nabla_{d}\pb_{e}+S^{(\text{c})}\right)\,;
\label{equation:systemUpperCaseMHDvhat}
\end{align}
the magnetic field,
\begin{align}
\gamu_{ab}&\gBi^{b c}\nabla_u \pb_c =2\gamu_{ab}\gamu^{
  [b}{}_c b^{d]}\gBi^{c e}\nabla_{d}\vhat_{e} +S^{(\mathbf{\pb})}_a
\non \\ &+\omega^{(\pb)}_a\left(
\gamu^{d}{}_c\gBi^{c e}\nabla_{d}\pb_{e}+S^{(\text{c})}\right)\,;
\label{equation:systemUpperCaseMHDperpb}
\end{align}
and finally the specific internal energy,
\begin{align}
\nabla_u \varepsilon=& -\frac{ p}{\rho_0} \gamu^{d}{}_c\gBi^{c e} \nabla_d
\vhat_e +S^{(\varepsilon)}\non \\
&+\omega^{(\varepsilon)}\left( \gamu^{d}{}_c\gBi^{c e}
\nabla_{d}\pb_{e}+S^{(\text{c})}\right)\,.\label{equation:systemUpperCaseMHDeps}
\end{align}
By Eq.~\eqref{equation:MWGLConstraint}, we also obtain the Gauss constraint
\begin{align}
\gamu^{ac}\nabla_a
b_c=&\gamu^{d}{}_c\gBi^{c e}\nabla_{d}\pb_{e}+S^{(\text{c})}.\label{equation:MWGLConstraintperpb}
\end{align}
The sources are given by 
\begin{align*}
&S^{(p)} =-\cs^2 W \rho_0 h \gamu^{d}{}_c\gBi^{c e} \nabla_d n_e\,,\non \\
&S^{(\mathbf{\vhat})}_a= -W\gamu_{a b}\gBi^{b e}
\nabla_u n_e+ \frac{2 W}{\rho_0 h^*} \gamu^{[b}{}_{a}b^{e]}
V_bb^d\nabla_d n_e\,, \non \\
&S^{(\mathbf{\pb})}_a=2W \gamu_{ab}\gamu^{
  [b}{}_c b^{d]}\gBi^{c e}\nabla_d n_e \non\\\
  &\qquad \qquad+ 2W \gamu^{e}_{\ [a} V_{b]}b^b
\nabla_u n_e\,, \non \\
&S^{(\varepsilon)}=-\frac{W p}{\rho_0} \gamu^{d}{}_c\gBi^{c e} \nabla_d n_e\,,\non\\
&S^{(\text{c})}=\left(WV^db^e-W(b^cV_c)\gamu^{de}\right)\nabla_dn_e\,.\non
\end{align*}
The auxiliary magnetic vector~$\pb_c$ is defined by the relation
\begin{align}
\gamu_{a c}\gBi^{c d}\nabla_{b}\pb_{d}:= &\
\gamu_{a c}\gBi^{c d}\nabla_{b}\hat{b}_{d}\non\\
&+V_a b_d \gBi^{de}\nabla_{b}\vhat_{e}\,.
\label{equation:DefinitionPerpb}
\end{align}
As usual, square brackets around indices denote antisymmetrization,
so that~$2\vhat^{[a}b^{b]}=\vhat^{a}b^{b}-\vhat^{b}b^{a}$. In the
system~(\ref{equation:systemUpperCaseMHDp})--(\ref{equation:systemUpperCaseMHDeps}),
we already added multiples of the Gauss
constraint~\eqref{equation:MWGLConstraintperpb} connected to
coefficients~$\omega^{(p)},\omega^{(\vhat)}_a,\omega^{(\pb)}_a,$
and~$\omega^{(\varepsilon)}$.

%%%%%%%%%%%%%%%%%%%%%%%%%%%%%%%%%%%%%%%%%%%%%%%%%%%%%%%%%%%%%
\subsection{Prototype algebraic constraint free formulation}
\label{section:PrototypeConstraintfreesystemGRMHD}
%%%%%%%%%%%%%%%%%%%%%%%%%%%%%%%%%%%%%%%%%%%%%%%%%%%%%%%%%%%%%

In the following subsections, we proceed with the characteristic
analysis for the prototype algebraic constraint free formulation of
Eqs.~(\ref{equation:systemUpperCaseMHDp})--(\ref{equation:systemUpperCaseMHDeps})
by setting~$\omega^{(p)}=0$, $\omega^{(\vhat)}_a= 0$,
$\omega^{(\pb)}_a=0$, and~$\omega^{(\varepsilon)}=0$. The resulting
system is connected to the augmented system of equations
in Ref.~\cite{AniPen87} as follows: 
take the equations of Ref.~\cite{AniPen87},
project the momentum equation and the evolution equation for the
magnetic field with~$\gamu^a{}_b$ orthogonal to the four-velocity of
the fluid, change the evolved variables
to~$(p,\vhat_a,\pb_a,\varepsilon)$, and replace the derivative of~$p$
in the evolution equation for the magnetic field using the evolution
equation for~$p$. After this, one obtains our principal symbol. The
fact that Anile and Pennisi~\cite{AniPen87} work exclusively in RMHD
is of no consequence, since in our notation the principal symbol in
GRMHD is fundamentally the same as that of RMHD.

As previously mentioned, the equations become very lengthy in the
lowercase frame. As such, we were not able to find a choice of
variables in which the principal symbol takes a nice and easy
form. Nevertheless, by applying the strategy of
Sec.~\ref{section:RecoverEVandEVecsGeneral}, we were able 
to derive for the
prototype system all lowercase characteristic quantities, such as
eigenvalues, eigenvectors, and characteristic variables, which are
displayed in Sec.~\ref{section:LowerCaseGRMHD}, including a
discussion of degeneracies that may occur. Our analysis of the
flux-balance law formulation of GRMHD is given afterward in
Sec.~\ref{section:FluxConsGRMHD}.

%%%%%%%%%%%%%%%%%%%%%%%%%%%%%%%%%%%%%%%%%%%%%%%%%%%%%%%%%%%%%
\subsection{Uppercase formulation}\label{section:UpperCaseGRMHD}
%%%%%%%%%%%%%%%%%%%%%%%%%%%%%%%%%%%%%%%%%%%%%%%%%%%%%%%%%%%%%

Writing Eqs.~(\ref{equation:systemUpperCaseMHDp})--(\ref{equation:systemUpperCaseMHDeps}) with~$\omega^{(p)}=0$,
$\omega^{(\vhat)}_a= 0$, $\omega^{(\pb)}_a=0$, and
$\omega^{(\varepsilon)}=0$ in a vectorial form with state
vector~$\mathbf{U}=(p,\vhat_a,\pb_a,\varepsilon)^T$,
\begin{align}
\mathbf{B}^{\textrm u} \nabla_u \mathbf{U}=\mathbf{B}^p \nabla_p \mathbf{U} +
\boldsymbol{\mathcal{S}}\,,\label{equation:systemUpperCaseMHDMatrix}
\end{align} 
we identify
\begin{align}
\mathbf{B}^{\textrm u}=&\begin{pmatrix}
1 & 0 & 0 & 0 \\
0 & \gamu_{a b}\gBi^{b c} & 0 & 0 \\
0 &  0 & \gamu_{a b}\gBi^{b c} & 0 \\
0 & 0 & 0 & 1 \\
\end{pmatrix},
\end{align}
and the uppercase spatial part
\begin{align}
\mathbf{B}^p =&\begin{pmatrix}
0 & -\cs^2 \rho_0 h \gamu^{p}{}_c\gBi^{c e}  & 0 & 0\\
f^{ p}{}_a &0 & l^{pe}{}_a & 0\\
 0 & 2\gamu_{ab}\gamu^{
  [b}{}_c b^{p]}\gBi^{c e} & 0 & 0 \\
0 & -\frac{ p}{\rho_0} \gamu^{p}{}_c\gBi^{c e} & 0  & 0\\
\end{pmatrix},
\end{align}
with shorthands
\begin{align}
l^{pe}{}_a&=\frac{2}{\rho_0 h^*}\gamu^{[
    b}{}_{a}b^{p]}\gamu_{bc}\gBi^{c e}\,,\nonumber\\
f^{ p}{}_a&=-\left( \frac{b^pb_a}{\rho_0^2 h h^*}+\frac{\gamu^{p}{}_a}{\rho_0 h^*}\right)\,
\end{align}
and source vector~$\boldsymbol{\mathcal{S}}=
(S^{(p)},S^{(\mathbf{\vhat})}_a,S^{(\mathbf{\pb})}_a,S^{(\varepsilon)})^T$.
A straightforward calculation shows that~$\mathbbm{1}+\mathbf{B}^V$ is
invertible for all~$V^aV_a<1$.

%%%%%%%%%%%%%%%%%%%%%%%%%%%%%%%%%%%%%%%%%%%%%%%%%%%%%%%%%%%%%
\subsubsection{2+1 decomposition}
%%%%%%%%%%%%%%%%%%%%%%%%%%%%%%%%%%%%%%%%%%%%%%%%%%%%%%%%%%%%%

Let~$S_a$ be an arbitrary unit spatial $1$-form and~$\perpQ_{\ a}^{
  b}$ be the associated orthogonal projector. Let~$\s_a$
and~$\perpq_{\ a}^b$ be their lowercase projected versions (see
Tables~\ref{tab:spatialvectors}
and~\ref{tab:RelationSpatialVectors}). Decomposing $\gamu^{b}{}_{a}$
and $\gamma^{b}{}_{a}$ against~$S_a$ and~$\s_a$, respectively,
Eq.~\eqref{equation:systemUpperCaseMHDMatrix} can be written as
\begin{align}
\left(\nabla_u\mathbf{U}\right)_{\shat,\,\ortAhat}\simeq\mathbf{P}^{S}
\left(\nabla_S\mathbf{U}\right)_{\shat,\,\ortBhat},
\label{equation:systemUpperCaseMHDMatrixPS}
\end{align}
with the principal symbol~$\mathbf{P}^S=\mathbf{B}^S=$
\begin{align}
\begin{pmatrix}
0&-\cs^2 \rho_0 h & 0^{\ortb} & 0&  0^{\ortb}& 0\\
-\frac{\left(b^S\right)^2+\rho_0 h }{ \rho_0^2 hh^*}&0&0^{\ortb} & 0
& -\frac{b^{\ortb}}{\rho_0 h^*}& 0\\
-\frac{b^Sb_{\orta} }{ \rho_0^2 hh^*}& 0_{\orta}&0^{\ortb}{}_{\orta}&0_{\orta} &
\frac{b^{S}}{\rho_0 h^*}\perpQ^{\ortb}\!\!_{\orta}&0_{\orta}\\
0& 0&0^{\ortb}& 0&0^{\ortb}& 0 \\
0_{\orta}& -b_{\orta}&b^{S}\perpQ^{\ortb}\!\!_{\orta} & 0_{\orta}
& 0^{\ortb}{}_{\orta} & 0_{\orta} \\
0& -\frac{p}{\rho_0}& 0^{\ortb} & 0& 0^{\ortb} & 0\\ 
\end{pmatrix}\,.\label{equation:PrincipalSymbolUpperCaseMHD}
\end{align}
The characteristic polynomial~$P_{\lambda}$ for the principal
symbol~\eqref{equation:PrincipalSymbolUpperCaseMHD} can be written as
\begin{align}
&P_{\lambda}=\frac{\lambda^2}{ (\rho_0h^*)^2} P_{\text{Alfv\'en}} P_{\text{mgs}}\,,
\label{equation:CharPolUpperCaseMHD}
\end{align}
with the quadratic polynomial for Alfv\'en waves
\begin{align}
P_{\text{Alfv\'en}}=-\left(b^S\right)^2+\lambda^2\rho_0h^*
\label{equation:CharPolAlfvenUpperCaseMHD}
\end{align}
and the quartic polynomial for magnetosonic waves
\begin{align}
P_{\text{mgs}}=\left(\lambda^2-1\right)
\left(\lambda^2b^2-\left(b^S\right)^2\cs^2\right)
+\lambda^2\left(\lambda^2-\cs^2
\right)\rho_0h\,.\label{equation:CharPolmgsUpperCaseMHD}
\end{align}
Solving~\eqref{equation:CharPolUpperCaseMHD} provides us with
different kinds of speeds of waves propagating in the
$S^a$-direction. All speeds are real, and the system is strongly
hyperbolic, as will be seen later. The entropic waves have speed
\begin{align}
\lambda_{(\text e)}=0\,.\label{equation:EigenvalueMHDUpperCaseEntropy}
\end{align}
The constraint waves have the same speed, given by
\begin{align}
\lambda_{(\text{c})}=0\,.\label{equation:EigenvalueMHDUpperCaseConstraint}
\end{align}
The Alfv\'en waves are given by solving~$P_{\text{Alfv\'en}}=0$, which
results in the two different speeds
\begin{align}
\lambda_{(\text{a}\pm)}=\pm\frac{b^S}{\sqrt{\rho_0h^*}}\,.
\label{equation:EigenvalueMHDUpperCaseAlfven}
\end{align}
Solving the quartic equation~$P_{\text{mgs}}=0$, we obtain four
different speeds of the magnetosonic waves, two slow magnetosonic
waves,
\begin{align}
\lambda_{(\text{s}\pm)}=\pm\sqrt{\zeta_{\text{S}}-\sqrt{\zeta_{\text{S}}^2
-\xi_{\text{S}}}}\,,
\label{equation:EigenvalueMHDUpperMgsSlow}
\end{align}
and two fast magnetosonic waves,
\begin{align}
\lambda_{(\text{f}\pm)}=\pm\sqrt{\zeta_{\text{S}}
+\sqrt{\zeta_{\text{S}}^2-\xi_{\text{S}}}}\,,
\label{equation:EigenvalueMHDUpperCaseMgsFast}
\end{align} 
where we employ the shorthands
\begin{align}
\zeta_{\text{S}}=\frac{\left(
b^2+\cs^2\left[\left(b^S\right)^2+\rho_0h\right] \right)}{2\rho_0
h^*},\ \ \xi_{\text{S}}=\frac{\left(b^S\right)^2\cs^2 }{\rho_0 h^*}\,.
\end{align}
Please note that the index~$``\text S"$ in~$\zeta_{\text{S}}$
and~$\xi_{\text{S}}$ is not a contraction with a vector but rather a
reminder that we used for~$2+1$ decomposition the
vector~$S^a$. Since~$(b^S)^2\leq b^2$ and $\cs^2\leq 1$, all
eigenvalues have absolute value smaller than or equal than one, and
relation~$|\lambda_{\text u}||V|<1$, required for application of the
formalism of Sec.~\ref{section:RecoverEVandEVecsGeneral}, is
satisfied for all boost velocities. Thus, we are allowed to use the
recovering procedure for arbitrary boost velocities.

The left eigenvectors corresponding to~$\lambda_{(\text{e})}, \lambda_{(\text{c})},
\lambda_{(\text{a}\pm)}$ and~$\lambda_{(\text{m}\pm)}$
with~$\text{m}=\text{s}, \text{f}$ being
\begin{align}
&\begin{pmatrix}
-\frac{p}{\cs^2 \rho_0^2 h} & 0 & 0^{\orta} & 0 & 0^{\orta} & 1
\end{pmatrix}, \
\begin{pmatrix}
0 & 0 & 0^{\orta} & 1 & 0^{\orta} & 0
\end{pmatrix},\non \\
&\begin{pmatrix}
0 & 0 & \mp\epsuS^{\orta \ortc} b_{\ortc}\sqrt{\rho_0 h^*}  & 0 & -\epsuS^{\orta \ortc} b_{\ortc} & 0
\end{pmatrix},\non \\
&\begin{pmatrix}
\frac{\rho_0 h^*\left(\lambda_{(\text{m}\pm)}\right)^2 -b^2 }{\cs^2 \rho_0 h} &
\frac{\left(b^S\right)^2-\rho_0 h^*\left(\lambda_{(\text{m}\pm)}\right)^2}{\lambda_{(\text{m}\pm)}}
&\frac{b^S b^{\orta}}{\lambda_{(\text{m}\pm)}}  & 0 & b^{\orta} & 0
\end{pmatrix},\label{equation:LeftEigenvectorsMHDUpperCaseS}
\end{align}
respectively. We defined the antisymmetric uppercase two- 
and three-Levi-Civit\`a tensors as~${\epsuS^{\orta \ortb}=S_d\epsu^{d \orta
\ortb}=u_cS_d\,\perpQ^{\orta}\!_a\perpQ^{\ortb}\!_b\epsilon^{cdab}}$.
 The
right eigenvectors can be obtained by inverting of the matrix of left
eigenvectors or by solving the eigenvalue problem and can be
expressed as
\begin{align}
\begin{pmatrix}
0\\
0\\
0_{\ortb}\\
0\\
0_{\ortb}\\
1
\end{pmatrix}\,,\
\begin{pmatrix}
0\\
0\\
0_{\ortb}\\
1\\
0_{\ortb}\\
0
\end{pmatrix}\,,\
\begin{pmatrix}
0\\
0\\
\mp  \frac{\epsuS_{\ortb \ortc}}{\sqrt{\rho_0 h^*}} b^{\ortc}\\
0\\
-\epsuS_{\ortb \ortc} b^{\ortc}\\
0
\end{pmatrix},
\end{align}
for entropy, constraint, and Alfv\'en waves, and
\begin{align}
\begin{pmatrix}
\frac{\cs^2 \rho_0^2 h}{p}\\ - \frac{\rho_0
  \lambda_{(\text{m}\pm)}}{p}\\ \frac{\rho_0 \lambda_{(\text{m}\pm)}}{p
  b^{S} b_{\perp}^2}\left[\left(b^S\right)^2+\rho_0
  h^*\left((\lambda_{(\text{m}\pm)})^2-2\zeta_{\text S}\right) \right]
b_{\ortb}\\ 0\\ \frac{\rho_0}{ b_{\perp}^2p} \left[b^2+\rho_0
  h^*\left((\lambda_{(\text{m}\pm)})^2-2\zeta_{\text S}\right) \right]b_{\ortb}\\
1
\end{pmatrix}\,\label{equation:RightEigenvectorsMHDUpperCaseS}
\end{align}
for the four magnetosonic waves with~$\text{m}=\text{s},\text{f}$.
We introduced in the magnetosonic eigenvectors the orthogonal magnetic
field vector~$b_{\perp}^a=\perpQ_{\ b}^{a} b^b$
with~$b_{\perp}^2=b_{\perp}^a b^{\perp}_a=b^\orta b_\orta$. For the
moment, we have a complete set of eigenvectors for real
eigenvalues. Nevertheless, we have to check if any of the eigenvalues
may change their multiplicity and, if so, whether or not a complete
set of eigenvectors is still available. The situation where 
{\it a priori}
distinct eigenvalues coincide and their multiplicity change is called 
a degenerate state or for short a degeneracy. To show strong
hyperbolicity of the system, we have to show that for each possible
degenerate state a complete set of eigenvectors still exists. For the
augmented system of RMHD, this was already described
in Refs.~\cite{AniPen87,Ani90a,Kom99,Bal01a}. 
A full account was furthermore
given by Ref.~\cite{AntMirMar10}. We also want to mention that in the
Appendix of Ref.~\cite{Kom99} the eigenvalues and right 
eigenvectors in
the fluid rest frame are given for seven variables in a
one-dimensional analysis of RMHD. They are obtained by explicitly
setting (locally) the spatial entries of the four-velocity to zero,
which is ultimately quite similar to our approach.

%%%%%%%%%%%%%%%%%%%%%%%%%%%%%%%%%%%%%%%%%%%%%%%%%%%%%%%%%%%%%
\subsubsection{Degeneracy analysis of the uppercase}
%%%%%%%%%%%%%%%%%%%%%%%%%%%%%%%%%%%%%%%%%%%%%%%%%%%%%%%%%%%%%

For the prototype algebraic constraint free formulation of GRMHD, just
as in the augmented system of Ref.~\cite{AniPen87}, 
two different types
of degeneracies can occur. For degeneracy type I,~$b^S$ is equal to
zero, whereas for degeneracy type II, the magnetic field
 four-vector is
parallel to~$S^a$, so that~$b_{\perp}^a=\perpQ^a_{\ b}b^b=0$ holds. To
describe the different situations properly, we write the magnetic
field four-vector as
\begin{align}
b^a=b^SS^a+b_{\perp}^a\,,\qquad b^2=(b^S)^2+b_{\perp}^2\,.
\end{align} 
First, we note that the polynomials~\eqref{equation:CharPolAlfvenUpperCaseMHD}
and~\eqref{equation:CharPolmgsUpperCaseMHD} have solutions
\begin{align}
\left.\frac{b^S}{\lambda}\right|_{(\text a \pm)}=&\pm \sqrt{\rho_0 h^*},\\ 
\left.\frac{b^S}{\lambda}\right|_{(\text m \pm)}=&\pm
\sqrt{\left(\rho_0 h+\frac{b^2}{\cs^2}\right)
+\rho_0h\left(1-\frac{1}{\cs^2}\right)\frac{\lambda_{(\text m \pm)}^2}{1-\lambda_{(\text m \pm)}^2}}\non \\
=&\pm \sqrt{(b^S)^2+\left(\rho_0 h+\frac{b^2}{\cs^2}\right)-\rho_0h^*\frac{\lambda_{(\text m \pm)}^2}{\cs^2}}\,,
\end{align}
which are well defined even for degeneracies.

For a type I degeneracy in the uppercase where~$b^S=0$
and~$b^2=b_{\perp}^2$, the eigenvalues become
\begin{align}
\lambda_{(\text{e})}=\lambda_{(\text c)}=\lambda_{(\text a \pm)}=\lambda_{(\text s
  \pm)}=0,\ \lambda_{(\text f \pm)}=\pm \frac{\sqrt{b^2+\cs^2 \rho_0 h
}}{\sqrt{\rho_0h^*}}
\end{align}
and
\begin{align}
\left.\frac{b^S}{\lambda}\right|_{(\text s \pm)}=\pm \sqrt{\rho_0
h+\frac{b^2}{\cs^2}},\qquad \left.\frac{b^S}{\lambda}\right|_{(\text
f \pm)}=0
\end{align}
hold.

For type II degeneracy, namely when~$b_{\perp}^a=0$ and~$b^2=(b^S)^2$,
we have
\begin{align}
\lambda_{(\text s \pm)}=\lambda_{(\text a )}^{\pm}
=\pm\frac{|b^S|}{\sqrt{\rho_0 h^*}},\ \lambda_{(\text f \pm)}=\pm \cs,
\,\,\, \text{if}\ \ \cs^2>\frac{(b^S)^2}{\rho_0 h^*}\,, \non\\
\lambda_{(\text f \pm)}=\lambda_{(\text a )}^\pm=\pm\frac{|b^S|}{\sqrt{\rho_0 h^*}},\
\lambda_{(\text s \pm)}=\pm
\cs, \,\,\, \text{if}\ \ \cs^2<\frac{(b^S)^2}{\rho_0 h^*}\,,
\end{align}
and get 
\begin{align}
\left.\frac{b^S}{\lambda}\right|_{(\text m \pm\neq\text a \pm)}
=\pm \frac{b^S}{\cs}\,,\qquad 
\left.\frac{b^S}{\lambda}\right|_{(\text m \pm=\text a \pm)}
=\pm \sqrt{\rho_0 h^*}\,.
\end{align}

To classify the corresponding waves with  equal speed properly (see
Refs.~\cite{Kom99,AntMirMar10}), we 
defined~$\lambda_{(\text a )}^{\pm}$ 
with~$\lambda_{(\text a )}^{+}\geq\lambda_{(\text a )}^{-}$ 
such that~$\lambda_{(\text a )}^{\pm}=\lambda_{(\text a \pm)}$ 
for $b^S\geq 0$ 
or~$\lambda_{(\text a )}^{\pm}=\lambda_{(\text a \mp)}$
 for $b^S< 0$ holds.
The special case~$(b^S)^2=\cs^2\rho_0 h^*$ is called a type II$'$
degeneracy where~$\lambda_{(\text s \pm)}=\lambda_{(\text
  a)}^\pm=\lambda_{(\text f \pm)}=\pm \cs$. Note that 
  type I and type II
degeneracies may occur simultaneously, in which case we 
recover the pure
GRHD decoupled from the magnetic field evolution as a limiting
case. On the other hand, since we insist that~$\cs>0$, it is not
possible for type I and type II$'$ degeneracies to occur
 simultaneously.

%%%%%%%%%%%%%%%%%%%%%%%%%%%%%%%%%%%%%%%%%%%%%%%%%%%%%%%%%%%%%
\subsubsection{Renormalized uppercase left eigenvectors}
%%%%%%%%%%%%%%%%%%%%%%%%%%%%%%%%%%%%%%%%%%%%%%%%%%%%%%%%%%%%%

We rescale the Alfv\'en and magnetosonic eigenvectors in a way 
analogous to~\cite{AntMirMar10}. The procedure can also be found
 in the provided notebook~\cite{HilWebsite_aastex}. 
 The rescaled eigenvectors are
\begin{align}
\text{entropy:}\ \ &\begin{pmatrix}
-\frac{p}{\cs^2 \rho_0^2 h} & 0 & 0^{\orta} & 0 & 0^{\orta} & 1
\end{pmatrix},\non \\
\text{constraint:}\ \ &\begin{pmatrix}
0 & 0 & 0^{\orta} & 1 & 0^{\orta} & 0
\end{pmatrix},\non \\
\text{Alfv\'en:}\ \ &\begin{pmatrix} 0 & 0 & \pm\epsuS^{\orta \ortc} 
\sqrt{\rho_0 h^*}\frac{b^{\perp}_{\ortc}}{|b_{\perp}|} & 0
  & \epsuS^{\orta \ortc} \frac{b^{\perp}_{\ortc}}{|b_{\perp}|} 
  & 0
\end{pmatrix};
\end{align}
the magnetosonic left eigenvectors which have eigenvalues closer to
the Alfv\'en eigenvalues,
\begin{align} 
&\begin{pmatrix} \frac{\mathcal{H}(\lambda^2-1)}{\rho_0
     h} & (1-\cs^2)\mathcal{H}\lambda & \left(\frac{b^S
   }{\lambda}\right)\frac{b_{\perp}^{\orta}}{|b_{\perp}|}
    & 0 & \frac{b_{\perp}^{\orta}}{|b_{\perp}|}  & 0
\end{pmatrix}_{(\text m \pm)};\label{equation:UppercaseRescaledLeftm1}
\end{align}
and the other two magnetosonic left eigenvectors,
\begin{align} 
&\begin{pmatrix} \frac{1}{\cs^2\rho_0 h} &
   \frac{(1-\cs^2)\lambda}{\cs^2(\lambda^2-1)} & \left(\frac{b^S
   }{\lambda}\right)\mathcal{F}^{\orta} & 0 &
   \mathcal{F}^{\orta} & 0
\end{pmatrix}_{(\text m \pm)}\,;\label{equation:UppercaseRescaledLeftm2}
\end{align}
with abbreviations
\begin{align}
\mathcal{H}=&\frac{|b_{\perp}|}{\cs^2 -\lambda_{(\text m \pm)}^2 }\,,\\
\mathcal{F}^{\orta}=&
\frac{b_{\perp}^{\orta}}{(\rho_0 h^*\lambda_{(\text m \pm)}^2-b^2)}\,,
\end{align}
where for type II and even for type II$'$ degeneracy we take~$\Qa^a$
and~$\Qb^a$ such that in the degenerate limit we have
\begin{align}
\frac{b^{\perp}_{\ortc}}{|b_{\perp}|}  =
&\frac{1}{\sqrt{2}}(\Qa_{\ortc}+\Qb_{\ortc})\,,
\label{equation:degenerateperpb} \\
\mathcal{H}=&0\,,\\
\mathcal{F}^{\orta}=&0^{\orta}\,.\label{equation:degenerateFa}
\end{align}
Here, some comments are in order. In
Eqs.~(\ref{equation:degenerateperpb})--(\ref{equation:degenerateFa})
we are just making a canonical choice for how to represent the
complete set of eigenvectors under a type II or type II$'$ degenerate
limit. Note that for type II degeneracies~$\mathcal{H}$
and~$\mathcal{F}^\orta$ vanish automatically. For type II$'$
degeneracies, depending on how the limit is taken, their values may
not vanish but the form~\eqref{equation:UppercaseRescaledLeftm1}
and~\eqref{equation:UppercaseRescaledLeftm2}
with~$\mathcal{H}=\mathcal{F}^\orta=0$ can nevertheless be obtained by
taking appropriate linear combinations of the resulting eigenvectors.

%%%%%%%%%%%%%%%%%%%%%%%%%%%%%%%%%%%%%%%%%%%%%%%%%%%%%%%%%%%%%
\subsubsection{Renormalized uppercase right eigenvectors}
%%%%%%%%%%%%%%%%%%%%%%%%%%%%%%%%%%%%%%%%%%%%%%%%%%%%%%%%%%%%%

The right eigenvectors are obtained in the same way and with the same
abbreviations. The entropy, constraint, and Alfv\'en eigenvectors are
given by
\begin{align}
\begin{pmatrix}
0\\
0\\
0_{\ortb}\\
0\\
0_{\ortb}\\
1
\end{pmatrix},\
\begin{pmatrix}
0\\
0\\
0_{\ortb}\\
1\\
0_{\ortb}\\
0
\end{pmatrix},\
\begin{pmatrix}
0\\
0\\
\pm \epsuS_{\ortb \ortc}\frac{b^{\ortc}}{|b_{\perp}|} \\
0\\
\epsuS_{\ortb \ortc}\sqrt{\rho_0 h^*} \frac{b^{\ortc}}{|b_{\perp}|}\\
0
\end{pmatrix};
\end{align}
the magnetosonic eigenvectors corresponding to the eigenvalues closer
to the Alfv\'en eigenvalues are
\begin{align}
\begin{pmatrix}
\cs^2 \rho_0 h \mathcal{H}\\
- \mathcal{H} \lambda\\
-\left(\frac{b^S }{\lambda}\right)\frac{b^{\perp}_{\ortb}}{|b_{\perp}|} \\
0\\
\frac{\rho_0h}{(\lambda^2-1)} \frac{b^{\perp}_{\ortb}}{|b_{\perp}|}\\
\frac{p}{\rho_0}\mathcal{H}
\end{pmatrix}_{(\text{m}\pm)};
\end{align}
and the other two magnetosonic eigenvectors are
\begin{align}
\begin{pmatrix}
\cs^2 \rho_0 h \\
- \lambda\\
\cs^2(1-\lambda^2)\left(\frac{b^S }{\lambda}\right) \mathcal{F}_{\ortb}\\
0\\
\cs^2\rho_0h \mathcal{F}_{\ortb}\\
\frac{p}{\rho_0}
\end{pmatrix}_{(\text{m}\pm)}\,;
\end{align}
respectively.

%%%%%%%%%%%%%%%%%%%%%%%%%%%%%%%%%%%%%%%%%%%%%%%%%%%%%%%%%%%%%
\subsubsection{Characteristic variables}
%%%%%%%%%%%%%%%%%%%%%%%%%%%%%%%%%%%%%%%%%%%%%%%%%%%%%%%%%%%%%

The characteristic variables valid for all degeneracies are
\begin{align}
\hat{\text{U}}_{\text{e}}=&\delta \varepsilon-\frac{p}{\cs^2 \rho_0^2 h}\delta
p\,, \qquad \hat{\text{U}}_{\text c}=(\delta
\pb)_{\hat{\s}}\,,\non\\ \hat{\text{U}}_{\text a \pm}=&\pm\epsuS^{\orta
  \ortc} \sqrt{\rho_0 h^*}\frac{b^{\perp}_{\ortc}}{|b_{\perp}|}
(\delta \vhat)_{{\ortAhat}}+\epsuS^{\orta \ortc}
\frac{b^{\perp}_{\ortc}}{|b_{\perp}|} (\delta \pb)_{{\ortAhat}}\,, \non
\\ \hat{\text{U}}_{\text m_1 \pm}=&\frac{\mathcal{H}(\lambda_{(\text
    m_1 \pm)}^2-1)}{\rho_0 h}\delta p+
(1-\cs^2)\mathcal{H}\lambda_{(\text m_1 \pm)}(\delta\vhat)_{\hat{\s}}\non
\\ &+ \left(\frac{b^S
}{\lambda}\right)_{(\text{m}_1\pm)}\frac{b_{\perp}^{\orta}}{|b_{\perp}|}
(\delta \vhat)_{\ortAhat}+ \frac{b_{\perp}^{\orta}}{|b_{\perp}|}
(\delta \pb)_{\ortAhat}\,,\non\\ \hat{\text{U}}_{\text m_2
  \pm}=& \frac{1}{\cs^2\rho_0 h}\delta p+
\frac{(1-\cs^2)\lambda_{(\text m_2 \pm)}}{\cs^2(\lambda_{(\text m_2
    \pm)}^2-1)}(\delta\vhat)_{\hat{\s}} \non\\ &+ \left(\frac{b^S
}{\lambda}\right)_{(\text{m}_2\pm)}\mathcal{F}^{\orta}(\delta
\vhat)_{\ortAhat} + \mathcal{F}^{\orta}(\delta \pb)_{\ortAhat}\,,
\end{align}
with~$\{\text{m}_1,\text{m}_2\}$ equal to $\{\text{s},\text{f}\}$ 
or $\{\text{f},\text{s}\}$.
 Note that, since the resulting similarity transform
matrix~$\mathbf{T}_S$ and inverse~$\mathbf{T}_S^{-1}$ always exist and
have bounded components, the regularity
condition~\eqref{equation:regularitycondition} is fulfilled. This
shows that the prototype algebraic constraint free system is in the
uppercase strongly hyperbolic. Since all the eigenvalues have
absolute values smaller than or equal to 1, the system must also be
strongly hyperbolic in the lowercase frame.

%%%%%%%%%%%%%%%%%%%%%%%%%%%%%%%%%%%%%%%%%%%%%%%%%%%%%%%%%%%%%
\subsection{Lowercase formulation}
\label{section:LowerCaseGRMHD}
%%%%%%%%%%%%%%%%%%%%%%%%%%%%%%%%%%%%%%%%%%%%%%%%%%%%%%%%%%%%%

We know already that the prototype algebraic constraint free
formulation is strongly hyperbolic. Nevertheless the lowercase
eigenvalues and eigenvectors would be important if we were to employ
the system numerically, and therefore we derive them in this
subsection.

%%%%%%%%%%%%%%%%%%%%%%%%%%%%%%%%%%%%%%%%%%%%%%%%%%%%%%%%%%%%%
\subsubsection{Recovering the lowercase quantities}
\label{section:RecoverEVandEVecsGRMHD}
%%%%%%%%%%%%%%%%%%%%%%%%%%%%%%%%%%%%%%%%%%%%%%%%%%%%%%%%%%%%%

To obtain the lowercase eigenvalues and eigenvectors as well as the
characteristic variables, we use the procedure described in
Sec.~\ref{section:RecoverEVandEVecsGeneral}. The recovery will
be done in several steps.

\textit{Step one.} First of all, we take the calculated uppercase
eigenvalues~(\ref{equation:EigenvalueMHDUpperCaseEntropy})--(\ref{equation:EigenvalueMHDUpperCaseMgsFast})
and replace the vector~$S^a$
by~$S_{\lambda}^{a}=(S^a-W(\lambda-WV^S))/N$, whereby we obtain the
new uppercase eigenvalues
\begin{align}
\lambda^{\text u}_{(\text e)}&=0\,, \label{equation:EigenvalueMHDUpperCaseEntropySlambda}\\
\lambda^{\text u}_{(\text{c})}&=0\,,\label{equation:EigenvalueMHDUpperCaseConstraintSlambda}\\
\lambda^{\text u}_{(\text{a}\pm)}&=\pm\frac{b^{S_{\lambda}}}{\sqrt{\rho_0h^*}}\,, \label{equation:EigenvalueMHDUpperCaseAlfvenSlambda}\\
\lambda^{\text u}_{(\text{s}\pm)}&=\pm\sqrt{\zeta_{\text{S}_{\lambda}}-\sqrt{\zeta_{\text{S}_{\lambda}}^2-\xi_{\text{S}_{\lambda}}}}\,,\label{equation:EigenvalueMHDUpperCaseMgsSlowSlambda}\\
\lambda^{\text u}_{(\text{f}\pm)}&=\pm\sqrt{\zeta_{\text{S}_{\lambda}}+\sqrt{\zeta_{\text{S}_{\lambda}}^2-\xi_{\text{S}_{\lambda}}}}\,,\label{equation:EigenvalueMHDUpperCaseMgsFastSlambda}
\end{align}
where we used the shorthands
\begin{align}
\zeta_{\text{S}_{\lambda}}=\frac{\left( b^2+\cs^2\left[\left(b^{S_{\lambda}}\right)^2+\rho_0h\right] \right)}{2\rho_0 h^*}\,,\ \ 
\xi_{\text{S}_{\lambda}}=\frac{\left(b^{S_{\lambda}}\right)^2\cs^2 }{\rho_0 h^*}\,,
\end{align}
and the magnetic field vector in the new 
direction~$S_{\lambda}^a$ becomes
\begin{align}
b^{S_{\lambda}}=&b^a S^{\lambda}_a=\frac{1}{N}\left(b^S-W(b^aV_a)(\lambda-WV^{S}) \right)\,,\\
N=&\sqrt{(W\lambda-W^2 V^S)^2+1+(V^S)^2W^2-\lambda^2}\,.
\end{align}
We want to reiterate that the relation $WV^S=-v^{\s}$ holds and is
used at several points in this paper.

\textit{Step two.} We calculate now the lowercase eigenvalues by use
of Eq.~\eqref{equation:frameconnectionEigenvalues}, that is,
\begin{align}
\frac{1}{N} W( \lambda-WV^S) = \lambda_{\text u}[S^a_{\lambda}]\,.
\label{equation:recoverlowercaseeigenvalues}
\end{align}
For example, taking~$\lambda_{\text u}[S_{\lambda}^a]=\lambda^{\text
u}_{(\text e)}=0$, we arrive with the lowercase entropy wave
speed~$\lambda_{(\text e)}=WV^S$, the normalization factor~$N$ becomes
unity, and~$S^a$ and~$S_{\lambda}^{a}$ are identical.

\textit{Step three}. We now transform the uppercase left eigenvectors
for~$S_{\lambda}^{a}$ in the lowercase left eigenvectors for the
state vector,
\begin{align*}
(\delta \mathbf{U})_{\shat,\,\ortAhat}=(\delta p,(\delta\vhat)_{\shat}, (\delta\vhat)_{\ortAhat}, (\delta \pb)_{\shat},
(\delta\pb)_{\ortAhat}, \delta\varepsilon)^T.
\end{align*} The transformation
is~$\lambda$ dependent and therefore has to be done in each eigenspace
independently. We take
Eq.~\eqref{equation:ObteinLeftEigenvectors}, that is,
\begin{align}
\left.\mathbf{l}_{\lambda}^{\text n}\right|_{\s}=
\left.\mathbf{l}_{\lambda_{\text u}}^{\text
  u}[S^a_{\lambda}]\right|_{\mathbf{S}_{\lambda}}
\left(\mathbbm{1}+\left.\mathbf{B}^V
\right|_{\mathbf{S}_{\lambda}}\right)\mathbf{T}_{\lambda}\,,
\end{align}
where we use the eigenvectors~$\left.\mathbf{l}_{\lambda_{\text
 u}}^{\text u}[S^a]\right|_{\mathbf{S}}$ explicitly written
in~\eqref{equation:LeftEigenvectorsMHDUpperCaseS} and replace all
basis vectors with the ones associated with~$S_{\lambda}^a$. The
matrices~$\left(\mathbbm{1}+\left.\mathbf{B}^V
\right|_{\mathbf{S}}\right)$ and~$\left(\mathbbm{1}+\left.\mathbf{B}^V
\right|_{\mathbf{S}_{\lambda}}\right)$ for bases~$\mathbf{S}=(S^a,\Qa^a,\Qb^a)$
and~$\mathbf{S}_{\lambda}=(S^a_{\lambda},\Qa^a_{\lambda},\Qb^a_{\lambda})$ can be found in
the notebook.

To obtain the basis transformation~$\mathbf T_{\lambda}$, we need to
give a little more details: writing~$S_{\lambda}^{ a}$ in the
basis~$\mathbf{S}$, we get
\begin{align}
S_{\lambda}^{\ a}=c_SS^a +c_1\Qa^a+ c_2\Qb^a\,,\non\\\
c_S=\frac{1+(W^2V^S-W \lambda)V^S}{N}\,,\non\\
c_1=\frac{(W^2V^S-W \lambda)V^{\Qa}}{N}\,,\non\\
c_2=\frac{(W^2V^S-W \lambda)V^{\Qb}}{N}\,.
\end{align}
This relation defines a rotation of the basis,
so we are able to build a transformation matrix which is an element of
SO(3). By denoting~$\Qa_{\lambda}^{a}$ and $\Qb_{\lambda}^{a}$ as
rotated basis vectors~$\Qa^a$ and~$\Qb^a$, respectively, the rotation
matrix is given by
\begin{align}
\mathbf{R}=\begin{pmatrix}
c_S& c_1&c_2\\ 
-c_1&\frac{c_S c_1^2+c_2^2}{c_1^2+c_2^2}&\frac{(c_S-1)c_1 c_2}{c_1^2+c_2^2}\\ 
-c_2&\frac{(c_S-1)c_1 c_2}{c_1^2+c_2^2}&\frac{ c_1^2+c_Sc_2^2}{c_1^2+c_2^2}\\
\end{pmatrix},
\end{align}
such that
\begin{align}
\begin{pmatrix}
S_{\lambda}^{ a}\\
\Qa_{\lambda}^{ a}\\
\Qb_{\lambda}^{ a}
\end{pmatrix}=
\mathbf{R}
\begin{pmatrix}
S^a\\
\Qa^a\\
\Qb^a
\end{pmatrix}\label{equation:TransformationUpperCaseBasis}.
\end{align}
Since~$\mathbf R \in \text{SO}(3)$, we may transpose to
invert~$\mathbf{R}^T=\mathbf{R}^{-1}$. The associated lowercase bases
obey the same transformation, since we just have to multiply
Eq.~\eqref{equation:TransformationUpperCaseBasis}
with~$\gamma^b{}_a$. The transformation matrix is taken to
be~$\mathbf{T}_{\lambda}=\text{diag}(1,\mathbf{R},\mathbf{R},1)$.  The
derivative of the state vector transforms like
\begin{align}
\mathbbm{1}(\nabla_{z} \mathbf U)_{\mathbf S}
=\mathbf T_{\lambda}^T  (\nabla_{z} \mathbf U)_{\mathbf{S}_\lambda}
\end{align}
for any vector~$z^a$.

\textit{Step four}. For a last step, we have to calculate the right
eigenvectors by Eq.~\eqref{equation:ObteinRightEigenvectors}, so
we arrive with
\begin{align}
\left.\mathbf{r}_{\lambda}^{\text n}\right|_{\s}=
\mathbf{T}_{\lambda}^T \left.\mathbf{r}_{\lambda_{\text u}}^{\text
 u}[S^a_{\lambda}]\right|_{\mathbf{S}_{\lambda}}.
\end{align}
For this, we will take the right
eigenvectors~$\left.\mathbf{r}_{\lambda_{\text u}}^{\text
  u}[S^a]\right|_{\mathbf{S}} $ given
in~\eqref{equation:RightEigenvectorsMHDUpperCaseS} and replace the
basis vectors.

%%%%%%%%%%%%%%%%%%%%%%%%%%%%%%%%%%%%%%%%%%%%%%%%%%%%%%%%%%%%%
\subsubsection{Definitions and formulas}
%%%%%%%%%%%%%%%%%%%%%%%%%%%%%%%%%%%%%%%%%%%%%%%%%%%%%%%%%%%%%

Let us first define some new relations and quantities:
\begin{align}
a:=&\,N \lambda_{\text u}=W\lambda-W^2 V^S=W\lambda+\vhat^{\s}\,,\nonumber\\
\mathcal B:=& N b^a S^{\lambda}_a=b^S-(b^aV_a)a\non \\
=&\,b^S+(b^aV_a)W(V^S W-\lambda)\,,\nonumber\\
\mathcal{G}:=&\,1+(V^S)^2W^2-\lambda^2\,,&\nonumber\\
N^2=&\,a^2+\mathcal{G}\,.
\end{align}
These definitions are motivated by those
in Refs.~\cite{AniPen87,AntMirMar10} 
in regard to the covariant approach of
characteristic analysis shown by
Eq.~\eqref{equation:ConnectionPhiToLiterature}.

In analogy to the uppercase, we write the magnetic field four-vector
as
\begin{align}
b^a=b^{S_{\lambda}}S_{\lambda}^a+b_{\perp}^a\,,\qquad
 b^2=(b^{S_{\lambda}})^2+b_{\perp}^2\,,
\end{align}
with
\begin{align}
|b_{\perp}|^2=b^2-(b^{S_{\lambda}})^2=b_{\perp}^ab^{\perp}_a\,.
\end{align} 
Please note that we nevertheless still use capital letters for
contraction with~$\perpQ$,
e.g.,~$b_{\perp}^{\orta}={\perpQ^\orta\!_a}b_{\perp}^a$. In
general,~$b_{\perp}^S\neq 0$ is not vanishing. These definitions are
taken for all lowercase characteristic
quantities. Since~$b_{\perp}^a$ is orthogonal to~$S_{\lambda}^a$, we
use the relation~$b_{\perp}^S=a (b_{\perp}^aV_a)$ several times.

%%%%%%%%%%%%%%%%%%%%%%%%%%%%%%%%%%%%%%%%%%%%%%%%%%%%%%%%%%%%%
\subsubsection{Entropy wave}
%%%%%%%%%%%%%%%%%%%%%%%%%%%%%%%%%%%%%%%%%%%%%%%%%%%%%%%%%%%%%

Taking~$\lambda_{\text u}=0$ as
in~\eqref{equation:EigenvalueMHDUpperCaseEntropySlambda}, we arrive
at the lowercase eigenvalue
\begin{align}
\lambda_{(\text e)}=WV^S\,.
\end{align}
In this case, we have~$N=1$ and~$S_{\lambda}^a=S^a$, and the left and
right eigenvectors for entropy waves remain the same,
\begin{align}
&\begin{pmatrix}
-\frac{p}{\cs^2\rho_0^2h} & 0 & 0^{\orta} &  0 &  0^{\orta} & 1
\end{pmatrix},\\
&\begin{pmatrix}
0 & 0 & 0_{\ortb} &  0 &  0_{\ortb} & 1
\end{pmatrix}^{T}.
\end{align}

%%%%%%%%%%%%%%%%%%%%%%%%%%%%%%%%%%%%%%%%%%%%%%%%%%%%%%%%%%%%%
\subsubsection{Constraint wave}
%%%%%%%%%%%%%%%%%%%%%%%%%%%%%%%%%%%%%%%%%%%%%%%%%%%%%%%%%%%%%

Taking~$\lambda_{\text u}=0$ as
in~\eqref{equation:EigenvalueMHDUpperCaseConstraintSlambda}, we arrive
at the lowercase eigenvalue
\begin{align}
\lambda_{(\text c)}=WV^S.
\end{align}
In this case, we have~$N=1$ and~$S_{\lambda}^a=S^a$, and the left and
right eigenvectors for constraint waves become
\begin{align}
&\begin{pmatrix}
0& (b^{\ortc}V_{\ortc}) & -b^SV^{\orta} &  1 &  0^{\orta} & 0
\end{pmatrix},\\
&\begin{pmatrix}
0 & 0 & 0_{\ortb} &  1 &  0_{\ortb} & 0
\end{pmatrix}^{ T},
\end{align}
respectively.

%%%%%%%%%%%%%%%%%%%%%%%%%%%%%%%%%%%%%%%%%%%%%%%%%%%%%%%%%%%%%
\subsubsection{Alfv\'en waves}
%%%%%%%%%%%%%%%%%%%%%%%%%%%%%%%%%%%%%%%%%%%%%%%%%%%%%%%%%%%%%

For Alfv\'en waves, we obtain by
taking~\eqref{equation:EigenvalueMHDUpperCaseAlfvenSlambda} the
 lowercase eigenvalues
\begin{align}
\lambda_{(\text a \pm)}=& \frac{b^S+V^SW^2
  \left[ (b^aV_a)\pm\sqrt{\rho_0 h^*}\right]}{W
  \left[ (b^aV_a)\pm\sqrt{\rho_0 h^*}\right]}\,.
\end{align}
They coincide up to a minus sign and factor~$W$ (due to our choice of
the spatial vector) with the literature~\cite{AntMirMar10}. The
already rescaled left and right eigenvectors to~$\lambda_{(\text a
  \pm)}$ are
  \begin{align}
&\begin{pmatrix}
\pm\frac{\epsuS_{\ortb\ortc}}{\sqrt{\rho_0 h^*}} \frac{V^{\ortb} 
b_{\perp}^{\ortc}}{|b_{\perp}|}  \\
b^S \epsuS_{\ortb\ortc}\frac{b_{\perp}^{\ortb}V^{\ortc}}{|b_{\perp}|}\\
\left( (b^aV_a)\pm \sqrt{\rho_0h^*}\right)\epsu^{\orta}{}_{ b c}
\frac{N S_{\lambda_{(\text a \pm)}}^b b_{\perp}^{c}}{|b_{\perp}|}\\
0\\
-\epsuS^{\orta}{}_{\ortb}
\left(\frac{b_{\perp}^{\ortb}}{|b_{\perp}|}\pm\frac{|b_{\perp}|V^{\ortb}}{\sqrt{\rho_0 h^*}}\right)\\
0
\end{pmatrix}^{T}
\end{align}
and
  \begin{align}
&\begin{pmatrix}
0\\
\frac{b^S}{\sqrt{\rho_0 h^*}}  \epsuS_{\orta\ortc}
\frac{b_{\perp}^{\orta}V^{\ortc}}{|b_{\perp}|}\\
\frac{(b^bV_b)\pm \sqrt{\rho_0h^*}}{\sqrt{\rho_0 h^*}}\epsu_{\ortb a c}
\frac{N S_{\lambda_{(\text a \pm)}}^a b_{\perp}^{c}}{|b_{\perp}|}\\
\pm b^S \epsuS_{\orta\ortc}\frac{b_{\perp}^{\orta}V^{\ortc}}{|b_{\perp}|}\\
\left(\sqrt{\rho_0h^*}\pm(b^bV_b) \right)\epsu_{\ortb a c}
\frac{N S_{\lambda_{(\text a \pm)}}^a b_{\perp}^{c}}{|b_{\perp}|}\\
0
\end{pmatrix},
\end{align}
respectively. 

%%%%%%%%%%%%%%%%%%%%%%%%%%%%%%%%%%%%%%%%%%%%%%%%%%%%%%%%%%%%%
\subsubsection{Magnetosonic waves}
%%%%%%%%%%%%%%%%%%%%%%%%%%%%%%%%%%%%%%%%%%%%%%%%%%%%%%%%%%%%%

The uppercase slow and fast magnetosonic eigenvalues are defined
in~\eqref{equation:EigenvalueMHDUpperCaseMgsSlowSlambda}
and~\eqref{equation:EigenvalueMHDUpperCaseMgsFastSlambda}. Inserting
one of these eigenvalues into
Eq.~\eqref{equation:recoverlowercaseeigenvalues} one can show
after some manipulations (given in the notebook~\cite{HilWebsite_aastex}) that the lowercase
magnetosonic eigenvalues are solutions of the quartic equation
\begin{align}
\mathcal{N}_4 =\rho_0 h \left(\frac{1}{\cs^2}-1\right)a^4
-\left(\rho_0 h+\frac{b^2}{\cs^2}\right)a^2 \mathcal{G}
+\mathcal{B}^2 \mathcal{G} =0\,,
\end{align}
where~$\mathcal{N}_4$ is the same polynomial as obtained
by Ref.~\cite{AniPen87}. We have computed analytic expressions for the
magnetosonic eigenvalues. Explicitly written out, however, they are
rather long, and hence a numerical computation relying on the
characteristic information may be better served by using some
root finder.

The rescaled left and right magnetosonic eigenvectors with eigenvalues
closer to the Alfv\'en speeds can be expressed as
\begin{align}
\begin{pmatrix}
-\frac{\mathcal{G}}{\rho_0 h}\frac{(b_{\perp}^aV_a)}{|b_{\perp}|}\left(\frac{\mathcal{B}}{a}\right)-\frac{(1-aV^S)\mathcal{G}}{\rho_0 h}\mathcal{F}\\
a(a^2+ \mathcal{G})\left((1-\cs^2)\mathcal{F}+\left[\left(\frac{\mathcal{B}}{a}\right)+(b^aV_a) \right]\frac{(b_{\perp}^aV_a)}{|b_{\perp}|}\right)\\
(a^2 +\mathcal{G})\left[\left(\frac{\mathcal{B}}{a}\right)+(b^aV_a) \right]\frac{b_{\perp}^{\orta}}{|b_{\perp}|}\\
0\\
a\frac{b_{\perp}^S V^{\orta}}{|b_{\perp}|}+(1-aV^S)\frac{b_{\perp}^{\orta}}{|b_{\perp}|}\\
0
\end{pmatrix}^{T}_{(\text m \pm)}
\end{align}
and
\begin{align}
\begin{pmatrix}
-\cs^2 \rho_0 h \mathcal{G}(a^2 +\mathcal{G})\mathcal{F}\\
\mathcal{G}\left(\frac{\mathcal{B}}{a}\right)\frac{b_{\perp}^S}{|b_{\perp}|}+a(1-aV^S)\mathcal{G}\mathcal{F}\\
\mathcal{G}\left(\frac{\mathcal{B}}{a}\right)\frac{b^{\perp}_{\ortb}}{|b_{\perp}|}-a^2\mathcal{G}\mathcal{F}V_{\ortb}\\
(a^2 +\mathcal{G})\frac{\rho_0 h}{|b_{\perp}|}b_{\perp}^S\\
(a^2 +\mathcal{G})\frac{\rho_0 h}{|b_{\perp}|}b^{\perp}_{\ortb}\\
-\frac{p}{\rho_0}(a^2 +\mathcal{G})\mathcal{G}\mathcal{F}
\end{pmatrix}_{(\text m \pm)}.
\end{align} 
The remaining two left and right magnetosonic lowercase eigenvectors
are given by
\begin{align}
&\begin{pmatrix}
-\frac{\mathcal{G}}{\rho_0 h}\left(\frac{\mathcal{B}}{a}\right)(\mathcal{C}^aV_a)+\frac{(1-aV^S)}{\cs^2\rho_0 h(a^2+\mathcal{G})}\\
\left(1-\frac{1}{\cs^2}\right)\frac{a}{\mathcal{G}}+a(a^2 +\mathcal{G})\left[\left(\frac{\mathcal{B}}{a}\right)+(b^aV_a) \right](\mathcal{C}^bV_b)\\
(a^2 +\mathcal{G})\left[\left(\frac{\mathcal{B}}{a}\right)+(b^aV_a) \right]\mathcal{C}^{\orta}\\
0\\
aC^S V^A+(1-aV^S)\mathcal{C}^A\\
0
\end{pmatrix}_{(\text m \pm)}^{T}
\end{align}
and
\begin{align}
&\begin{pmatrix}
 \rho_0 h \\
\left(\frac{\mathcal{B}}{a}\right)\mathcal{G}\mathcal{C}^S-\frac{a(1-aV^S)}{\cs^2(a^2+\mathcal{G})}\\
\left(\frac{\mathcal{B}}{a}\right)\mathcal{G}\mathcal{C}_{\ortb}+\frac{a^2}{\cs^2(a^2+\mathcal{G})}V_{\ortb}\\
(a^2 +\mathcal{G})\rho_0 h\mathcal{C}^S\\
(a^2 +\mathcal{G})\rho_0 h\mathcal{C}_{\ortb}\\
\frac{p}{\cs^2\rho_0}
\end{pmatrix}_{(\text m \pm)}.
\end{align}
Here, we took the definitions 
\begin{align}
\mathcal{C}^a=\frac{b_{\perp}^a}{a^2 \rho_0 h-\mathcal{G}b^2}\,,\non \\
\mathcal{F}=\frac{|b_{\perp}|}{\cs^2(a^2+\mathcal{G})-a^2}\,,
\end{align}
where for type II and type II$'$ degeneracies we take
\begin{align}
\mathcal{C}^a=0\,, \qquad \mathcal{F}=0\,,
\end{align}
and
\begin{align}
\frac{b^{\perp}_{\ortc}}{|b_{\perp}|} 
=\frac{1}{\sqrt{2}}(\Qa^{\lambda}_{\ortc}+\Qb^{\lambda}_{\ortc})\,.
\end{align}

%%%%%%%%%%%%%%%%%%%%%%%%%%%%%%%%%%%%%%%%%%%%%%%%%%%%%%%%%%%%%
\subsubsection{Degeneracies in lowercase GRMHD}\label{section:LowerCaseGRMHDdegenAna}
%%%%%%%%%%%%%%%%%%%%%%%%%%%%%%%%%%%%%%%%%%%%%%%%%%%%%%%%%%%%%

In the lowercase frame, the degeneracy analysis is performed just as
in the uppercase setting. One has only to replace the vector~$S^a$
by~$S^a_{\lambda}$ and the corresponding orthogonal basis vectors as
well. We then have for type~$\text{I}$ degeneracy 
that~$b^aS_a^{\lambda}$ is equal to zero. In this case, the entropic
wave, the constraint wave, the two Alfv\'en waves, and the two slow
magnetosonic waves propagate at the same speed
($\lambda_{(\text e)}=\lambda_{(\text c)}=\lambda_{(\text a \pm)}=\lambda_{(\text s
  \pm)}=-v^{\s}$). For type~$\text{II}$ degeneracy, the tangential
magnetic field vector, $b_{\perp}^a=\perpQl_{\ b}^{ a} b^b $,
$\perpQl_{\ b}^{ a}=\gamu_{\ b}^{ a}-S^a_{\lambda}S_b^{\lambda} $,
vanishes. In this case, one of the Alfv\'en waves and one of the
magnetosonic waves of the appropriate class (here denoted by a
 superscript as in Ref.~\cite{AntMirMar10}) have the same speed
($\lambda^+_{(\text a )}=\lambda^+_{(\text s )}$
 or~$\lambda^-_{(\text a  )}=\lambda^-_{(\text s )}$ 
 or~$\lambda^+_{(\text a )}=\lambda^+_{(\text f )}$
or $\lambda_{(\text a )}^-=\lambda^-_{(\text f )}$). In the type II$'$
degeneracy, one Alfv\'en wave and the slow and fast magnetosonic waves
of the appropriate class travel at the same speed ($\lambda_{(\text a
  )}^+=\lambda^+_{(\text s )}=\lambda^+_{(\text f )}$ 
  or 
  $\lambda^-_{(\text a )}=\lambda^-_{(\text s )}=\lambda^-_{(\text f )}$).
   In the uppercase, we
have for type~$\text{II}$ and type~$\text{II}'$ degeneracies that both
Alfv\'en speeds are degenerate at the same time. Replacing~$S^a$
by~$S^a_{\lambda}$ leads to different SO(3)-transformations for
different values of~$\lambda$. Therefore, in the lowercase, this 
cannot be fulfilled in general. A more detailed description and
derivation can be found in Ref.~\cite{Sch18}.

%%%%%%%%%%%%%%%%%%%%%%%%%%%%%%%%%%%%%%%%%%%%%%%%%%%%%%%%%%%%%
\subsubsection{Characteristic variables}
\label{section:LowerCaseGRMHDdegenAna}
%%%%%%%%%%%%%%%%%%%%%%%%%%%%%%%%%%%%%%%%%%%%%%%%%%%%%%%%%%%%%

The characteristic variables valid for all degeneracies are
\begin{align}
\hat{\text{U}}_0=&\delta \varepsilon-\frac{p}{\cs^2 \rho_0^2 h}\delta
p\,, \non\\
%%%%%%%%%%%%%%%%%%%%%%%%%%%%%
 \hat{\text{U}}_{\text c}=&(\delta
\pb)_{\shat}+(b^AV_A)(\delta\vhat)_{\shat}-b^SV^A(\delta\vhat)_{\ortAhat}\,,\non\\ 
%%%%%%%%%%%%%%%%%%%%%%%%%%%%%%%
\hat{\text{U}}_{\text a \pm}=&\pm\frac{\epsuS_{\ortb\ortc}}{\sqrt{\rho_0 h^*}} \frac{V^{\ortb} 
b_{\perp}^{\ortc}}{|b_{\perp}|}\delta p +b^S \epsuS_{\ortb\ortc}\frac{b_{\perp}^{\ortb}V^{\ortc}}{|b_{\perp}|}(\delta\vhat)_{\hat{\s}}\non\\
+&\left( (b^aV_a)\pm \sqrt{\rho_0h^*}\right)
\frac{N S_{\lambda_{(\text a \pm)}}^b b_{\perp}^{c}}{|b_{\perp}|}\epsu^{\orta}{}_{ b c}(\delta \vhat)_{{\ortAhat}}\non\\
-&\left(\frac{b_{\perp}^{\ortb}}{|b_{\perp}|}\pm\frac{|b_{\perp}|V^{\ortb}}{\sqrt{\rho_0 h^*}}\right)\epsuS^{\orta}{}_{\ortb}(\delta \pb)_{{\ortAhat}}
\,, \non \\
\end{align}
for entropy, constraint, and Alfv\'en waves, and
%%%%%%%%%%%%%%%%%%%%%%%%%%%%%%% 
\begin{align}
\hat{\text{U}}_{\text m_1 \pm}&=-\left(\frac{\mathcal{G}}{\rho_0 h}\frac{(b_{\perp}^aV_a)}{|b_{\perp}|}\left(\frac{\mathcal{B}}{a}\right)+\frac{(1-aV^S)\mathcal{G}}{\rho_0 h}\mathcal{F}\right)\delta p \non\\
&+a(a^2+ \mathcal{G})(1-\cs^2)\mathcal{F}(\delta\vhat)_{\hat{\s}}\non\\
&+N^2\left[\left(\frac{\mathcal{B}}{a}\right)+(b^aV_a) \right]\left( \frac{b_{\perp}^S}{|b_{\perp}|}(\delta\vhat)_{\hat{\s}} +\frac{b_{\perp}^{\orta}}{|b_{\perp}|}(\delta \vhat)_{{\ortAhat}} \right)\non\\
& +\left(a\frac{b_{\perp}^S V^{\orta}}{|b_{\perp}|}+(1-aV^S)\frac{b_{\perp}^{\orta}}{|b_{\perp}|}\right)(\delta \pb)_{{\ortAhat}}\,,\non\\
%%%%%%%%%%%%%%%%%%%%%%%%%%%%%%%
 \hat{\text{U}}_{\text m_2 \pm}&=\left(\frac{(1-aV^S)}{\cs^2\rho_0 h(a^2+\mathcal{G})}-\frac{\mathcal{G}}{\rho_0 h}\left(\frac{\mathcal{B}}{a}\right)(\mathcal{C}^aV_a)\right)\delta p\non\\
& +\left(1-\frac{1}{\cs^2}\right)\frac{a}{\mathcal{G}}(\delta\vhat)_{\hat{\s}}\non\\
&+N^2\left[\left(\frac{\mathcal{B}}{a}\right)+(b^aV_a) \right]\left(C^{S}(\delta \vhat)_{{\shat}}+C^{\orta}(\delta \vhat)_{{\ortAhat}}\right) \non\\
&+\left(aC^S V^A+(1-aV^S)\mathcal{C}^A\right)(\delta \pb)_{{\ortAhat}}\,,
\end{align}
for magnetosonic waves, with~$\{\text{m}_1,\text{m}_2\}$ equal to $\{\text{s},\text{f}\}$ or $\{\text{f},\text{s}\}$. 
The functions on the right-hand side 
of~$\hat{\text{U}}_{\text m \pm}$ are evaluated with the 
corresponding eigenvalue.

%%%%%%%%%%%%%%%%%%%%%%%%%%%%%%%%%%%%%%%%%%%%%%%%%%%%%%%%%%%%%
\subsection{Weak hyperbolicity of the flux-balance law formulation of GRMHD}\label{section:FluxConsGRMHD}
%%%%%%%%%%%%%%%%%%%%%%%%%%%%%%%%%%%%%%%%%%%%%%%%%%%%%%%%%%%%%

We want now to analyze whether or not the flux-balance law formulation
of GRMHD as in Ref.~\cite{AntZanMir05} is strongly hyperbolic. 
To do so, we need to find the values
for the formulation parameters such that a linear combination of the
Eqs.~(\ref{equation:systemUpperCaseMHDp})--(\ref{equation:systemUpperCaseMHDeps})
is equal to the system in the form of 
Refs.~\cite{CerDurFon08,AntMirMar10},
up to the use of the same evolved variables. In fact, several
flux-balance law formulations exist, but remarkably, in our variables,
they differ only by a linear combination of the conservation of
particle number equation.

To reproduce the flux-balance law formulation given
in Ref.~\cite{IbaCorCar15}, we worked in computer algebra 
and found the
linear combination of our equations that reproduced the flux-balance
ones. This was done ignoring all derivatives of the normal
vector~$n^a$. In our analysis, we may ignore all derivatives of the
normal vector anyway since they only contribute to the source
vector and do not affect the principal part.  The coefficients are
then
\begin{align}
\omega^{(p)}&=\frac{\kappa}{\rho_0} (b^cV_c)\,, & \omega^{(\vhat)}_a
&=\frac{1}{\rho_0h}b_a\,,\non \\
\omega^{(\pb)}_a&=-V_a\,,
&\omega^{(\varepsilon)}&=\frac{1}{\rho_0}(b^cV_c)\,.
\end{align}
Proceeding in the same way as for the previous formulation, the
principal symbol~$\mathbf{P}^S$ becomes
\begin{align}
\begin{pmatrix}
0&-\cs^2 \rho_0 h & 0^{\ortb} & \frac{\kappa}{\rho_0}(b^cV_c)&  0^{\ortb}& 0\\
-\frac{\left(b^S\right)^2+\rho_0 h }{ \rho_0^2 hh^*}&0&0^{\ortb}
& \frac{b^S}{\rho_0h} & -\frac{b^{\ortb}}{\rho_0 h^*}& 0\\
-\frac{b^Sb_{\orta} }{ \rho_0^2 hh^*}& 0_{\orta}&0^{\ortb}\!_{\orta}
&\frac{b_{\orta}}{\rho_0h}
&\frac{b^{S}}{\rho_0 h^*}\perpQ^{\ortb}\!\!_{\orta}&0_{\orta}\\
0& 0&0^{\ortb}& -V^S&0^{\ortb}& 0 \\
0_{\orta}& -b_{\orta}&b^{S}\perpQ^{\ortb}\!\!_{\orta} & -V_{\orta}
& 0^{\ortb}\!_{\orta} & 0_{\orta} \\
0& -\frac{p}{\rho_0}& 0^{\ortb} & \frac{1}{\rho_0}(b^cV_c)& 0^{\ortb} & 0\\ 
\end{pmatrix}\label{equation:PrincipalSymbolFluxGRMHD}
\end{align}
the characteristic polynomial is then of the form
\begin{align}
P_{\lambda}=\frac{1}{(\rho_0h^*)^2}\lambda(\lambda+V^S)
P_{\text{Alfv\'en}} P_{\text{mgs}}\,,
\end{align}
where~$P_{\text{Alfv\'en}}$ and~$P_{\text{mgs}}$ coincide with the
polynomials given earlier in
Eqs.~\eqref{equation:CharPolAlfvenUpperCaseMHD}
and~\eqref{equation:CharPolmgsUpperCaseMHD}. As expected, the
eigenvalue associated with the constraint has changed from zero, in
the previous formulation, to~$-V^S$. Therefore, new degeneracies have
to be considered, for example, when the constraint and entropic speeds
collide. This occurs when~$V^S=0$, in which case we find that the
principal symbol is not diagonalizable. Hence, the system is only
weakly hyperbolic and has an ill-posed IVP. To get
an intuitive idea of what precisely goes wrong, we may consider the
left eigenvectors associated with the entropy and constraint waves in
generic directions and then consider a limiting direction
with~$V^S\to0$. These are
\begin{align}
\begin{pmatrix}
-\frac{p\rho_0}{\cs^2 \rho_0^2 h-\kappa p}\frac{V^S}{(b^cV_c)} & 0
& 0^{\orta} & 1 & 0^{\orta} & \frac{\cs^2 \rho_0^3 h}{\cs^2 \rho_0^2 h-\kappa p}
\frac{V^S}{(b^cV_c)}
\end{pmatrix}\,\non
\end{align}
and
\begin{align}
\begin{pmatrix}
0 & 0 & 0^{\orta} & 1 & 0^{\orta} & 0
\end{pmatrix}\,,\non
\end{align}
respectively, with eigenvalues~$\lambda_{(\text e)}=0$ and~$\lambda_{(\text c)}=-V^S$.
Both right eigenvectors can be found in our scripts but are
suppressed here because the constraint eigenvector is quite
lengthy. Taking the limit~$V^S\rightarrow0$, we immediately arrive at
the conclusion that the geometric multiplicity is only 1 as the two
vectors become coincident. The eigenvector can not be rescaled as for
the earlier degeneracies since only some entries in the left entropy
eigenvector become zero; the limit of the principal symbol is truly
problematic. This degeneracy was unfortunately overlooked
in Ref.~\cite{IbaCorCar15}, although there the focus was rather on the
convexity of the system as opposed to hyperbolicity. Nevertheless, we
have explicitly checked in our notebooks~\cite{HilWebsite_aastex} 
that, taking the lowercase matrices from Ref.~\cite{IbaCorCar15} 
and deriving the left eigenvectors of
the entropy and constraint waves, the exact same problem is
present. Deriving the right constraint eigenvector in the lowercase
frame is much worse than in the uppercase, however, so we only
evaluated the left ones. We want to stress that using the matrices
of Ref.~\cite{IbaCorCar15} is a completely independent calculation and
underlines the weak hyperbolicity of the system. Somewhat
interestingly, in the Newtonian limit, the flux-balance formulation,
see, for example,~\cite{JefTan64,BriWu88}, suffers from the same
degeneracy and is also only weakly hyperbolic. 

It should be explicitly noted that in more than one spatial dimension
the condition~$V^S=0$ will certainly be satisfied everywhere in space
for some~$S_a$. One should therefore avoid thinking that the breakdown
of hyperbolicity happens only on a set of measure zero in
spacetime. Rather the generic situation is that when the flow is
nontrivial there are specific bad directions everywhere in spacetime
which obstruct the well-posedness of the IVP. The fact
that only specific directions are problematic may make the effect in
numerical work hard to identify. In particular many tests of GRMHD are
focused on one-dimensional (nonsmooth) solutions, and by construction,
such experiments are insensitive to the breakdown identified
here. This will be studied in greater detail in future work.

We stress again that the result does not automatically apply to
formulations evolving the magnetic
four-potential~\cite{GiaRezBai10,EtiPasHaa15} nor systems with
divergence cleaning~\cite{ZanFamDum15}. It would naturally be
desirable to perform a similar analysis for those systems also.

%%%%%%%%%%%%%%%%%%%%%%%%%%%%%%%%%%%%%%%%%%%%%%%%%%%%%%%%%%%%%
\section{Hyperbolicity of RGRMHD}\label{section:RGRMHD}
%%%%%%%%%%%%%%%%%%%%%%%%%%%%%%%%%%%%%%%%%%%%%%%%%%%%%%%%%%%%%

In this section, we want to investigate the evolution equations used in
the literature for
RRMHD~\cite{TakIno11,Kom07,DumZan09,PalLehReu09,Miz13}
and~\cite{Pal12,BucZan13,QiaFenNob17,DioAliPal13,DioAliRez15}
describing RGRMHD and show that these two systems are weakly
hyperbolic and therefore have ill-posed IVPs. In this section, we will
use the lowercase frame exclusively. As in
Sec.~\ref{section:GRMHD}, we use Lorentz-Heaviside units where
vacuum permittivity and vacuum permeability are equal to 1. We start
by deriving the equations of motion for the state vector~$\textbf{U}$.

%%%%%%%%%%%%%%%%%%%%%%%%%%%%%%%%%%%%%%%%%%%%%%%%%%%%%%%%%%%%%
\subsection{Equations of RGRMHD}
%%%%%%%%%%%%%%%%%%%%%%%%%%%%%%%%%%%%%%%%%%%%%%%%%%%%%%%%%%%%%

As with earlier, we want to derive the evolution equations and are primarily
concerned with their mathematical structure. Interesting physical
facts, particularly those related to Ohm's law, will be sidelined
in our discussion.

%%%%%%%%%%%%%%%%%%%%%%%%%%%%%%%%%%%%%%%%%%%%%%%%%%%%%%%%%%%%%
\subsubsection{Augmented Maxwell equations}
%%%%%%%%%%%%%%%%%%%%%%%%%%%%%%%%%%%%%%%%%%%%%%%%%%%%%%%%%%%%%

As in the beginning of the last section about GRMHD, we take the
following definition of the field strength tensor for a generic
Eulerian observer with four-velocity~$n^a$,
\begin{align}
  F^{ab}&=n^a  E^b-n^b E^a+\epsilon^{abcd} n_c  B_d\,,
  \label{equation:DefinitionFieldStrengthTensorResMHD}\\
  {^*F^{ab}}&=n^a  B^b-n^b  B^a-\epsilon^{abcd} n_c  E_d\,,
  \label{equation:DefinitionDualFieldStrengthTensorResMHD}
\end{align}
with the Levi-Civit\`a tensor,
\begin{align}
\epsilon^{abcd}=-\frac{1}{\sqrt{-g}}\left[ a b c d \right]\,,
\end{align}
the Levi-Civit\`a symbol~$\left[ a b c d \right]$; $\left[ 0 1 2 3
 \right]=1$ and
\begin{align}
\epsilon^{abcd}n_a=\epsilon^{bcd}=\frac{1}{\sqrt{\gamma}}\left[ b c d
\right]\,,
\end{align}
where we follow the definition and convention
by Ref.~\cite{AlcDegSal09}. Please note that in this
convention~$2{^*F^{ab}}=-\epsilon^{abcd}F_{cd}$ holds.

To control the constraints during the evolution, the augmented scalar
fields~$\psi$ and~$\phi$ are introduced, see for
example~\cite{Kom07,PalLehReu09,DioAliRez15}, and hence the Maxwell
equations become
\begin{align}
\nabla_b\left( F^{ab}- g^{ab}\psi\right)&=\mathcal{J}^a-
\frac{1}{\tau} n^a \psi\,,\label{equation:AugMaxwellResMHDE}\\
\nabla_b\left({^*F^{ab}}- g^{ab}\phi\right)&=- \frac{1}{\tau} n^a
\phi\,. \label{equation:AugMaxwellResMHDB}
\end{align}
Note that in the literature the notation~$\kappa=\tau^{-1}$ is
normally employed. The electric four-current is split against~$n^a$
and~$\gamma^b{}_{a}$ defined by
\begin{align}
\mathcal{J}^a:= q n^a+J^a\,,\qquad
n_aJ^a=0\,. \label{equation:FourCurrentResMHD}
\end{align}

Proceeding with a~$3+1$ decomposition of~\eqref{equation:AugMaxwellResMHDE}
and~\eqref{equation:AugMaxwellResMHDB}
using~\eqref{equation:FourCurrentResMHD}, we arrive at the equations
\begin{align}
\gamma^a{}_b\Lie_n E^b&=
\epsilon^{abc}D_b B_c-\gamma^{ab}D_b\psi+S_{{\! (\boldsymbol{\mathrm{E}})}}^a\,,
\label{equation:LieERGRMHD}\\
\gamma^a{}_b\Lie_n B^b&= -\epsilon^{abc}D_b E_c-\gamma^{ab}D_b\phi
+S_{{\! (\boldsymbol{\mathrm{B}})}}^a\,,
\label{equation:LieBRGRMHD}\\
\Lie_n\psi&=-D_aE^a-\frac{1}{\tau}\psi+q\,,\label{equation:LiePsiRGRMHD}\\
\Lie_n\phi&=-D_aB^a-\frac{1}{\tau}\phi\,,\label{equation:LiePhiRGRMHD}
\end{align}
with sources
\begin{align*}
S_{{\!(\boldsymbol{\mathrm{E}})}}^a&=\frac{1}{\alpha}B_c
\epsilon^{abc}D_b \alpha+\Kn E^a-J^a\,,\\
S_{{\! (\boldsymbol{\mathrm{B}})}}^a&=-\frac{1}{\alpha}E_c
\epsilon^{abc}D_b\alpha+\Kn B^a\,.
\end{align*}
The constant~$\tau$ is the timescale for the exponential driving of
Eqs.~\eqref{equation:LiePsiRGRMHD}
and~\eqref{equation:LiePhiRGRMHD} toward the constraints
\begin{align}
D_aE^a=q\,,\\ 
D_aB^a=0\,, 
\end{align} 
respectively. The three-current~$J^a$ is given by generalized Ohm's
law; see below. We must to stress that although~$J^a$ is inside the
`source' term, it {\it could} contain derivatives of the evolved
variables. Such terms would then of course contribute to the principal
part.

As a consequence of the antisymmetry of the field strength tensor, we
have additionally a conservation law for the electric
charge,~$\nabla_a \mathcal{J}^a=0$, that is in the~$3+1$ language
\begin{align}
\Lie_n q=- \gamma^{ab}D_a J_b-\frac{1}{\alpha}J^bD_b
\alpha+\Kn\,q\,.\label{equation:LieQRGRMHD}
\end{align}

%%%%%%%%%%%%%%%%%%%%%%%%%%%%%%%%%%%%%%%%%%%%%%%%%%%%%%%%%%%%%
\subsubsection{Energy-momentum tensor}
%%%%%%%%%%%%%%%%%%%%%%%%%%%%%%%%%%%%%%%%%%%%%%%%%%%%%%%%%%%%%

The energy-momentum tensor~$T^{ab}$ of RGRMHD contains an ideal fluid
component,
\begin{align}
T^{a b}_{\rm{mat}}=\rho_0 h u^au^b+ p g^{ab}\,,
\end{align}
plus the standard electromagnetic energy-momentum tensor,
\begin{align}
T^{ab}_{\rm{em}}=F^{ac}F^b{}_c-\frac{1}{4}g^{ab}F_{cd}F^{cd}\,,
\end{align}
with a field strength tensor defined
in~\eqref{equation:DefinitionFieldStrengthTensorResMHD}. Writing~$F^{ab}$
in terms of~$E^a$ and~$B^a$, we obtain
\begin{align}
  T^{ab}_{\rm{em}}=& \frac{1}{2}\left(B_cB^c+E_cE^c\right)
  \left(\gamma^{ab}+n^a n^b \right)\non \\
  &-B^aB^b-E^aE^b+\left(n^{a}\epsilon^{bcd}
  +n^{b}\epsilon^{acd}\right)E_cB_d\,.
\end{align}

%%%%%%%%%%%%%%%%%%%%%%%%%%%%%%%%%%%%%%%%%%%%%%%%%%%%%%%%%%%%%
\subsubsection{Generalized Ohm's law}
%%%%%%%%%%%%%%%%%%%%%%%%%%%%%%%%%%%%%%%%%%%%%%%%%%%%%%%%%%%%%

The generalized Ohm's law provides us with an expression for the spatial
current~$J^a$. Explanations about the physical validity and form
of~$J^a$ can be found in the literature~\cite{Mei04,DioAliRez15}.
We consider here an equation for~$J^a$ which is of the form
\begin{align}
J^a=q v^a+\tilde{J}^a,\qquad
\tilde{J}^a=\tilde{J}^a(p,v_b,\varepsilon,E_c,B_d)\,,
\label{equation:DefinitionJRGRMHD}
\end{align}
where~$\tilde{J}^a$ contains no derivatives of the matter and
electromagnetic variables nor second-order or higher derivatives of
the metric tensor. This fairly general choice of~$J^a$ includes the
particular form used in the literature mentioned above, that is,
\begin{align}
J^a=q v^a +W \sigma \left( E^a+\epsilon^{abc}v_bB_c-(v_bE^b)v^a
\right)\,,\label{equation:DefinitionJRGRMHDLiterature}
\end{align}
where~$\sigma$ is the conductivity of the fluid and is permitted to be
an arbitrary function of the evolved variables besides the charge
density~$q$.

%%%%%%%%%%%%%%%%%%%%%%%%%%%%%%%%%%%%%%%%%%%%%%%%%%%%%%%%%%%%%
\subsubsection{Hydrodynamical equations}
%%%%%%%%%%%%%%%%%%%%%%%%%%%%%%%%%%%%%%%%%%%%%%%%%%%%%%%%%%%%%

To obtain the evolution equations for~$p$,~$v_a$, and~$\varepsilon$ we
take the conservation of the number of particles and the conservation
of energy momentum,
\begin{align}
\nabla_{a}(\rho_0 u^{a})=0\,, \label{equation:ConservationOfParticlesRGRMHD}\\
\nabla_{a}(T^{a b})=0\,, \label{equation:ConservationOfEnergyMomentumRGRMHD}
\end{align}
and proceed with the~$3+1$ split. After combining the equations, using
Maxwell equations and introducing the speed of sound, we arrive at
the evolution equation for the pressure,
\begin{align}
\Lie_n p=& (\cs^2-1)v^p \Wcs^2 D_p p- \cs^2 \rho_0 h \Wcs^2 \gamma^{p c}D_pv_c \non\\
&-c_2(E^b v_b)\gamma^{pc}D_pE_c -c_2(B^b v_b)\gamma^{pc}D_pB_c\non\\
&+\left(c_1E^p  -c_2 \epsilon^{bdp}B_bv_d\right) D_p \psi\non\\
&+\left(c_1B^p  +c_2\epsilon^{bdp}E_bv_d\right)D_p \phi+S^{(p)}\,;
\label{equation:EvolutionpRGRMHD}
\end{align}
for the fluid velocity,
\begin{align}
\gamma_{\ a}^b\Lie_n v_b=&-\frac{1}{W^2\rho_0 h}\left(\gamma^p{}_a
+(\cs^2-1)\Wcs^2v^pv_a \right)D_p p\non\\
&+\left(\frac{\cs^2\Wcs^2}{W^2}v_a \gamma^{pc}-v^p \gamma^c{}_a\right)D_p v_c\non\\
&+\frac{1}{W^2\rho_0 h}\left(E_a+c_2(E^bv_b)v_a \right)\gamma^{pc}D_pE_c\non\\
&+\frac{1}{W^2\rho_0 h}\left(B_a+c_2(B^bv_b)v_a \right)\gamma^{pc}D_pB_c\non\\
&+\frac{1}{W^2\rho_0 h}\left(\gamma_{ad}+c_2v_av_d \right)\epsilon^{bdp}B_b D_p \psi\non\\
&-\frac{1}{W^2\rho_0 h}\left(\gamma_{ad}+c_2v_av_d \right)\epsilon^{bdp}E_b D_p \phi\non\\
&-c_5v_aE^p D_p \psi-c_5v_aB^p D_p \phi+S^{(\textbf{v})}_a\,;\label{equation:EvolutionvRGRMHD}
\end{align}
 and for the internal specific energy,
\begin{align}
\Lie_n \varepsilon=&\frac{p \Wcs^2}{W^2 \rho_0^2 h}v^pD_p p
-\frac{p \Wcs^2}{\rho_0}\gamma^{p c}D_p v_c-v^pD_p\varepsilon\non\\
&-c_4(E^bv_b)\gamma^{pc}D_pE_c-c_4(B^bv_b)\gamma^{pc}D_pB_c\non\\
&+\left(c_3E^p  -c_4 \epsilon^{bdp}B_bv_d\right) D_p \psi\non\\
&+\left(c_3B^p  +c_4\epsilon^{bdp}E_bv_d\right)D_p \phi+S^{(\varepsilon)}\,;
\label{equation:EvolutionepsilonRGRMHD}
\end{align}
 with sources 
\begin{align}
S^{(p)}=& c_1(E^b J_b) +c_2 \epsilon^{bcd}B_bJ_cv_d \non \\
&+\Wcs^2\cs^2 \rho_0 h \gBi^{bc}\Kn_{bc}\,,\non\\
S^{(\textbf{v})}_a=&\frac{1}{W^2\rho_0 h}\left(\gamma_{ad}+c_2v_av_d \right)\epsilon^{bde}B_bJ_e\non\\
&-c_5(E^dJ_d)v_a-\frac{1}{\alpha}\gBi^c{}_aD_c\alpha\non\\
&- \cs^2 \frac{\Wcs^2}{W^2}\gBi^{bc} K_{bc} v_a-\Kn_{bc}v^bv^cv_a\,,\non\\
S^{(\varepsilon)}=&c_3(E^b J_b) +c_4 \epsilon^{bcd}B_bJ_cv_d \non \\
&+\frac{\Wcs^2p}{\rho_0} \gBi^{bc}\Kn_{bc}\,
\end{align}
where we have employed the shorthands,
\begin{align}
c_1=&\frac{\Wcs^2}{W^2 \rho_0}\left(\kappa W^2 +\cs^2 (W^2-1)\rho_0 \right)\,,\non\\
c_2=&\Wcs^2\left(\frac{\kappa}{\rho_0} +\cs^2\right)\,, \non\\
c_3=&\frac{\Wcs^2}{W^2 \rho_0^2 h}\left(p(W^2-1)+(\chi-\chi W^2+h W^2)\rho_0 \right)\,,\non\\
c_4=&\frac{\Wcs^2}{W^2 \rho_0^2 h}\left(pW^2+(\chi-\chi W^2+h W^2)\rho_0 \right)\,,\non\\
c_5=&\frac{\Wcs^2}{W^2 \rho_0^2 h}\left(\kappa  +\rho_0 \right)\,.
\end{align}
The system of Eqs.~\eqref{equation:EvolutionpRGRMHD},
\eqref{equation:EvolutionvRGRMHD},
\eqref{equation:EvolutionepsilonRGRMHD}, \eqref{equation:LieERGRMHD},
\eqref{equation:LieBRGRMHD}, \eqref{equation:LiePsiRGRMHD},
\eqref{equation:LiePhiRGRMHD}, and~\eqref{equation:LieQRGRMHD} is
identical to the system of evolution equations in 
Ref.~\cite{DioAliRez15},
as was explicitly checked up to source terms.

%%%%%%%%%%%%%%%%%%%%%%%%%%%%%%%%%%%%%%%%%%%%%%%%%%%%%%%%%%%%%
\subsection{Analysis with evolution of~$q$}
%%%%%%%%%%%%%%%%%%%%%%%%%%%%%%%%%%%%%%%%%%%%%%%%%%%%%%%%%%%%%

In this subsection, we want to analyze the characteristic 
structure of
equations used in Refs.~\cite{Kom07,DumZan09,PalLehReu09,Pal12,Miz13}. 
As always,
we~$2+1$ decompose the equations, this time using an arbitrary unit
spatial 1-form~$s_a, s_as^a=1, s_a n^a=0$ and denote the orthogonal
projector by~${\perpqs^{ b}\!_{ a}}:={\gamma^b{}_a}-s^bs_a$. Taking the
state vector to
be~$\boldsymbol{\rm{U}}=(p,v_a,\varepsilon,q,E_a,B_a,\psi,\phi)^T$, we
write the Eqs.~\eqref{equation:EvolutionpRGRMHD},
\eqref{equation:EvolutionvRGRMHD},
\eqref{equation:EvolutionepsilonRGRMHD}, \eqref{equation:LieQRGRMHD},
\eqref{equation:LieERGRMHD}, \eqref{equation:LieBRGRMHD},
\eqref{equation:LiePsiRGRMHD}, and \eqref{equation:LiePhiRGRMHD} for
the~$14$ components of~$\boldsymbol{\rm{U}}$ in matrix form:
\begin{align}
\mathbf{A}^{\textrm n} \Lie_n\mathbf{U}=\mathbf{A}^p D_p \mathbf U
+ \boldsymbol{\mathcal{S}}\,.
\end{align}
The form of the matrices is easily obtained from the
system of equations and so is not explicitly given. A simple~$2+1$
decomposition of this equation yields the principal symbol in the form
\begin{align}
\mathbf{P}^{s}=\mathbf{A}^s=
\begin{pmatrix}
\mathbf A_{6\times6}  &\mathbf B_{6\times8} \\
\mathbf  0_{8\times6} & \mathbf  C_{8\times8}\\
\end{pmatrix}\,,\label{equation:PrincipalSymbolRGRMHD}
\end{align}
where~$\mathbf B_{6\times8}$ contains the coefficients of spatial
derivatives with respect to the variables~$(E_a$,$B_a$,$\psi$,$\phi)$
in the time evolution of~$(p$,$v_a$,$\varepsilon$,$q)$. The
matrix~$\mathbf C_{8\times8}$ is the submatrix of the electromagnetic
variables~$(E_a, B_a, \psi, \phi)$. The matrix~$\mathbf{A}_{6\times6}$
can be written as
\begin{align}
\mathbf A_{6\times6} =\begin{pmatrix}
\mathbf A_{5\times5} & \mathbf 0_{5\times1}  \\
\mathbf A_{1\times5} & \mathbf  -v^s
\end{pmatrix}\,,
\end{align}
with~$\mathbf A_{5\times5}=\mathbf P^s_{\rm{HD}}$ the principal symbol
of the pure hydrodynamical sector, explicitly given
by~\eqref{equation:PrincipalSymbolLowerCaseHDsAppendix} and
\begin{align}
\mathbf A_{1\times5}=\begin{pmatrix} -\frac{\p J^s}{\p p} &
-s_c\frac{\p J^s}{\p v_c}&-\perpqs^{B}\!\!_A\frac{\p J^s}{\p v_A} &
-\frac{\p J^s}{\p \varepsilon}
\end{pmatrix}\,.
\end{align}
Since the principal symbol~\eqref{equation:PrincipalSymbolRGRMHD} is
block triangular, the eigenvalues are given by those 
of~$\mathbf{A}_{6\times6}$ and~$\mathbf{C}_{8\times8}$, these are
\begin{align} 
\mathbf A_{6\times6}:\qquad \lambda&=-v^{s}\,, \ \textrm{(multiplicity
  4})\,,\non\\ \lambda&=\lambda_{(\pm)}\,, \ \ \ \textrm{[see
  \eqref{equation:EigenvaluesForvsHD}]}\,,\\ \mathbf C_{8\times8}:\qquad
\lambda&=\pm 1\,, \ \ \ \textrm{(multiplicity 4})\, .
\end{align}
Continuing the characteristic analysis, it can be shown that only~$13$
eigenvectors exist. The eigenspace of the eigenvalue~$\lambda=-v^{s}$,
with algebraic multiplicity 4, has only geometric multiplicity
3. For example, the linearly independent right eigenvectors can be
chosen as
\begin{align}
\begin{pmatrix}
0 \\
0 \\
0_B \\
1 \\
0 \\
\mathbf  0_{8\times1} 
\end{pmatrix},\qquad
\begin{pmatrix}
0 \\
0 \\
0_B \\
0 \\
1 \\
\mathbf  0_{8\times1} 
\end{pmatrix}, \qquad
\begin{pmatrix}
0 \\
0 \\
\epsns_{\ortb \ortc} \perpqs^{\ortc}\!\!_{\orta}\frac{\p J^s}{\p v_{\orta}} \\
0 \\
0 \\
\mathbf  0_{8\times1} 
\end{pmatrix}\,,
\end{align}
where we defined the antisymmetric lowercase two-Levi-Civit\`a tensor
for~$s_a$ as~${\epsns^{\orta\ortb}=n_cs_d\,\perpqs^{\orta}\!_a\perpqs^{\ortb}\!_b\epsilon^{cdab}}$.
This result is contrary to an earlier analysis presented
in Ref.~\cite{CorCerIba12b}. 
The earlier analysis is erroneous since the
three vectors called~$r_{{\lambda_H}_0}$ corresponding to~$\lambda
=-v^s$ are not eigenvectors. The explicit error is that the ninth
component of these vectors may not be zero, since they produce 
cross-terms with the $A_{qH}$ (corresponding 
to our $\mathbf{A}_{1 \times
  5}$ part of the principal symbol). To substantiate our result, we
performed a Jordan decomposition of the principal
symbol~\eqref{equation:PrincipalSymbolRGRMHD}. The Jordan normal
form~$\mathbf J[\mathbf P^s]$
of~\eqref{equation:PrincipalSymbolRGRMHD} can be written as
\begin{align}
\mathbf J[\mathbf P^s]=\text{diag}(\lambda_{(+)} ,
\lambda_{(-)},\mathbf{J}_{v^s},-\mathbbm{1}_{2 }
v^s,-\mathbbm{1}_{4},\mathbbm{1}_{4})\,,
\end{align}
with
\begin{align}
\mathbf{J}_{v^s}=\begin{pmatrix}
-v^s & 1 \\
0 & -v^s
\end{pmatrix}\,,
\end{align}
and confirms that~$\mathbf{P}^s$ is not diagonalizable. Therefore, the
system of equations is weakly hyperbolic and has an ill-posed IVP.

It should be mentioned that for the special
subcase~$\tilde{J}^a\equiv0$ the system is strongly hyperbolic.  More
generally, if~$\tilde{J}^a$ does not depend on~$v^b$ (more precisely,
if~$\frac{\p J^s}{\p v_\orta}$
vanishes identically), then the system is strongly hyperbolic. For the
current in Eq.~\eqref{equation:DefinitionJRGRMHDLiterature}, these
two cases coincide.

%%%%%%%%%%%%%%%%%%%%%%%%%%%%%%%%%%%%%%%%%%%%%%%%%%%%%%%%%%%%%
\subsection{Analysis without evolution of~$q$}
%%%%%%%%%%%%%%%%%%%%%%%%%%%%%%%%%%%%%%%%%%%%%%%%%%%%%%%%%%%%%

Next, we consider the system but suppress the~$q$ variable. This
analysis is for the system of equations used
in Refs.~\cite{BucZan13,QiaFenNob17,DioAliPal13,DioAliRez15}. 
We set~$\psi$
to zero, the set of equations reduce to~$12$ evolution
equations~\eqref{equation:EvolutionpRGRMHD},~\eqref{equation:EvolutionvRGRMHD},~\eqref{equation:EvolutionepsilonRGRMHD},~\eqref{equation:LieERGRMHD},~\eqref{equation:LieBRGRMHD},
and~\eqref{equation:LiePhiRGRMHD} for the components of the state
vector~$\boldsymbol{\rm{U}}=(p, v_a, \varepsilon, E_a, B_a, \phi)^T$,
and Eq.~\eqref{equation:LiePsiRGRMHD} becomes the standard Gauss
constraint~$D_aE^a=q$. This equation is not a constraint in the PDE
sense; it is now rather the definition used to obtain~$q$.

Since now we do not evolve~$q$ by the conservation of charge
equation~\eqref{equation:LieQRGRMHD}, we have to replace all~$q$'s
by~$D_aE^a$. Therefore, in
Eqs.~\eqref{equation:EvolutionpRGRMHD},~\eqref{equation:EvolutionvRGRMHD},~\eqref{equation:EvolutionepsilonRGRMHD},
and~\eqref{equation:LieERGRMHD}, we replace~$J^a$ by use of
Eq.~\eqref{equation:DefinitionJRGRMHD} with
\begin{align}
J^a=v^a\gamma^{pc}D_pE_c +\tilde{J}^a\,,
\end{align}
where the first term will contribute to the principal symbol.  

Writing the system of equations in matrix form and decomposing
against~$s_a$, $s_as^a=1$, and~$\perpqs_{\ a}^{b}$, we obtain
\begin{align}
\mathbf{A}^{\textrm n} \Lie_n\mathbf{U}=\mathbf{A}^p D_p \mathbf U + \boldsymbol{\mathcal{S}}\,,
\end{align}
with the principal symbol
\begin{align}
\mathbf{P}^s=\mathbf{A}^s=
\begin{pmatrix}
\mathbf A_{5\times5}  &\mathbf B_{5\times7} \\
\mathbf  0_{7\times5} & \mathbf  C_{7\times7}\\
\end{pmatrix}.\label{equation:PrincipalSymbol12x12RGRMHD}
\end{align}
Again,~$\mathbf B_{5\times7}$ contains the coefficients of spatial
derivatives with respect to the variables~$(E_a,B_a,\phi)$ in the time
evolution of~$(p, v_a, \varepsilon)$,
and~$\mathbf{A}_{5\times5}=\mathbf P^s_{\rm{HD}}$ is the principal
symbol of the pure hydrodynamical sector, explicitly given
in~\eqref{equation:PrincipalSymbolLowerCaseHDsAppendix}. The
matrix~$\mathbf{C}_{7\times7}$ is the submatrix of the
electromagnetic variables~$(E_a$,$B_a$,$\phi)$, explicitly given by,
\begin{align}
\mathbf C_{7\times7}=\begin{pmatrix}
-v^s &0 &0 &0 &0 &0 &0 \\
-v^{q_1}  &0 &0 &0 &0 &-1&0 \\
-v^{q_2} &0 &0 &0 &1 &0 &0 \\
0    &0 &0 &0 &0 &0 &-1 \\
0    &0 &1 &0 &0 &0 &0 \\
0    &-1&0 &0 &0 &0 &0 \\
0    &0 &0 &-1&0 &0 &0
\end{pmatrix}.
\end{align}

The 12 eigenvalues of~\eqref{equation:PrincipalSymbol12x12RGRMHD}
are given by the ones of~$\mathbf A_{5\times5}$
and~$\mathbf{C}_{7\times7}$; these are
\begin{align}
\mathbf A_{5\times5}:\qquad \lambda&=-v^{s}\,,\ \textrm{(multiplicity 3})\,,\non\\  
\lambda&=\lambda_{(\pm)}\,,\ \ \ \textrm{[see \eqref{equation:EigenvaluesForvsHD}]}\,,\\
\mathbf C_{7\times7}:\qquad \lambda&=\pm 1\,, \ \ \textrm{(multiplicity 3})\,,\non\\
\lambda&=-v^s\,,\ \textrm{(multiplicity 1})\,.
\end{align}

As in the previous case, the eigenspace of the
eigenvalue~$\lambda=-v^{s}$ with algebraic multiplicity 4 has only
geometric multiplicity 3. A set of right eigenvectors is
\begin{align}
\begin{pmatrix}
\mathbf  0_{2\times1} \\
1 \\
\mathbf  0_{9\times1} 
\end{pmatrix},\qquad
\begin{pmatrix}
\mathbf  0_{3\times1} \\
1 \\
\mathbf  0_{8\times1} 
\end{pmatrix},\qquad
\begin{pmatrix}
\mathbf  0_{4\times1} \\
1 \\
\mathbf  0_{7\times1}
\end{pmatrix} \,.
\end{align}
The Jordan normal form~$\mathbf J[\mathbf P^s]$
of~\eqref{equation:PrincipalSymbol12x12RGRMHD} is given by
\begin{align}
\mathbf J[\mathbf P^s]=\text{diag}(\lambda_{(+)} ,
\lambda_{(-)},-\mathbbm{1}_{2 }
v^s,\mathbf{J}_{v^s},-\mathbbm{1}_{3},\mathbbm{1}_{3})\,,
\end{align}
with
\begin{align}
\mathbf{J}_{v^s}=\begin{pmatrix}
-v^s & 1 \\
0 & -v^s
\end{pmatrix}\,.
\end{align}
Therefore, the system of equations is also only weakly hyperbolic when
the charge density variable~$q$ is not evolved. The result also holds
for~$\phi=0$, so that equation~\eqref{equation:LiePhiRGRMHD} reduces
to the usual constraint~$D_aB^a=0$, and we evolve the 11
variables~$(p, v_a, \varepsilon, E_a, B_a)$. In this case, a pair of
eigenvalues~$\lambda=\pm 1$ changes to the single
eigenvalue~$\lambda=0$. 

For the special subcase~$\tilde{J}^a\equiv0$, the system is strongly
hyperbolic. This happens because in that case~$q$ is algebraically
related to the rest mass density~$\rho_0$ and may thus be seen as a
source term. Then, the algebraic multiplicity of~$\lambda=-v^s$ changes
to 3, and a complete set of eigenvectors can be found. From the
physical point of view, the relevance of this model to compact
binaries is, however, unclear to us. Note that we have not considered in
this section general formulations of RGRMHD and that our calculations
apply only to those formulations implemented.  It is possible that
these systems can be cured by a carefully chosen constraint addition.

A final comment is reserved for the special case of charged dust. In
this model,~$p=\varepsilon=0$, and the charge density is proportional
to the mass density with constant of proportionality equal to the
specific charge. The system of equations for
variables~$(\rho_0,v_i,E_i,B_i)$ decouples into two parts: first, the
evolution equations for~$(\rho_0,v_i)$, which were already found to be
weakly hyperbolic in Sec.~\ref{section:Dust}, and second the
electromagnetic equations, which can be given in a symmetric
 hyperbolic
form; see~\cite{AlcDegSal09}. The whole system is thus only weakly
hyperbolic. In Ref.~\cite{PerCar10}, it is shown that a different
formulation of charged dust using~$(v_i,E_i,B_i)$ as variables is
strongly hyperbolic. In the authors' system,~$\rho_0$ 
is obtained by the Gauss
constraint equation relating the divergence of the electric field
 with
the charge density. Under this treatment, however, the minimal
 coupling
condition with the gravitational field equations, see
Eq.~\eqref{equation:principalsymbolEinstein}, breaks. Therefore,
away from the Cowling approximation, the full coupled system must be
considered fresh.

%%%%%%%%%%%%%%%%%%%%%%%%%%%%%%%%%%%%%%%%%%%%%%%%%%%%%%%%%%%%%
\section{Conclusion}\label{section:Conclusion}
%%%%%%%%%%%%%%%%%%%%%%%%%%%%%%%%%%%%%%%%%%%%%%%%%%%%%%%%%%%%%

Motivated by applications in numerical relativity, and in particular
by the wish for the computation of accurate gravitational waveforms in
compact binary spacetimes, we have revisited hyperbolicity of several
popular relativistic fluid models. Our main technical achievement has
been to bring about the DF formalism~\cite{Hil15,HilHarBug16} to
these matter models in a systematic way. This allowed us to arrive at
a tractable form of even GRMHD, which is notorious for its complicated
characteristic structure. The key idea was to use a Lagrangian frame
in the analysis. In this frame the principal symbol takes the simplest
possible form and can be easily analyzed. Afterward, we could
translate the results into the desired frame using the developed
formalism.

Along the way, we arrived at several disconcerting results. That a
commonly used formulation of GRMHD, plus those of RGRMHD, is only
weakly hyperbolic is clearly a huge shortcoming that must be overcome
if we are ever to obtain numerical results with meaningful error
estimates for binary systems involving magnetic fields. One might
wonder why the problem has not been discovered earlier on the basis of
numerical work. The effect of ill-posedness on the errors in
approximation is a subtle issue, however, and without very careful
convergence testing can be easily overlooked, particularly when
considering very complicated data. One aspect of this is that
canonical test beds often focus on one-dimensional tests, which would
not be suitable for identifying 
the issue identified for GRMHD. That said,
it is important to realize that, although we are motivated 
by numerical
applications, the analysis presented here is for the continuum PDE
system. Thus, no numerical method, no matter how sophisticated, can
circumvent our results, and therefore the equations {\it do} have to
be altered. An obvious step in this direction would be to use our
prototype algebraic constraint free formulation of GRMHD, which is at
least strongly hyperbolic. This formulation cannot be written in
flux-balance law form, but it fails only by the addition of constraint
terms, so there is reason to be optimistic that existing codes can be
easily modified to overcome this worst possible problem of
ill-posedness of the IVP. There is hope that formulations using the
four-potential, or those with divergence cleaning, are strongly
hyperbolic. Thus, another possibility would be to affirm this and, if
so, move wholesale to such systems. For RGRMHD, more work is needed.

We started the paper by stressing the well-known fact that the stellar
surface is {\it also} a terrible problem in numerical relativity. Even
in the case of GRHD, which does not have the same problems as
flux-balance law GRMHD, the formally singular nature of the surface
prevents clean convergence in simulations of even the most simple
spacetimes. We expect that before this problem can be solved a much
deeper understanding of the underlying initial free boundary value
problem will be needed. So far, nothing in our treatment does anything
whatsoever to alleviate this. We do think, however, that by carefully
choosing the complete uppercase frame it may be possible to make
progress by building on the present work. Sadly, this 
remains a distant
goal.

%%%%%%%%%%%%%%%%%%%%%%%%%%%%%%%%%%%%%%%%%%%%%%%%%%%%%%%%%%%%%
\acknowledgments
%%%%%%%%%%%%%%%%%%%%%%%%%%%%%%%%%%%%%%%%%%%%%%%%%%%%%%%%%%%%%

We are grateful to S. Bernuzzi, I. Cordero-Carri\'on, T. Font,
H. Markakis and R. Meinel for helpful conversations. We particularly
want to thank K. Dionysopoulou both for helpful feedback and for her
patient explanation of RGRMHD. The majority of our calculations was
performed in {\it Mathematica} 10 using the~\texttt{xTensor}
package~\cite{xAct_web_aastex}. Our notebooks can be downloaded
from Ref.~\cite{HilWebsite_aastex}. This work was partially 
supported by
the DFG funded Graduiertenkolleg 1523/2, the FCT (Portugal) IF 
Program No. IF/00577/2015 and the GWverse COST action Grant
 No.~CA16104.

%%%%%%%%%%%%%%%%%%%%%%%%%%%%%%%%%%%%%%%%%%%%%%%%%%%%%%%%%%%%%
\appendix
%%%%%%%%%%%%%%%%%%%%%%%%%%%%%%%%%%%%%%%%%%%%%%%%%%%%%%%%%%%%%

%%%%%%%%%%%%%%%%%%%%%%%%%%%%%%%%%%%%%%%%%%%%%%%%%%%%%%%%%%%%%
\section{GRHD using the boost vector}\label{section:HDforv}
%%%%%%%%%%%%%%%%%%%%%%%%%%%%%%%%%%%%%%%%%%%%%%%%%%%%%%%%%%%%%

It holds that 
\begin{align}
\gamma^{b}{}_a\Lie_n \vhat_b=W \gB^{b}{}_a\Lie_n
v_b+\vhat_a\Kn_{cd}\vhat^c\vhat^d\,. 
%,\qquad \nabla_cn_d=-\Kn_{cd}-n_c a_d\, 
\end{align}
Using the state vector~$\textbf{U}=(p,v_a,\varepsilon)$ and an
arbitrary unit spatial 1-form~$s_a$ with~$s_as^a=1,\ s_a n^a=0$ and
denoting the orthogonal projector by~$\perpqs_{\ a}^{
  b}:=\gamma_{\ a}^{ b}-s^bs_a$, the system of equations reads
\begin{align}
(\Lie_n \mathbf{U})_{s,\,A}\simeq\mathbf{P}^{s} (D_s \mathbf{U})_{s,\,B}
\end{align}
and the principal symbol~$\mathbf{P}^{s}$ is given by
\begin{align}
\begin{pmatrix}
\Wcs^2(\cs^2-1)v^{s} &-\Wcs^2\cs^2 \rho_0 h & 0^{\ortb}& 0\\
-\frac{1+(\cs^2-1) (v^{s})^2 \Wcs^2}{W^2 \rho_0 h} &\Wcs^2(\cs^2-1)v^{s} &0^{\ortb}  & 0\\
\frac{(1-\cs^2)  \Wcs^2}{W^2 \rho_0 h}v^sv_{\orta}& \frac{\cs^2\Wcs^2}{W^2}v_{\orta}&
-v^{s}\,\perpqs^{\ortb}\!\!_{\orta}& 0_{\orta}\\
\frac{p \Wcs^2}{W^2 \rho_0^2 h} v^{s} &-\frac{p \Wcs^2}{ \rho_0} &0^{\ortb} & -v^{s}\\ 
\end{pmatrix}\,,\label{equation:PrincipalSymbolLowerCaseHDsAppendix}
\end{align}
with eigenvalues for material and acoustic waves
\begin{align}
\lambda_{(0,1,2)}&=-v^{s}\,,\non\\
 \lambda_{(\pm)}&=
-\frac{1}{1-\cs^2v^2}\left((1-\cs^2)v^{s}\right.\non\\
&\pm\left.\frac{\cs}{W}
\sqrt{(1- \cs^2v^2)-(1-\cs^2)(v^s)^2}\right),
\label{equation:EigenvaluesForvsHD}
\end{align}
respectively. They coincide with the
literature~\cite{FonIbaMar94}. The corresponding left eigenvectors are
given by
\begin{align}
&\begin{pmatrix}
-\frac{p}{\cs^2 \rho_0^2 h}&0 &0^{\orta} &1
\end{pmatrix},\non \\
&\begin{pmatrix}
\frac{1}{W^2 \rho_0 h}v_{\ortc}&v^sv_{\ortc} &(1-(v^s)^2)\,\perpqs^{\orta}\!_{\ortc} &0 
\end{pmatrix},\non \\
&\begin{pmatrix}
\pm \frac{\sqrt{(1- \cs^2v^2)-(1-\cs^2)(v^s)^2}}{\cs \rho_0 h W}&1 &0^{\orta} &0 
\end{pmatrix}.
\end{align}
For the same variables and order, the right eigenvectors are 
\begin{align}
\begin{pmatrix}
0 \\
0 \\
0_{\ortb} \\
1
\end{pmatrix},
\begin{pmatrix}
0 \\
0 \\
\perpqs^{\ortc}\!_{\ortb} \\
0
\end{pmatrix},
\end{align}
and
\begin{align}
\begin{pmatrix}
\frac{\cs^2 \rho_0^2 h}{p}(1-(v^s)^2) \\
\pm \frac{\cs \rho_0}{p W}\sqrt{(1- \cs^2v^2)-(1-\cs^2)(v^s)^2}(1-(v^s)^2) \\
-\frac{\cs \rho_0}{p W} \left(\frac{\cs}{W} \pm v^s
\sqrt{(1- \cs^2v^2)-(1-\cs^2)(v^s)^2}\right) v_{\ortb}\\
1-(v^s) ^2
\end{pmatrix}
\end{align}
in agreement with the ones given in Ref.~\cite{FonIbaMar94} up to the
chosen set of variables and the spatial vector~$s^a$. The
characteristic variables corresponding to the speeds~$\{
\lambda_{(0,1,2)}\text{,} \lambda_{(\pm)} \}$ are given by
\begin{align}
\hat{\text{U}}_0&= \delta \varepsilon-\frac{p}{\cs^2 \rho_0^2 h}\delta
  p,\non \\
  \hat{\text{U}}_A &=(\delta v)_{{\orta}}+v^s(v_{\orta}(\delta
  v)_s-v_s(\delta v)_{\orta})+\frac{1}{\rho_0 h W^2}\vhat_{\orta}\delta
  p,\non \\ \hat{\text{U}}_{\pm}&=(\delta v)_{s}\pm \frac{\sqrt{(1-
      \cs^2v^2)-(1-\cs^2)(v^s)^2}}{\cs \rho_0 h W}\delta p\,.
\end{align} 

%%%%%%%%%%%%%%%%%%%%%%%%%%%%%%%%%%%%%%%%%%%%%%%%%%%%%%%%%%%%%
\bibliographystyle{unsrt}
\bibliography{Hydro_DF.bbl}{}
%%%%%%%%%%%%%%%%%%%%%%%%%%%%%%%%%%%%%%%%%%%%%%%%%%%%%%%%%%%%%

%%%%%%%%%%%%%%%%%%%%%%%%%%%%%%%%%%%%%%%%%%%%%%%%%%%%%%%%%%%%%
\end{document}